\documentclass[fleqn,usenatbib,useAMS]{mnras}
\usepackage{graphicx}	
\usepackage{amsmath}	
\usepackage{amssymb}	
\usepackage{multicol}        
\usepackage{bm}		
\usepackage{pdflscape}	

\usepackage{verbatim,xspace}
\usepackage{import}

\usepackage[T1]{fontenc}
\usepackage{ae,aecompl}
\usepackage{newtxtext,newtxmath}

\newcommand{\ro}[1]{\ensuremath{\textrm{#1}}}
\newcommand{\ten}[1]{\ensuremath{\times 10^{#1}}}

\newcommand{\Msol}{\ensuremath{M_{\odot}}\xspace}

\newcommand{\df}{\ensuremath{~ \ro{d} }}
\newcommand{\dd}{\ensuremath{\ro{d} }}

\newcommand{\eagle}{\textsc{eagle}\xspace}

\newcommand{\Om}{\ensuremath{\Omega_\ro{m}}\xspace}
\newcommand{\Omr}{\ensuremath{\Omega_\ro{r}}\xspace}
\newcommand{\Omc}{\ensuremath{\Omega_\ro{c}}\xspace}
\newcommand{\Omb}{\ensuremath{\Omega_\ro{b}}\xspace}
\newcommand{\Omk}{\ensuremath{\Omega_\ro{k}}\xspace}
\newcommand{\OmL}{\ensuremath{\Omega_\Lambda}\xspace}

\newcommand{\rhoL}{\ensuremath{\rho_\Lambda}\xspace}
\newcommand{\rhob}{\ensuremath{\rho_\ro{b}}\xspace}
\newcommand{\rhoc}{\ensuremath{\rho_\ro{c}}\xspace}
\newcommand{\rhom}{\ensuremath{\rho_\ro{m}}\xspace}
\newcommand{\ngam}{\ensuremath{n_\gamma}\xspace}
\newcommand{\rhor}{\ensuremath{\rho_\ro{r}}\xspace}

\newcommand{\rhok}{\ensuremath{\rho_k}\xspace}

\newcommand{\rhobz}{\ensuremath{\rho_{\ro{b},0}}\xspace}
\newcommand{\rhocz}{\ensuremath{\rho_{\ro{c},0}}\xspace}
\newcommand{\rhomz}{\ensuremath{\rho_{\ro{m},0}}\xspace}
\newcommand{\ngamz}{\ensuremath{n_{\gamma,0}}\xspace}
\newcommand{\rhorz}{\ensuremath{\rho_{\ro{r},0}}\xspace}
\newcommand{\rhokz}{\ensuremath{\rho_{k,0}}\xspace}
\newcommand{\rhocrz}{\ensuremath{\rho_{\ro{crit},0}}\xspace}

\newcommand{\xib}{\ensuremath{\xi_\ro{b}}\xspace}
\newcommand{\xic}{\ensuremath{\xi_\ro{c}}\xspace}
\newcommand{\xim}{\ensuremath{\xi_\ro{m}}\xspace}

\title[Galaxy Formation and the Multiverse]{Galaxy Formation Efficiency and the Multiverse Explanation of the Cosmological Constant with EAGLE Simulations}

\author[L. Barnes et al.]{%
Luke A. Barnes,$^{1,2}$\thanks{E-mail: luke.barnes@sydney.edu.au}  
Pascal J. Elahi,$^{3,4}$ 
Jaime Salcido,$^5$ 
Richard G. Bower,$^5$ \newauthor
Geraint F. Lewis,$^1$
Tom Theuns,$^5$
Matthieu Schaller,$^5$
Robert A. Crain,$^6$
Joop Schaye$^7$
\\ \\
$^1$ Sydney Institute for Astronomy, School of Physics, A28, The University of Sydney, NSW 2006, Australia \\
$^2$ Western Sydney University, School of Computing, Engineering and Mathematics, Locked Bag 1797, Penrith, NSW, 2751, Australia \\
$^3$ International Centre for Radio Astronomy Research, The University of Western Australia, 35 Stirling Highway, Crawley WA 6009, Australia\\
$^4$ ARC Centre of Excellence for All Sky Astrophysics in 3 Dimensions (ASTRO 3D)\\
$^5$ Institute for Computational Cosmology, Department of Physics, University of Durham, South Road, Durham, DH1 3LE, UK\\
$^6$ Astrophysics Research Institute, Liverpool John Moores University, 146 Brownlow Hill, Liverpool L3 5RF, UK\\
$^7$ Leiden Observatory, Leiden University, P.O. Box 9513, 2300 RA Leiden, The Netherlands
}

\date{Last updated Today; in original form Yesterday}

\pubyear{2017}

\begin{document}
\label{firstpage}
\pagerange{\pageref{firstpage}--\pageref{lastpage}}
\maketitle

\begin{abstract}
Models of the very early universe, including inflationary models, are argued to produce varying universe domains with different values of fundamental constants and cosmic parameters. Using the cosmological hydrodynamical simulation code from the \eagle collaboration, we investigate the effect of the cosmological constant on the formation of galaxies and stars. We simulate universes with values of the cosmological constant ranging from $\Lambda = 0$ to $\Lambda_0 \times 300$, where $\Lambda_0$ is the value of the cosmological constant in our Universe. Because the global star formation rate in our Universe peaks at $t = 3.5$ Gyr, before the onset of accelerating expansion, increases in $\Lambda$ of even an order of magnitude have only a small effect on the star formation history and efficiency of the universe. We use our simulations to predict the observed value of the cosmological constant, given a measure of the multiverse. Whether the cosmological constant is successfully predicted depends crucially on the measure. The impact of the cosmological constant on the formation of structure in the universe does not seem to be a sharp enough function of $\Lambda$ to explain its observed value alone.
\end{abstract}



\section{Introduction} 
Cosmological inflation, it has been argued, naturally predicts a vast ensemble of varying universe domains\footnote{For simplicity, we call such regions ``universes''.}, each with different cosmic conditions and even different fundamental constants \citep[see the review of][]{Linde2017}. A typical mechanism for generating these universes is as follows \citep{2007JPhA...40.6811G}. The inflaton field undergoes quantum fluctuations, and so we might expect some parts of the universe to still be inflating while other parts have entered a post-reheating ``big bang'' phase. The universe as a whole consists of post big-bang universes filled with ordinary matter and radiation, surrounded by an ever-inflating background.

In evaluating such models, predicting what we would expect to observe is necessarily tied to where observers are formed in the multiverse. In this instance, anthropic reasoning is inevitable \citep{1974IAUS...63..291C,1979Natur.278..605C,1983PrPNP..10....1D,1986acp..book.....B}. With different cosmic and fundamental constants in different parts of the multiverse, the values we expect to observe are unavoidably tied to their ability to support the complexity required by life.

These multiverse models could successfully explain the fine-tuning of the universe for life: small changes in their values can suppress or erase the complexity upon which physical life as we know it, or can imagine it, depends. The scientific literature on the fine-tuning of the universe for life has been reviewed in \citet{2000RvMP...72.1149H,2012PASA...29..529B,2013RvMP...85.1491S,2014arXiv1409.2959M,Lewis2016}. For example, as pointed out by \citet{1981RSPSA.377..147D,Sakharov1984,Linde1984,1985NuPhB.249..332B,Linde1987,1987PhRvL..59.2607W,1989RvMP...61....1W}, only a small subset of values of the cosmological constant ($\Lambda$) permit structure to form in the universe at all. Universes in which the cosmological constant is large and positive will expand so rapidly that gravitational structures, such as galaxies, are unable to form. Large negative values will cause space to recollapse rapidly, also preventing the formation of galaxies.

If inflation creates a huge number of variegated universe domains, then a structure-permitting value of the cosmological constant will probably turn up \emph{somewhere}. Any observers will see a universe with at least some structure. In thus way, the seemingly improbable suitability of our universe for life is rendered more probable.

As \citet{1987PhRvL..59.2607W} noted, we can test a particular multiverse model via its prediction of the distribution of universe properties. Observers will inhabit universes drawn in a highly-biased way from the population of universes, but we can  calculate the typical properties of a universe that contains observers. In this way, we can calculate the likelihood of our observations, and so compare multiverse models. For example, a model in which 99\% of observers measure a value of the cosmological constant as large as our value should (other things being equal) be preferred over a model in which only 1\% of observers make such a measurement. Whether these consistency tests can give absolute (rather than just relative) support to the idea of a multiverse is the subject of some debate \citep{2014Natur.516..321E,2017arXiv170401680B}.

To test the relative merits of multiverse models in this way, we need to know how life, or at least the cosmic structures that are the likely preconditions for life, depend on the fundamental constants of nature and cosmic parameters. In the case of the cosmological constant, the large-scale structure of the universe is most directly affected. Galaxies are the sites of star formation, and stars provide both a steady source of energy and the heavier elements from which planets and life forms are made.

Within an anthropic approach, we can also shed light on the coincidence problem: we live at a time in the universe when the energy density of the cosmological constant and the energy density of matter are within a factor of two of each other \citep{2007ApJ...671..853L}. The coincidence problem has motivated a search to alternative modification to gravity that might explain the value of the cosmological constant more naturally. Although, alternative models, such as quintessence can explain why the relative densities of matter and cosmological constant densities track each other, fine tuning of the model parameters is still required to explain their observed similarity \citep{1999PhRvL..82..896Z,1999PhLB..459..570Z,2000PhRvL..85.5276D,2003PhRvD..67h3513C}.

Investigations of the effect of the cosmological constant on galaxy formation have thus far relied on analytic models of increasing levels of sophistication. \citet{1995MNRAS.274L..73E} located galaxies at the peaks of the smoothed density field of the universe, and found that --- assuming that the cosmological constant is positive, observers should expect to see $\OmL \approx 0.67 - 0.9$. \citet{2007MNRAS.379.1067P} extended this approach to negative values of the cosmological constant, finding a significant probability that $\OmL < 0$ is observed. These approaches have been extended by \citet{2000PhRvD..61h3502G,2000PhRvD..61b3503G,2006PhRvD..73b3505T,2009PhRvD..79f3506B,2010PhRvD..81f3524B,2016PhRvL.116h1301P,2017MNRAS.464.1563S,2017JCAP...03..021A}.

The modern approach to galaxy formation uses supercomputer simulations that incorporate the effects of gravity, gas pressure, gas cooling, star formation, black hole formation, and various kinds of feedback from stars and black hole accretion. It has been long known that feedback is very important to explaining the star formation history of our universe; models without feedback are too effective at forming stars, compared to observations \citep{1978MNRAS.183..341W,1986ApJ...303...39D,1991ApJ...379...52W,2015ARA&A..53...51S}. One of the key ingredients that has allowed this progress is the inclusion of realistic models for the impact of feedback from the growth of black holes. All successful models now demonstrate the need for Active Galactic Nuclei (AGN) as an additional source of feedback that suppresses the formation of stars in high-mass haloes \citep{2003ApJ...599...38B,2006MNRAS.365...11C,2006MNRAS.370..645B}. Although this idea was initially developed using semi-analytic models, this has now been confirmed in a wide range of numerical simulations \citep[eg.][]{2016MNRAS.463.3948D,2017MNRAS.465...32B,2017arXiv170302970P}.

Here, we will use the \eagle project's galaxy formation code to calculate the effect of the cosmological constant on the formation of structure in different post-inflation universes. Each of our models will be practically indistinguishable at early times, including nucleosynthesis and the epoch of recombination. Their histories diverge at later times due to the onset of cosmological constant-powered accelerating expansion. In Section \ref{S:code}, we describe the \eagle galaxy formation code and the suite of simulations that we have run. In Section \ref{S:global}, we describe the effect of changing the cosmological constant on the global accretion and star-forming properties of the universe. Section \ref{S:indiv} looks at the effect on an individual galaxy, and its relation to its environment. In Section \ref{S:multiverse}, we use our simulations to derive prediction from models of the multiverse.

\section{Galaxy Formation Simulation Code} \label{S:code}
The Virgo Consortium's \textsc{\eagle} project (\emph{Evolution and Assembly of GaLaxies and their Environment}) is a suite of hydrodynamical simulations that follow the formation of galaxies and supermassive black holes in cosmologically representative volumes of a standard $\Lambda$CDM universe.  The details of the code, and particularly the sub-grid models, are described in \citet{2015MNRAS.446..521S}, and are based on the models developed for OWLS \citep{2010MNRAS.402.1536S}, and used also in GIMIC \citep{2009MNRAS.399.1773C} and cosmo-OWLS \citep{2014MNRAS.441.1270L}. The simulations code models the effect of radiative cooling for 11 elements, star formation, stellar mass loss, energy feedback from star formation, gas accretion onto and mergers of supermassive black holes (BHs), and AGN feedback.

The initial conditions for the \eagle simulations were set up using a transfer function generated using CAMB \citep{2000ApJ...538..473L} and a power-law primordial power spectrum with index $n_s = 0.9611$. Particles were arranged in a glass-like initial configuration were displaced according to second-order Lagrangian perturbation theory \citep{2010MNRAS.403.1859J}.

Black holes are seeded in all dark matter haloes with masses greater than $10^{10} h^{-1} \Msol= 1.48 \ten{10} \Msol$. The halo finding algorithm is described in \citet{2015MNRAS.446..521S}; in short, the code regularly runs the friends-of-friends (FoF) finder \citep{Davis1985} with linking length 0.2 on the dark matter distribution. When analysing the simulations in following sections, we are interested in membership with any halo, rather than distinguishing substructures, so we use the FOF algorithm to identify haloes.

\subsection{Cosmological Parameters and Scale Factor} \label{Ss:cosmparam}

We need to choose the cosmological parameters for our simulation. The problem with the standard set of cosmological parameters $(\Omega_m,\Omega_\Lambda,\Omega_b,h)$ is that they are all time dependent. In the model universes that we will consider, there is no unique ``today'' at which we can compare sets of parameters. We follow \citet{2006PhRvD..73b3505T} by defining cosmological parameters that are constant in time. We use only one time-dependent parameter, which is cosmic time $t$. The constant parameters are listed in Table \ref{tab:params}. Note that the cosmological constant ($\Lambda$) and its associated energy density are related linearly, $\Lambda = 8 \pi G \rhoL / c^2$.

\begin{table*}
\begin{center}
\begin{tabular}{|l|ll|}
\hline
Parameter		& 				 & Measured value\\
\hline   
$\rhoL$	& Cosmological constant energy (mass) density				&$5.98 \ten{-27}$ kg m$^{-3}$\\
$\xib$	    & Baryon mass per photon $\rhob/\ngam$					&$1.01\ten{-36} $ kg m$^{-3}$\\
$\xic$	    & Cold dark matter mass per photon $\rhoc/\ngam$		&$5.43\ten{-36}$ kg m$^{-3}$\\
$\kappa$ & Dimensionless spatial curvature (in Planck units) $k/a^2 T_0^2$ 	    &$|\kappa| \lesssim 10^{-60} \approx 0$ \\
\hline
\end{tabular}
\caption{Free parameters in the FLRW model, defined so that they are constant in time, at least since very early times. The measured value derives from the \citet{2014A&A...571A..16P} cosmological parameters, as used by the \eagle project: $(\Omega_m,\Omega_\Lambda,\Omega_b,h,\sigma_8,n_s,Y) = (0.307,0.693,0.04825,0.6777,0.8288,0.9611,0.248)$.} \label{tab:params}
\end{center}
\end{table*}

How do we solve the Friedmann equations, given the dimensionless cosmological parameters in Table \ref{tab:params}, so that we can derive the usual cosmological parameters for the simulation? We have the freedom to choose ``today'', that is, we can rescale $a(t)$ to make $a(t_0) = 1$ for any time $t_0$. A useful way to proceed initially is to define $t_0$ to be the time at which the energy densities of the cosmological constant and matter are equal. Then, we calculate the matter densities,
\begin{align}
\rhomz &= \rhoL \\
\xim &\equiv \xib + \xic ~\Rightarrow \quad \rhobz = \frac{\xib}{\xim} \rhomz ~; \quad \rhocz = \frac{\xic}{\xim} \rhomz
\end{align}
Then, we calculate the photon number density at $t_0$, and from it the CMB temperature $(T_0)$ and the radiation (photons and neutrinos) energy density,
\begin{align}
\ngamz &= \rhobz / \xib = \rhocz / \xic = \rhomz / \xim \\
\ngamz &= \frac{2 \zeta(3)}{\pi^2} \left( \frac{k_B T_0}{\hbar c} \right)^3 \\
\rhorz &= g \frac{\pi^2}{30} k_B T_0 \left( \frac{k_B T_0}{\hbar c} \right)^3 \\
~ \textrm{where}\quad  g &= 2 + 2~\frac{7}{8}~3~\left( \frac{4}{3} \right)^{4/3} ~.
\end{align}
We can then solve the Friedmann equation,
\begin{align}
H^2 = \left( \frac{1}{a}\frac{\dd a}{\dd t} \right)^2 &= \frac{8 \pi G}{3} (\rhom + \rhor + \rhoL + \rhok) \\
 &= \frac{8 \pi G}{3} (\rhomz a^{-3} + \rhorz a^{-4} + \rhoL + \rhokz a^{-2}) \\
 \ro{where} \quad \rhok &= -\kappa \frac{3}{8 \pi G} \left( \frac{k_B T_0}{\hbar} \right)^2 ~.
\end{align}
We can calculate the critical density, $\rhocrz = 3 H_0^2/8 \pi G = \rhomz + \rhorz + \rhoL + \rhokz$ and then the usual cosmological parameters $\Om = \rhomz/\rhocrz$, and similarly for \Omr, \Omc, \Omb, \OmL,  and \Omk. With these parameters, FLRW codes can solve the Friedmann equations\footnote{There are two potential complications. If we consider a universe with no cosmological constant $(\rhoL = 0)$ then the choice of ``initial'' matter density is effectively arbitrary. Secondly, if the universe recollapses, then it may never reach the time at which $\rhomz = \rhoL$. The most general way to find some matter density at which we can apply the technique above is to write the Friedmann equation in terms of the CMB temperature $T_0$. We can then solve for the CMB temperature at turnaround $H(t) = 0$, and from this calculate the minimum matter density of the universe.}.

Having solved the Friedmann equations for $a(t)$, we can rescale to change the time of ``today'' to be any other time $(t_0')$: $a_\ro{new}(t) = a(t) / a(t_0')$, and recalculate the various density parameters appropriately. We will describe our choices for the normalisation of $a(t)$ in Sections \ref{Ss:testing}and \ref{Ss:sims}.

\subsection{Initial Conditions and Sub-Grid Physics}

We use the same initial conditions for each simulation. For the range of cosmological constants we consider here, there has been minimal effect on the evolution of the universe at the start of the simulation. Specifically, we use the same initial conditions for the SPH particles \emph{in physical coordinates}: in the \eagle code, like its GADGET ancestor, we need to convert code quantities into physical quantities taking into account the initial scale factor ($a_i$) and the Hubble parameter ($h$) of the original simulation: distance ($d_\ro{phys} = a_i h^{-1} d_\ro{code}$), velocity ($v_\ro{phys} = v_\ro{code} \sqrt{a_i}$), and mass ($m_\ro{phys} = h^{-1} m_\ro{code}$).

We must also be careful regarding parameters in our sub-grid physics recipes. The sub-grid physics of the \eagle code has been checked, and the necessary parameters rescaled as necessary to keep the same \emph{physical} values. We also discovered a few cases in which it was assumed that $\rhoL \neq 0$, which needed to be remedied for the test runs below.

Note the assumptions that we are making when we change the cosmological constant, but keep the physical parameters of the subgrid model unchanged. This is potentially worrisome, given that these parameters are often inferred, not from first principles, but by calibrating against observations of galaxy populations in our Universe. Our assumptions are twofold. First, we assume that the subgrid model is sufficiently sophisticated that it captures the relevant physics. For example, we assume that star formation in any cosmology occurs when the local density is sufficiently high. It is appropriate to apply such a model to other universes. Secondly, we assume that the parameters inferred from observations are the same as would be inferred from a first-principles calculation; they do not depend on the cosmological constant for such small-scale processes. For example, the \emph{local} matter density above which star formation occurs should only depend on conditions within 10-100pc scale molecular clouds, far below cosmological scales. We can plausibly use the same threshold for different cosmologies.

Using the same subgrid parameters would create a problem only if our overall cosmology is wrong, for it could be the case that we have inferred the wrong value of some subgrid parameter to partially compensate for an incorrect expansion history of the universe. In this case, of course, the entire \eagle simulation suite would need to be redone, as would almost every other cosmological simulation. We will leave that worry for another day.

\subsection{Testing our Modifications} \label{Ss:testing}

The freedom to choose ``today'' $t_0$ in our simulation gives us a way to confirm that our modifications are correct. Setting $\rhoL = 0$ and $\kappa = 0$, and noting that $\rhor$ is negligible for the time covered by the simulation, we  simulate structure formation in an Einstein de-Sitter (EdS) Universe. We can use this freedom to define three different sets of simulation initial conditions.
\begin{itemize}
\item[\textbf{A.}] The initial time of the simulation has the same scale factor as the corresponding Planck cosmology simulation, $a_{i,A} = 1/(1 + z_{i,A}) = 1/128$. We solve for the proper initial time $t_\ro{init}$ in the Planck cosmology, and then require that $a_\ro{EdS}(t)$ is normalised so that $a_\ro{Planck}(t_\ro{init}) = a_\ro{EdS}(t_\ro{init})$. This requires that we set Hubble parameter to $h_A = 0.375$.
\item[\textbf{B.}] We alter the initial redshift of the simulation so that `today' ($z = 0$, $a = 1$) is at $t_0 = 13.8$ Gyr. This requires that we set the initial redshift of the simulation to $z_{i,B} = 108$ and the Hubble parameter to $h_B = 0.4716$.
\item[\textbf{C.}] The Hubble parameter $h$ of the simulation has the same value as the corresponding Planck cosmology simulation, $h_\ro{Planck} = 0.6777$. Having found the time in the EdS universe when $h_C = 0.6777$, we normalise the scale factor so that $a_\ro{EdS} = 1$ at that time. This requires that we set the  initial redshift to $z_{i,C} = 85.4$.
\end{itemize}

The simulations A, B and C are trying to solve the same \emph{physical} problem, and should produce the same properties of the universe as a function of proper time. If we have not correctly accounted for factors of $h$ \citep{2013PASA...30...52C} or confused comoving/physical quantities in our calculations, then these two simulations should diverge. Inevitably, there will be numerical differences: because the ``time'' variable of the simulation is actually $\log a$, the time stepping is not identical.

\begin{figure} \centering
	\includegraphics[width=0.45\textwidth]{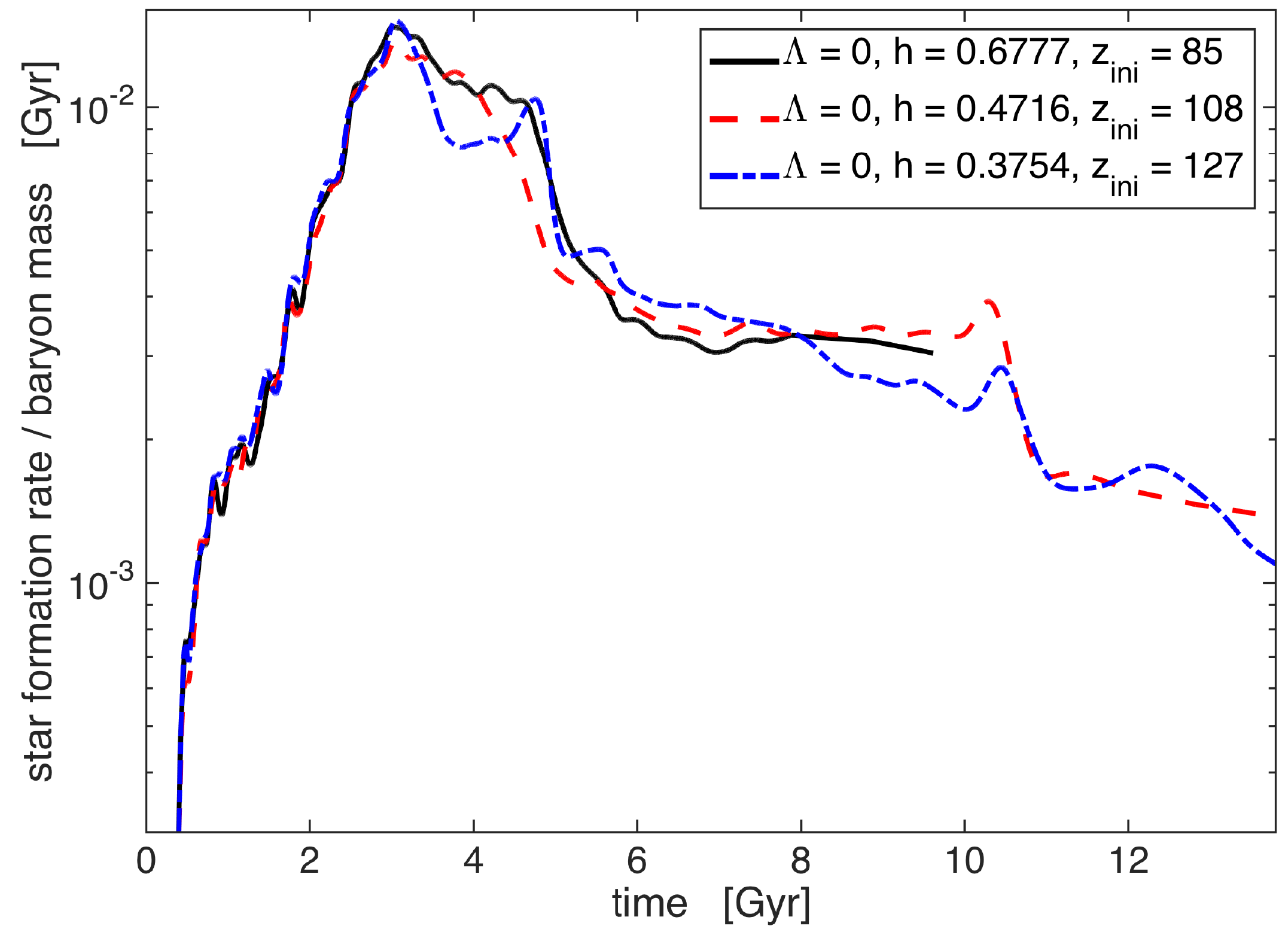}
	\caption{The star formation rate efficiency (that is, star formation rate divided by the total baryon mass in the simulation box), for three simulations with $\Lambda = 0$ but different choices for ``today'' (at which $a(t_0) = 1$). While there is scatter between the different simulations, they show an overall star formation history that is consistent. The scatter is comparible in magnitude to that caused by using a different seed for the random number generator associated with subgrid physics.} \label{fig:rate_star_EdS}
\end{figure}

Figure \ref{fig:rate_star_EdS} shows the star formation rate efficiency (that is, SFR divided by the total baryon mass in the simulation box), for three simulations (A, B and C) with $\Lambda = 0$. While there is scatter between the different simulations, they show an overall star formation history that is consistent. We have also run simulations that alter the seed for the random number generator. The scatter that this produces for a single set of parameters is similar in magnitude to the differences between the simulations A, B and C. We conclude that the code is functioning as expected.

In a companion paper \citep{Salcido2017}, we consider a more detailed comparison between the EdS cosmology and our universe, to quantify the effect of the cosmological constant on galaxy formation in our universe.

\subsection{Simulation Suite} \label{Ss:sims}

The \eagle reference simulations used cosmological parameters measured by the \citet{2014A&A...571A..16P}. We run seven \eagle simulations that modify the cosmological constant, while keeping the same baryon mass per photon (\xib), cold dark matter mass per photon (\xic), and spatial curvature ($\kappa = 0$) unchanged. We also use the same physical sub-grid parameters as the reference model. The values of the cosmological and numerical parameters used for the simulations are listed in Table \ref{tab:a}.

As noted in Section \ref{Ss:cosmparam}, we can solve the Friedmann equations for $a(t)$ with an arbitrary normalisation, and then rescale appropriately. For our cosmological simulations, we choose the initial scale factor (or equivalently, redshift $z_\ro{initial}$) to be the same for all values of $\Lambda$. In our universe, $z_\ro{initial} =  127$ corresponds to a proper time of $t_\ro{init} = 11.5$ Myr. Thus, for a given value of $\Lambda$ for which we have the scale factor $a(t)$ with any arbitrary normalisation, we rescale so that $a(t_\ro{initial}) = 1 / (1 + z_\ro{initial})$.

In fact, we can solve for the new cosmological parameters ($H'_0$, $\OmL'$, $\Om'$) in terms of their values in our universe ($H_0$, $\OmL$, $\Om$) analytically in this case. We require the expansion of the universe to be the same at early times, which implies that $H_0^2 \Om$ is equal for all universes. In addition, we increase the physical energy density of dark energy by a factor $f$: $\Lambda_\ro{new} = f \Lambda_0$, which implies that ${H'_0}^{2} \OmL' = f H_0^2 \OmL$. Combining these equations gives,
\begin{equation}
~~H'_0 = H_0 \sqrt{\Om + f \OmL} \qquad
\Om' = \frac{\Om}{\Om + f \OmL}
\end{equation}
Using these equations gives the cosmological parameters in Table \ref{tab:a}, as a function of $\Lambda$. 

\begin{table*}
\begin{tabular}{lccccccccccc}
\hline
Sim. Name            &  $L$   &        $N$       &  $h$   & \Om   &  \Omb   & \OmL & $\sigma_8$ \\
                     & [cMpc] &                  &        &       &         &       \\
\hline
EdS\_25 ($\Lambda=0$) &  25   & $2 \times 376^3$ & 0.3755 &  1.0  & 0.1572  & 0 & 0.6826 \\
Ref\_25 ($\Lambda_0$) &  25   & $2 \times 376^3$ & 0.6777 & 0.307 & 0.04825 & 0.693 & 0.8288 \\
$\Lambda_0 \times 3$    &  25   & $2 \times 376^3$ & 1.047  & 0.1287 & 0.0202 & 0.8713 & 0.8913 \\
$\Lambda_0 \times 10$   &  25   & $2 \times 376^3$ & 1.823  & 0.0424 & 0.00667 & 0.9576 & 0.8955 \\
$\Lambda_0 \times 30$   &  25   & $2 \times 376^3$ & 3.113  & 0.01455 & 0.00229 & 0.98545 & 0.8434 \\
$\Lambda_0 \times 100$  &  25   & $2 \times 376^3$ & 5.654  & 0.00441 & 6.93\ten{-4} & 0.99559 & 0.7476 \\
$\Lambda_0 \times 300$  &  25   & $2 \times 376^3$ & 9.779  & 0.00147 & 2.32\ten{-4} & 0.99853 & 0.6446 \\
\hline
\end{tabular}
\caption{Cosmological and numerical parameters for our simulations: Box-size (``comoving'', that is, the size of the box today in the Reference $\Lambda_0$ simulation), number of particles, and cosmic parameters $(h,\Om,\Omb,\OmL)$. (Larger 50 Mpc boxes were run and are analysed in more detail in \citet{Salcido2017}. For our purposes, their results were consistent with the 25 Mpc simulations we use here.) Note that these numbers use the convention for `today' defined in Section \ref{Ss:sims}: the proper age of the universe when $a_\ro{initial}  = 1/(1 + z_\ro{initial}$ is the same for all models. The parameter $\sigma_8$ is the rms linear fluctuation in the mass distribution on scales of 8 $h^{-1}$ Mpc, calculated using CAMB \citep{2000ApJ...538..473L}. Note that this parameter varies between cosmologies due to differences in the growth of matter fluctuations, and differences in the averaging scale due to the different values of $h$. As noted in Section \ref{Ss:sims}, we do \emph{not} run CAMB again to generate new initial conditions for each of our simulations; we use the same initial snapshot (particle positions and velocities) for each simulation. For all simulations, the initial baryonic and dark matter particle mass, ``comoving'' and Plummer-equivalent gravitational softening, and initial redshift are as follows: $m_\ro{gas} = 1.81\ten{6} \Msol$, $m_\ro{DM} = 9.70 \ten{6} \Msol$, $\epsilon_\ro{com} = 2.66$ kpc, $\epsilon_\ro{prop} = 0.70$ kpc, $z_\ro{initial} = 127$. Not listed are the three simulations used for the convergence test (Figure \ref{fig:rate_star_EdS}), which use smaller boxes: $L = 12.5$ cMpc, $N = 2 \times 188^3$. }\label{tab:a}
\end{table*}

We are interested in star formation across cosmic time, and so we want to run the simulation as far as possible into the future. This becomes increasingly difficult as the universe transitions into its era of accelerating expansion. The internal-time variable in the code is log$(a)$, and when $a$ begins to increase exponentially in cosmic time ($t$), it takes more and more internal-time steps to cover the same amount of cosmic time. Furthermore, because the internal spatial variable is \emph{comoving} distance, objects that have a constant proper size are shrinking in code units. In our experience, in the accelerating era, the densest particles in the simulations are assigned very short internal-time steps. The simulation slows to a crawl, spending inordinate amounts of CPU time on a small number of particles at the centres of isolated galaxies.

In future work, we will look for ways to overcome these problems. Here, we have been able to run the simulation far enough into the future that, particularly for large values of the cosmological constant, quantities such as the collapse fraction and the fraction of baryons in stars have approached constant values. The endpoint of the $\Lambda_0 \times 30$, $\Lambda_0 \times 100$ and $\Lambda_0 \times 300$ simulations can be seen in the figures in following sections. We have captured the initial burst of galaxy and star formation in these universes, and the accelerating expansion of space makes any future accretion negligible. Each galaxy becomes a separate `island universe'. Nevertheless, the far-future ($\gg 20$ Gyr) fate of baryons in haloes is not captured by our simulations. Very slow processes that are difficult to capture in any simulation (let alone one in a cosmological volume) become relevant: gas cooling on very long time scales, a trickle of star formation, rare supernovae in low density environments, accretion of diffuse gas onto stellar remnants and black holes. These processes could be relevant to our models of observer creation over all of cosmic time; we will return to these issues in Section \ref{Ss:extrap}.


\section{Changing the Cosmological Constant: Global Properties}  \label{S:global}

We vary the cosmological constant between zero and several hundred times larger than the value in our Universe. We do not consider negative values of the cosmological constant here, as it would require significant changes to the time stepping in the code to handle the transition from expansion to contraction.

\begin{figure} \centering
		\includegraphics[width=0.47\textwidth]{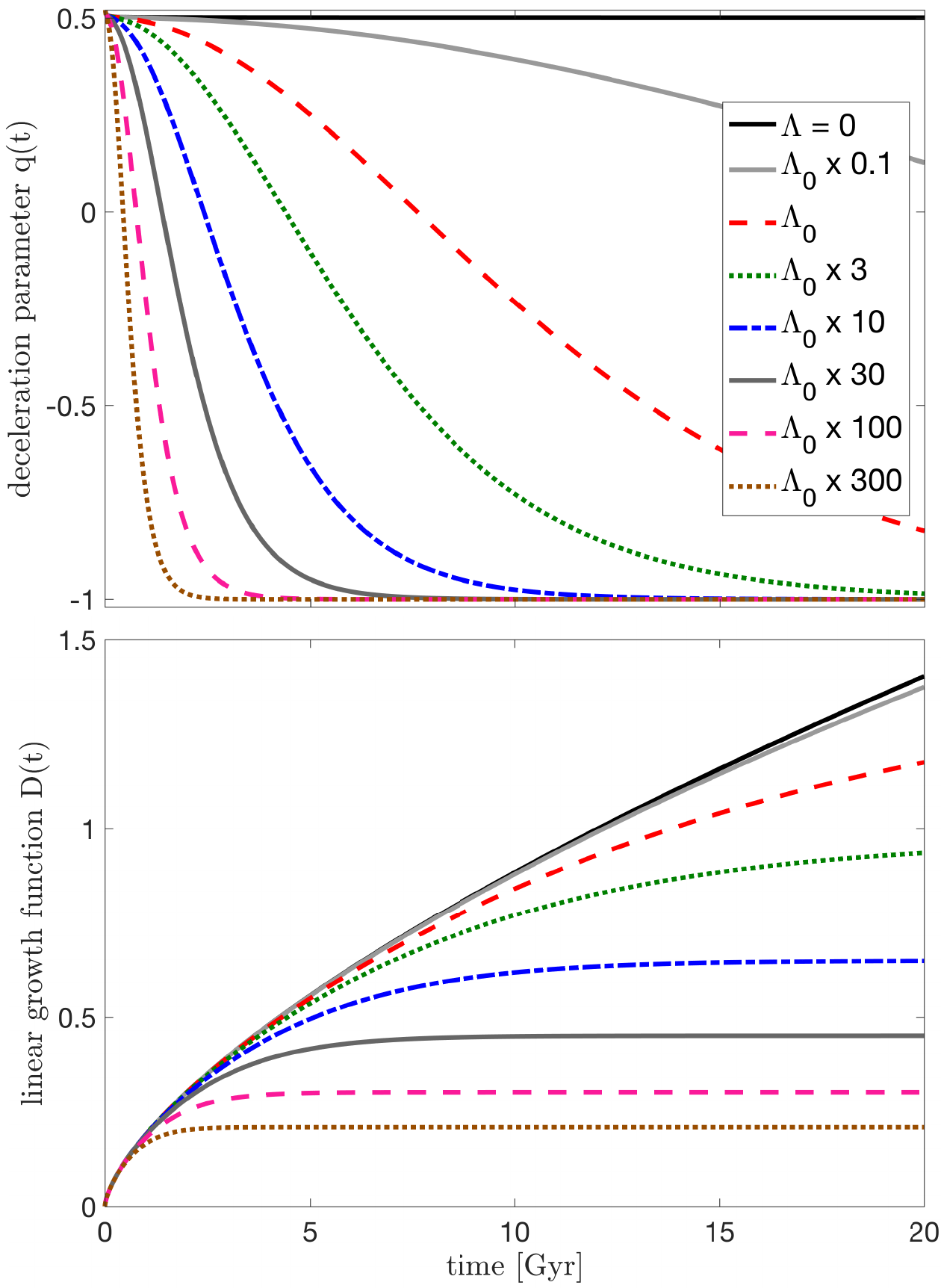}
			\caption{The deceleration parameter $q \equiv -\ddot{a} / (a H^2)$ (\emph{top}) and the linear growth factor $D(t)$ (\emph{bottom}) as a function of cosmic time, for different values of the cosmological constant. Note that $q = 1/2$ at all times for the $\Lambda = 0$ cosmology. As the cosmological constant increases, the time at which the expansion of the universe begins to accelerate $(q < 0)$ moves to earlier times as $t_\ro{accel} \sim 1 / \sqrt{G \rhoL}$. Once accelerated expansion begins, the formation of structure freezes and accretion stops, and $D(t)$ approaches a constant.} \label{fig:qGt}
\end{figure}

Figure \ref{fig:qGt} shows the deceleration parameter $q \equiv -\ddot{a} / (a H^2)$ and the linear growth factor $D(t)$ as a function of cosmic time, for different values of the cosmological constant.  As the cosmological constant increases, the time at which the expansion of the universe begins to accelerate $(q < 0)$ moves to earlier times as $t_\ro{accel} \sim 1 / \sqrt{G \rhoL}$. Once accelerated expansion begins, the formation of structure freezes and accretion stops. We can see this in linear perturbation theory, where all modes grow in proportion to the growth factor $D(t)$; we normalise $D(t)$ so the curves are equal at early times, and $D(t_0) = 1$ in our Universe today. We see that once the expansion of the universe begins to accelerate, the growth factor approaches a constant, and structures ceases to grow.

In this section we will characterise the details of structure formation in these universes. Ordinarily, one describes these properties using comoving quantities, such as the comoving halo number density and comoving star formation rate density. One immediate problem is that the term ``comoving'' is meaningless when different universes are being compared. There is no ``today'' that is common to all models, relative to which we can define comoving volumes, densities and the like. There is nothing special, cosmically speaking, about 13.8 Gyr or 2.725 K. We can arbitrarily change the comoving density of star formation, for example, by choosing a different cosmic time in a given universe to be ``today'', which makes the comparison of comoving densities meaningless.

To overcome this, we will calculate quantities relative to the  physical \emph{mass} (total or baryonic) in the simulation box\footnote{We could define comoving densities relative to the initial physical volume of the simulation box, which is the same in all models. But while this allows a meaningful comparison, the initial cosmic time is still arbitrary. Choosing an earlier time would increase all the ``comoving'' densities, which makes their value in a given universe difficult to interpret. Calculating specific (per unit mass) quantities overcomes this problem.}. This provides a meaningful comparison between the simulated universes, and like comoving densities it does not automatically scale with expansion of the universe. We can ask, for example, what fraction of the total baryonic mass in the universe is in the form of stars as a function of cosmic time? What fraction has been converted into metals?

\subsection{Mass accretion}

Formally, in a cold dark matter universe, every particle is in a dark matter halo of \emph{some} mass. That is, the collapse fraction of the universe is always unity; from \citet{1974ApJ...187..425P} theory,
\begin{equation}
F(>M | t) = \textrm{erfc}\left( \frac{\delta_\textrm{crit} (t)}{\sqrt{2} \sigma(M,t)} \right) ~,
\end{equation}
where $F(>M | t)$ is the fraction of matter at cosmic time $t$ that is part of a collapsed halo of mass greater than $M$, $\delta_\textrm{crit} (t)$ is the critical linear overdensity of a collapsed object, and $\sigma(M,t)$ is the standard deviation of the cosmic matter field when smoothed on a scale that encloses mass $M$. The matter variance  $\sigma(M,t) \rightarrow \infty$ as $M \rightarrow 0$, thus $F(> 0 | t) = 1$ at all times.

In the simulation, however, there is a minimum dark matter halo mass that can be resolved by the particles. Given that each dark matter particle has  mass $m_\ro{DM} = 9.7 \ten{6} \Msol$ and we require 32 particles to identify a halo, we can resolve haloes with mass greater than $m_\ro{min} = 3.1 \ten{8} \Msol$. Summing the total mass in these haloes, then, gives the collapse fraction for resolved haloes: $F(>m_\ro{min} | t)$. This approximately excludes haloes that are too small to form stars, so gives us the fraction of mass in the universe that resides in potentially star-forming haloes; the remainder can be considered as the inter-galactic medium.

\begin{figure*} \centering
	\begin{minipage}{0.45\textwidth}
		\includegraphics[width=\textwidth]{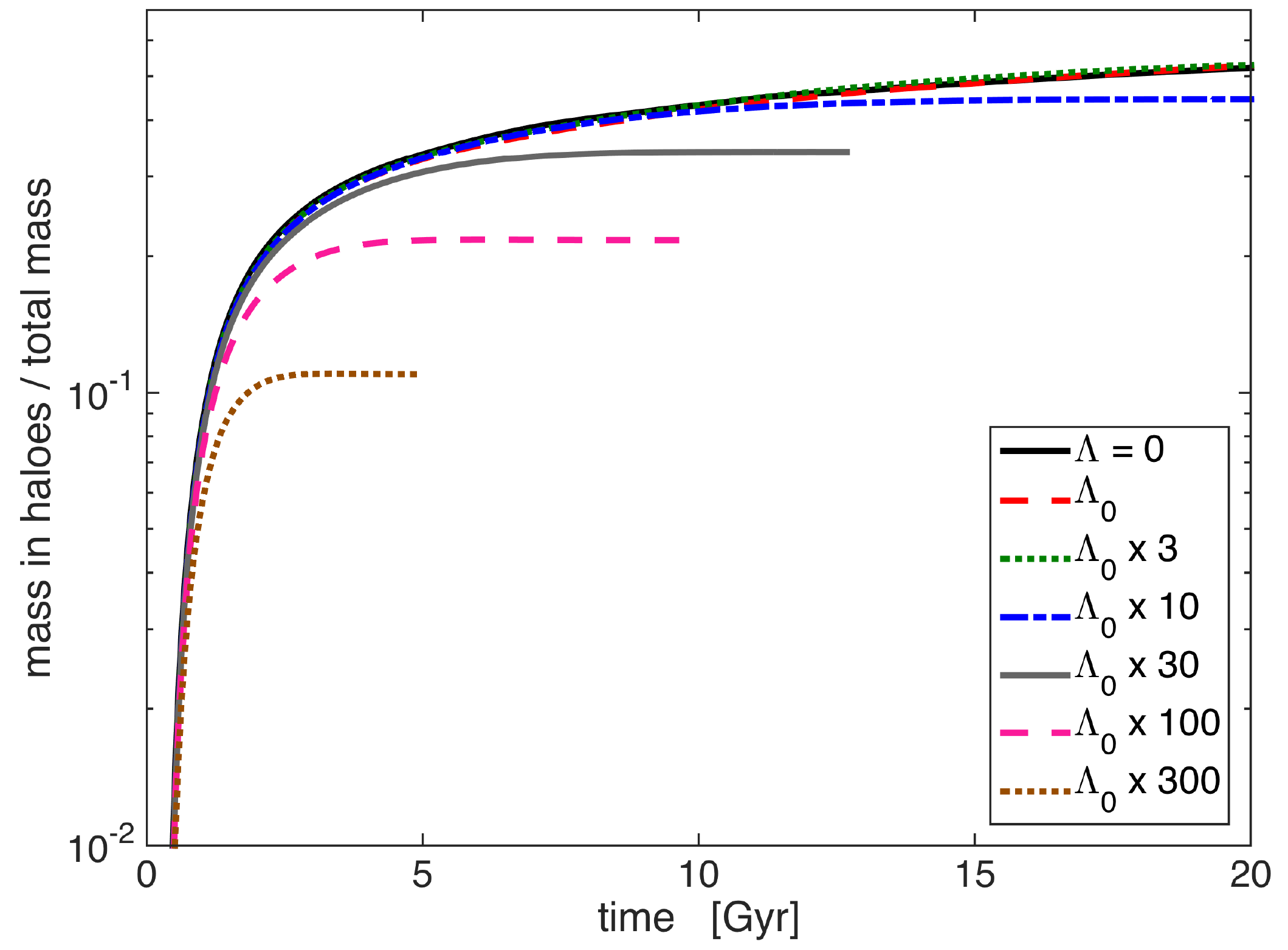}
	\end{minipage}
	\begin{minipage}{0.45\textwidth}
		\includegraphics[width=\textwidth]{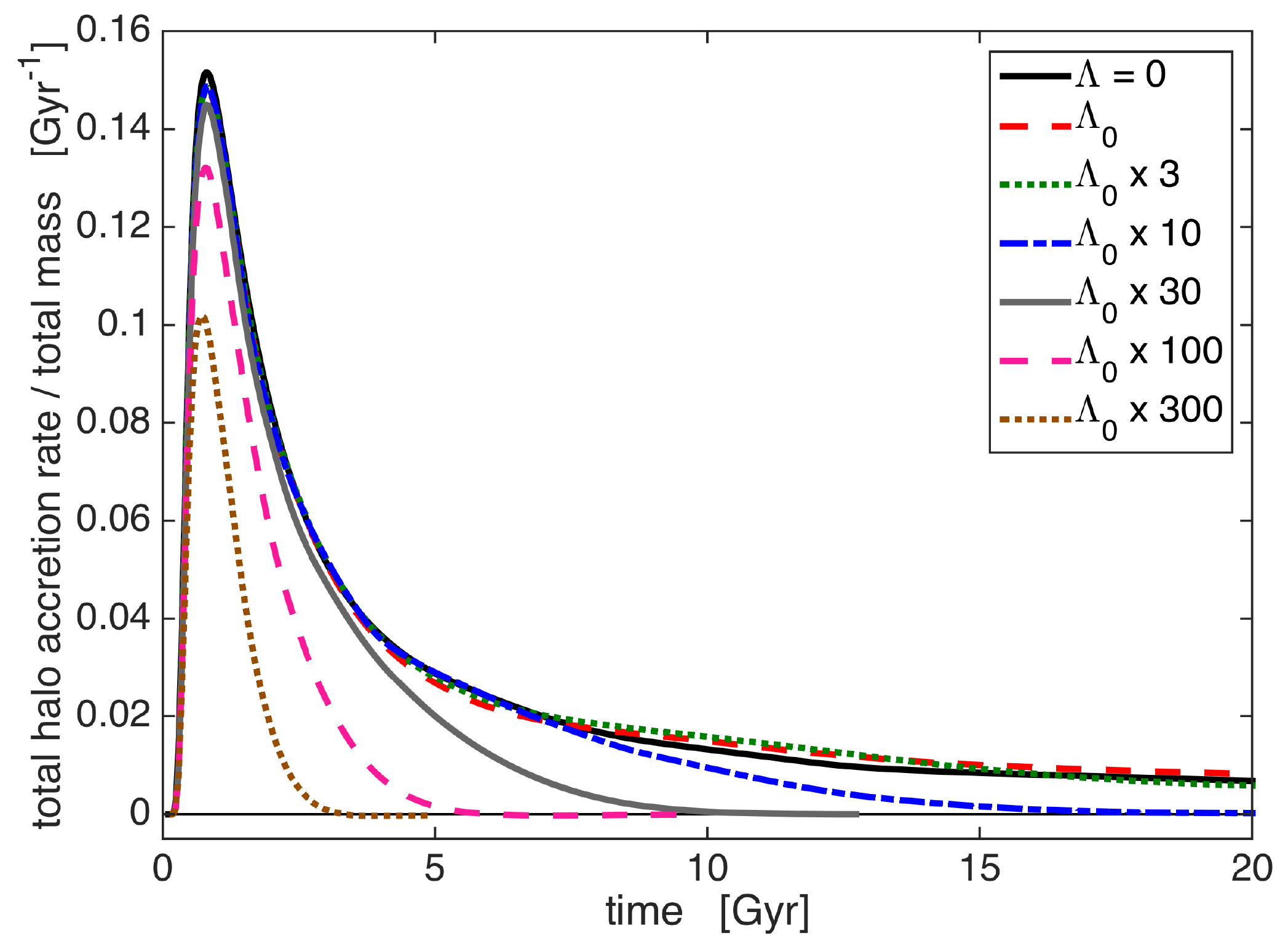}
	\end{minipage}
			\caption{\emph{Left:} The fraction of mass in each simulation that is part of a resolved halo: $F(>m_\ro{min} | t)$, where $m_\ro{min} = 3.1 \ten{8} \Msol$. This minimum mass is a consequence of the numerical resolution of the simulations, but is consistent across all of them and approximately excludes haloes that are too small to form stars. The result is a measure of the fraction of mass in the universe that resides in potentially star-forming haloes. \emph{Right:} The specific total halo accretion rate, that is, the time derivative of the left hand curve. The rate peaks at $t = 0.8$ Gyr in our Universe ($\Lambda = \Lambda_0$). Even for a universe with a cosmological constant ten times larger than ours ($\Lambda_0 \times 10$), there is minimal difference in total halo mass fraction even after 20 Gyrs, well into the accelerating phase of the universe's expansion.} \label{fig:accr_cuml}
\end{figure*}

Figure \ref{fig:accr_cuml} shows (\emph{left}) the fraction of the total mass in the universe that has collapsed into resolved haloes, and (\emph{right}) the specific total halo accretion rate, that is, the time derivative of the left hand curve. In this figure and those following, the time derivative is calculated after smoothing the accretion fraction. Even for a universe with a cosmological constant ten times larger than ours ($\Lambda_0 \times 10$), there is minimal difference in total halo mass fraction even after 20 Gyrs, well into the accelerating phase of the universe's expansion. The initial peak in the accretion rate at $t = 0.8$ Gyr remains largely unchanged even in a universe with a cosmological constant 30-100 times larger than ours. In a universe with $\Lambda = \Lambda_0 \times 100$, a fifth of the mass in the universe accretes into haloes with $m > m_\ro{min} = 3.1 \ten{8} \Msol$.

\subsection{Baryon flow}

\begin{figure*} \centering
	\begin{minipage}{0.45\textwidth}
		\includegraphics[width=\textwidth]{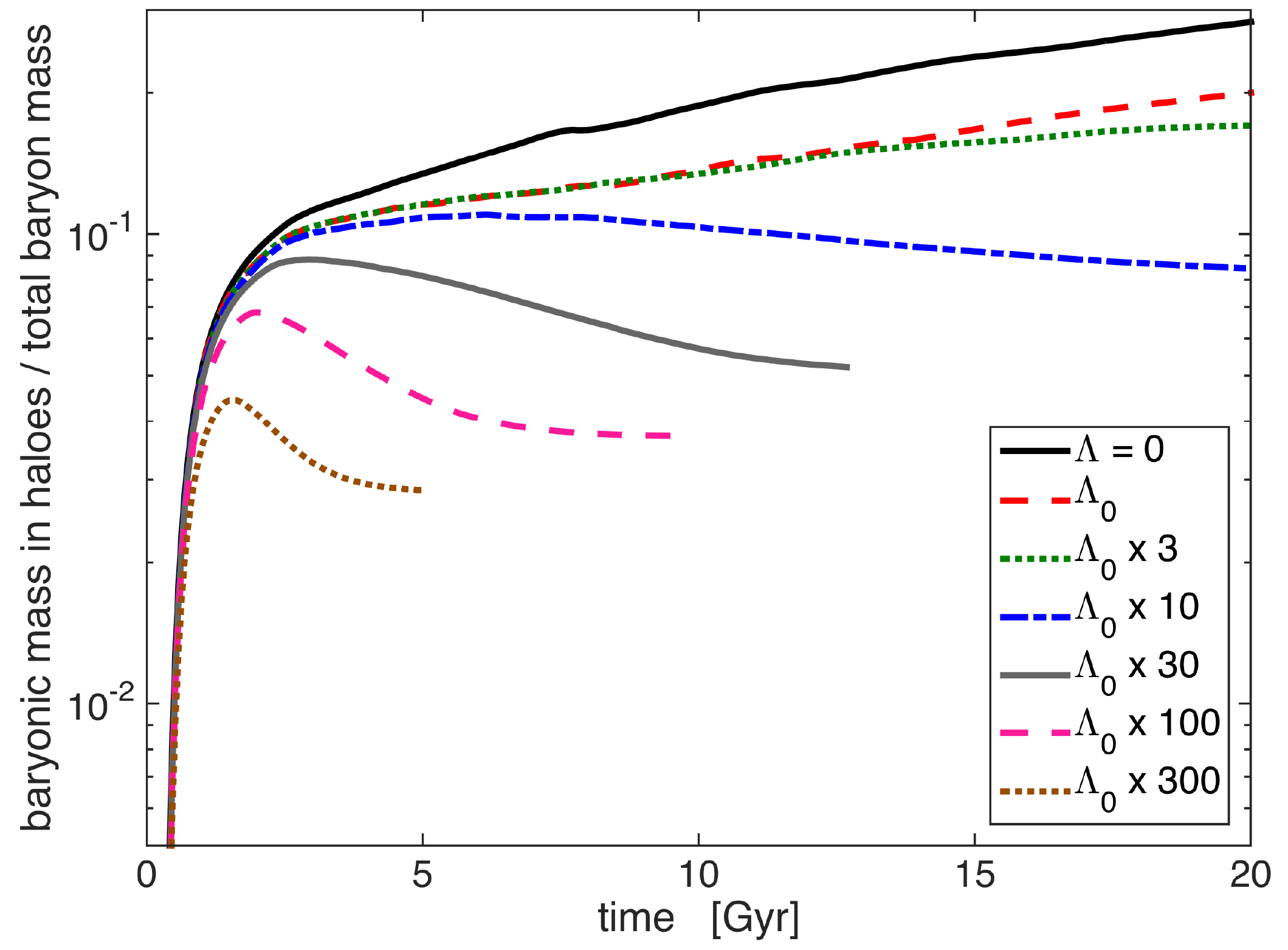}
	\end{minipage}
	\begin{minipage}{0.45\textwidth}
		\includegraphics[width=\textwidth]{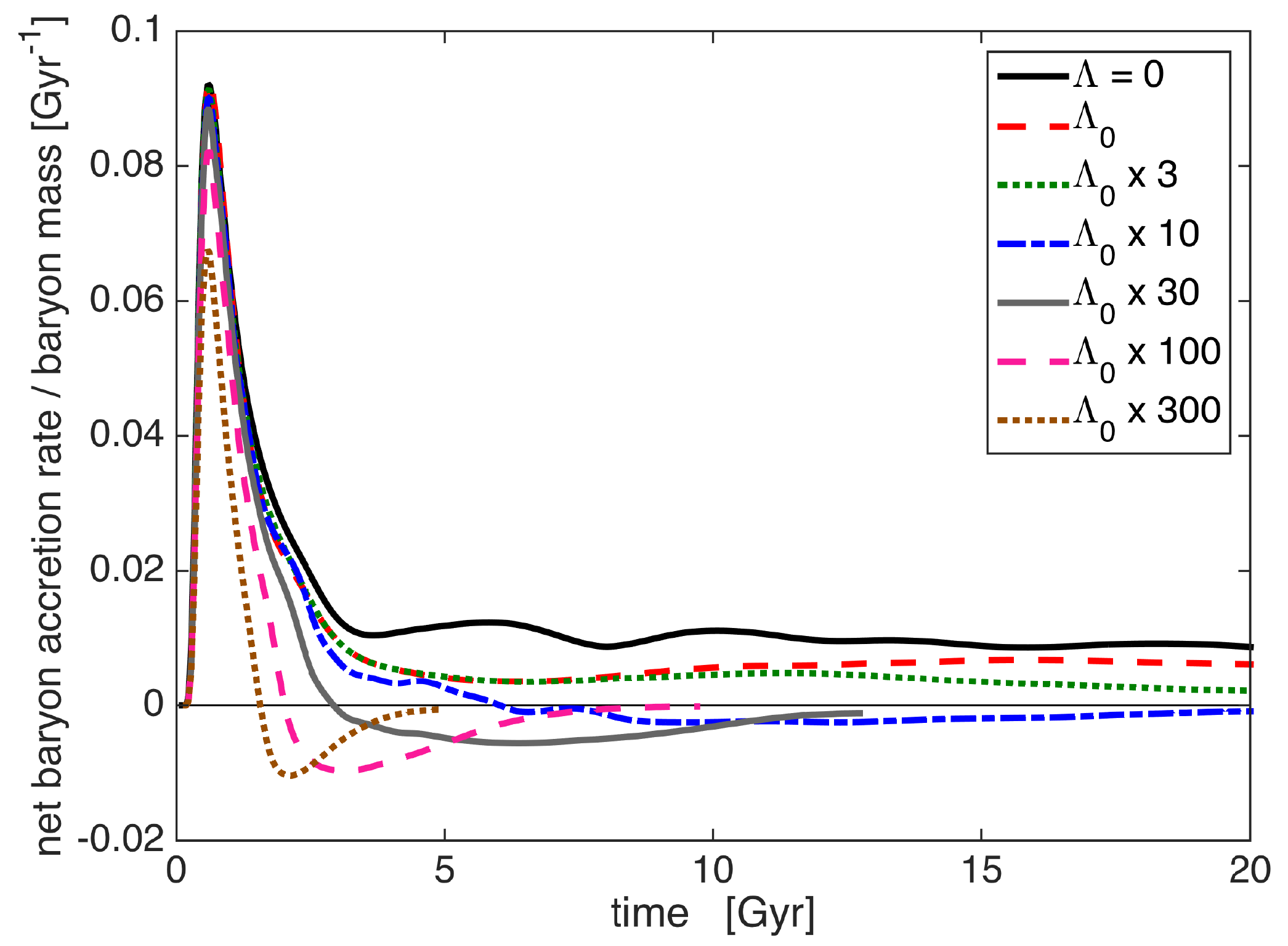}
	\end{minipage}
			\caption{\emph{Left} the fraction of the baryonic mass in the simulation that is inside dark matter haloes with $m > m_\ro{min} = 3.1 \ten{8} \Msol$ as a function of cosmic time, and \emph{Right} the specific rate of baryon accretion. The rate peaks at $t = 0.6$ Gyr in our Universe ($\Lambda = \Lambda_0$).} \label{fig:accr_gas_cuml}
\end{figure*}

Baryons are subject to physical forces other than gravity: the smoothing effect of gas pressure, cooling and heating from radiation, star formation, supernovae-driven galactic winds, black hole feedback and more. Figure \ref{fig:accr_gas_cuml} shows \emph{left} the fraction of the baryonic mass (in the form of stars and gas) in the simulation that is inside dark matter haloes with $m > m_\ro{min} = 3.1 \ten{8} \Msol$ as a function of cosmic time, and \emph{right} the specific rate of baryon accretion (i.e. per unit total baryon mass).

We see the same peak in the accretion rate at $t = 0.8$ Gyr, and when there is zero cosmological constant, the baryon accretion rate increases in a similar way to the total accretion rate (Figure \ref{fig:accr_cuml}). As the cosmological constant increases, it has a much larger effect on the baryons than the dark matter. In fact, for $\Lambda = \Lambda_0 \times 10$, the rate of baryon accretion becomes negative, as baryons are --- on average --- being ejected from galaxies.

We can understand this effect as follows. We can write the acceleration ($a_g$) of a test mass at distance $r$ from a large mass $M$ under the Newtonian gravitational force with a cosmological constant term,
\begin{equation}
a_g = -G\frac{M}{r^2} + \frac{\Lambda c^2}{3} r ~.
\end{equation}
If we consider a large collapsed mass, then the distance ($d_0$) at which the force on a test mass is balanced between attraction to the central mass and repulsion by the cosmological constant is found by setting $a_g = 0$,
\begin{equation}
d_0 = 1.1 \ro{ Mpc } \left( \frac{M}{10^{12} \Msol} \right)^{1/3}
\left( \frac{\Lambda}{\Lambda_0} \right)^{-1/3} ~,
\end{equation}
or equivalently in terms of the ratio $\rhoL / \rho_{\Lambda_0}$. In our universe, this is $\sim 4$ times larger than the virial radius of the halo (which also scales as the $1/3$ power of mass). In universes in which the cosmological constant is larger, these distances are comparable.

As seen in Figure \ref{fig:accr_gas_cuml}, this does not dramatically affect the growth of the dark matter halo. But baryonic matter ejected from galaxies in galactic winds or outflows, if it reaches the outer parts of the halo, is liable to be lost. Rather than raining back down on the galaxy after a delay of $\sim$ 1 Gyr \citep{2008MNRAS.387..577O,2010MNRAS.406.2325O,2017ASSL..430..301V}, this material is lost, drawn away into the expansion of the universe by the repulsive effect of the cosmological constant \citep{2006MNRAS.373..382B}.

As we will see in the next subsection, in universes for which ($\Lambda \gtrsim \Lambda_0 \times 10$), the initial burst of star formation in the universe occurs when the universe has begun to expand exponentially. This rapid star formation, combined with black hole feedback, launches outflows that are carried away by the accelerating expansion. This effect  overwhelms accretion by gravitational attraction, causing the net accretion rate to become negative. The result is that there is not a simple, linear relationship between dark matter halo growth and baryon accretion that holds for all values of the cosmological constant.

\subsection{Star formation}

\begin{figure*} \centering
	\begin{minipage}{0.45\textwidth}
		\includegraphics[width=\textwidth]{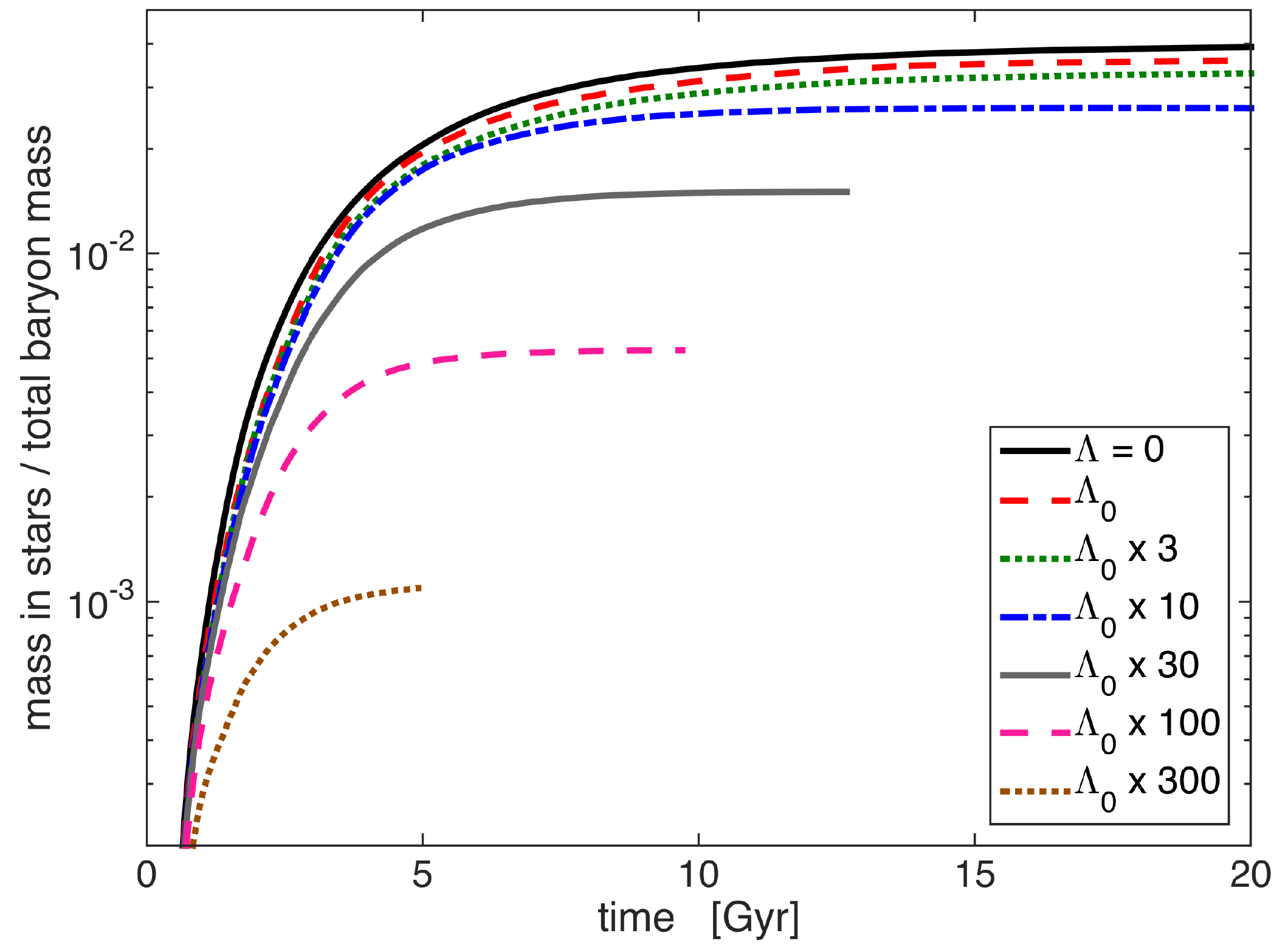}
	\end{minipage}
	\begin{minipage}{0.45\textwidth}
		\includegraphics[width=\textwidth]{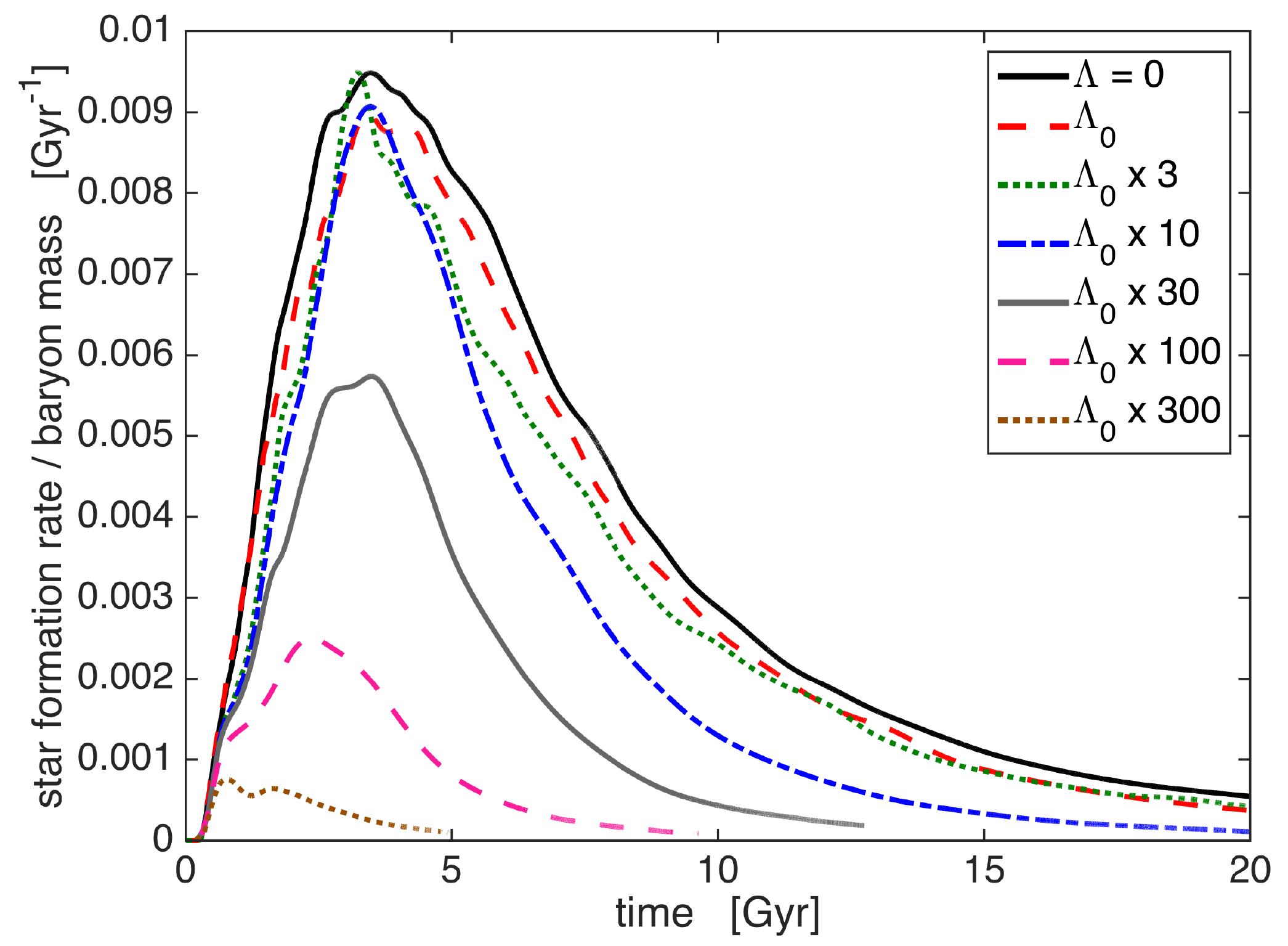}
	\end{minipage}
			\caption{\emph{Left:} the fraction of cosmic baryons by mass that are in the form of stars as a function of cosmic time. \emph{Right:} the star formation rate efficiency, which takes into account both star birth only. The rate peaks at $t = 3.5$ Gyr in our Universe ($\Lambda = \Lambda_0$).} \label{fig:accr_star_cuml}
\end{figure*}

Some of the baryons that accrete into haloes will form stars. Figure \ref{fig:accr_star_cuml} shows (\emph{left}) the fraction of cosmic baryons that are in the form of stars as a function of cosmic time, and (\emph{right}) the star formation rate efficiency, which takes into account star birth only. Note that, as it is commonly used in the galaxy formation literature, ``specific star formation'' refers to the star formation rate of a galaxy divided by its \emph{stellar} mass. To avoid confusion, we will call the star formation mass (rate) per unit total baryon mass the star formation (rate) efficiency.

The star formation rate efficiency peaks at $t \approx 3.5$ Gyr. This is a delayed consequence of the peak in the mass accretion rate at $t = 0.8$ Gyr. As the cosmological constant increases, the haloes are starved both by the cessation of fresh accretion from the intergalactic medium and the lack of recycling of outflowing gas, noted above. The result is a significant curtailing of the star formation rate efficiency. While the $\Lambda = 0$ universe has turned $\sim 4$\% of its baryons into stars by $t = 20$ Gyr, for $\Lambda_0 \times 100$, this fraction is essentially constant after 10 Gyr at $0.5$\%. This factor of 8 decrease contrasts with the factor of $2.4$ decrease in the total mass accretion.

\subsection{Cosmic metal production}
\begin{figure*} \centering
	\begin{minipage}{0.44\textwidth}
		\includegraphics[width=\textwidth]{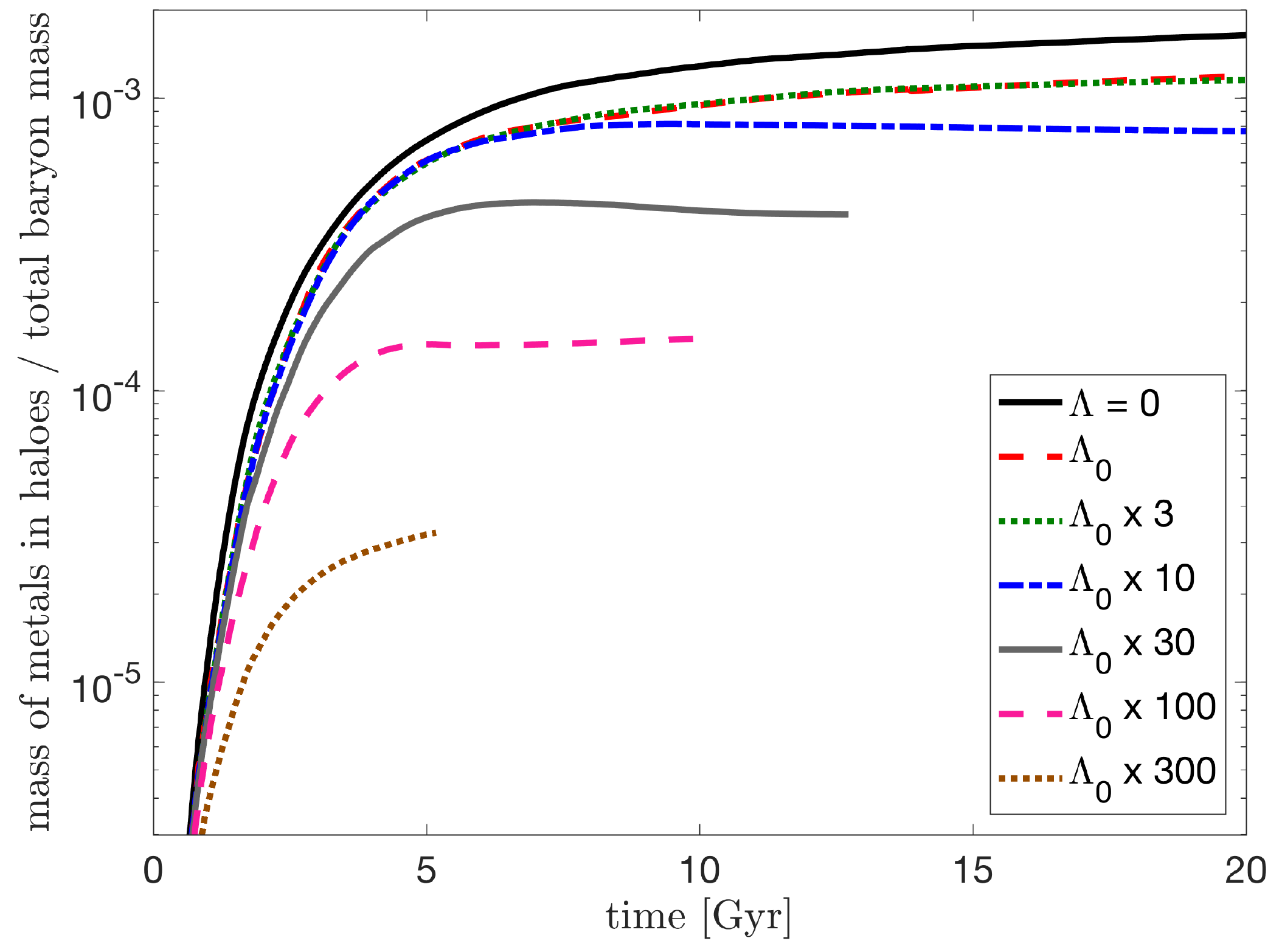}
	\end{minipage}
	\begin{minipage}{0.47\textwidth}
		\includegraphics[width=\textwidth]{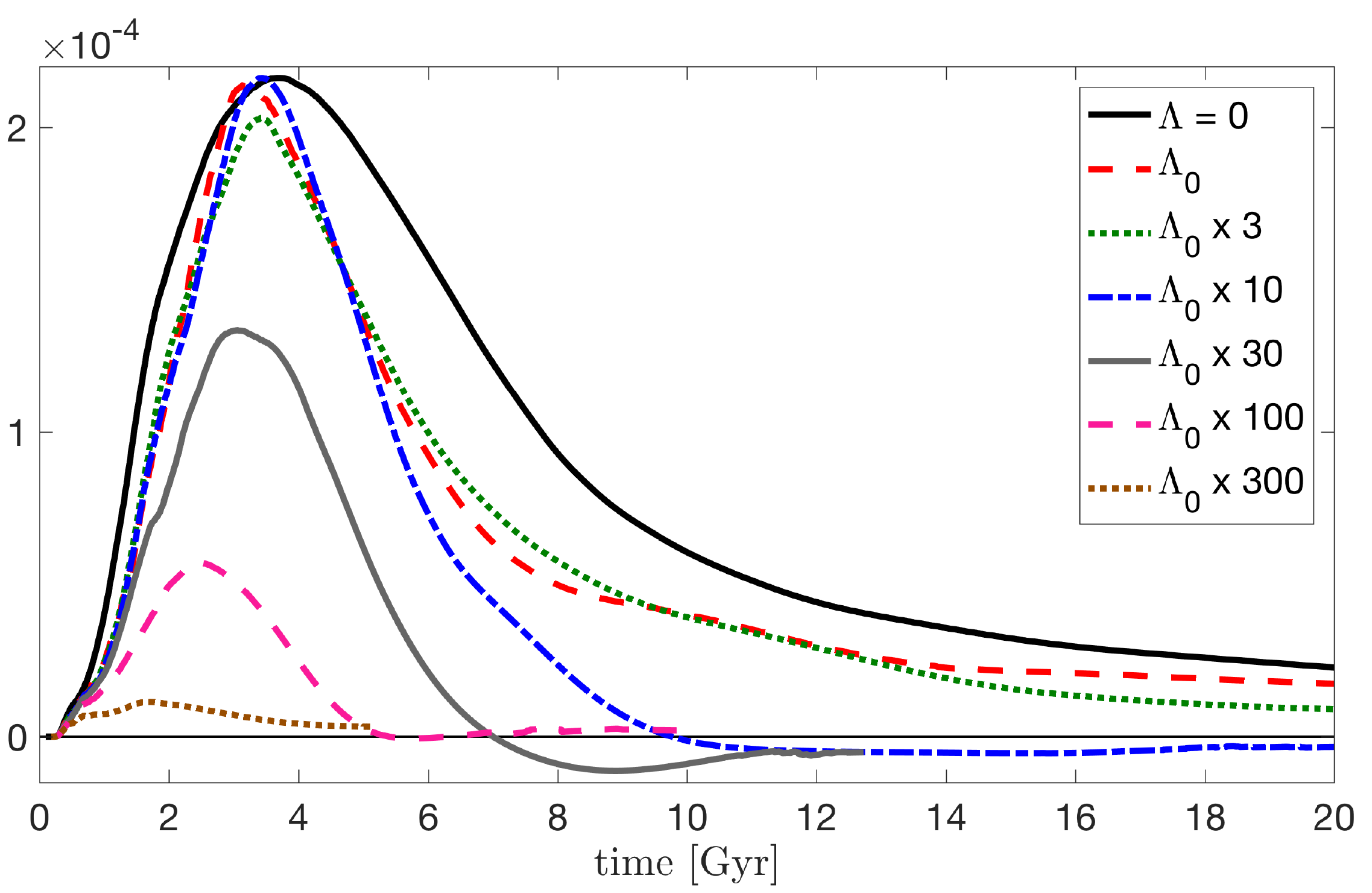}
	\end{minipage}
			\caption{\emph{Left:} The fraction of cosmic baryons that are in the form of metals in collapsed haloes. Right: the halo metal production rate. The rate peaks at $t = 3.2$ Gyr in our Universe ($\Lambda = \Lambda_0$), and the peak is steadily dimished as the cosmological constant increases.} \label{fig:metgas_cuml}
\end{figure*}

Planets and their occupants are formed from the products of stellar nucleosynthesis. The \eagle code, in addition to primordial H and He, follows 9 metals: C, N, O, Ne, Mg, Si, S, Ca, and Fe.  Figure \ref{fig:metgas_cuml} shows (\emph{left}) the fraction of cosmic baryons that are in the form of metals in collapsed haloes, and (\emph{right}) the halo metal production rate. Note that this includes metals in all phases: inside stars, in dense star-forming clouds, and in the hot, non-star forming interstellar gas. The halo metal production rate reflects the balance between metal production in stars, recycling back into the inter-stellar medium by winds and supernovae, re-incorporation into later generation stars, ejection from haloes in galactic winds, and reaccretion into haloes.

As star formation peaks (Figure \ref{fig:accr_star_cuml}), metals are being produced in stars and returned to the IGM in supernovae and planetary nebulae. This feedback also loads metals into the galactic winds that drive baryons out of haloes (Figure \ref{fig:accr_gas_cuml}). As with the baryon accretion rate, the net accretion rate becomes negative for certain values of $\Lambda$ as metals are ejected in winds at a higher rate than they are produced and reaccreted. Our universe turns approximately a fraction $1.2 \ten{-3}$ by mass of its baryons into halo metals by 20 Gyr, while for $\Lambda_0 \times 100$ the fraction asymptotes by 10 Gyr to $1.5 \ten{-4}$. This factor of 10 difference contrasts with the factor of 2.5 difference with regards to the total fraction of mass in haloes.

\section{Accretion and star formation in individual haloes} \label{S:indiv}

\begin{figure*} \centering
	\begin{minipage}{0.4\textwidth}
		\includegraphics[width=\textwidth]{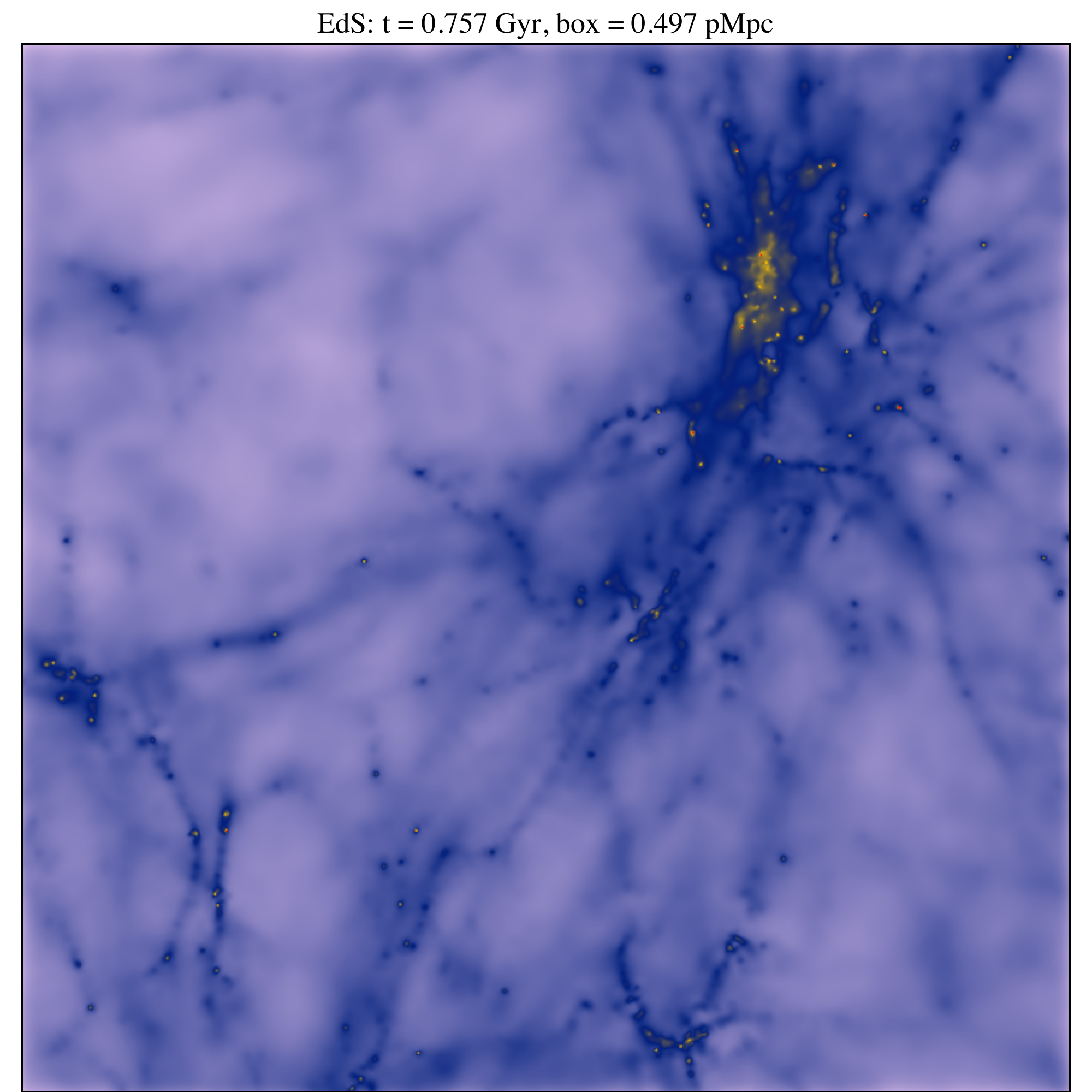}
	\end{minipage}
	\begin{minipage}{0.4\textwidth}
		\includegraphics[width=\textwidth]{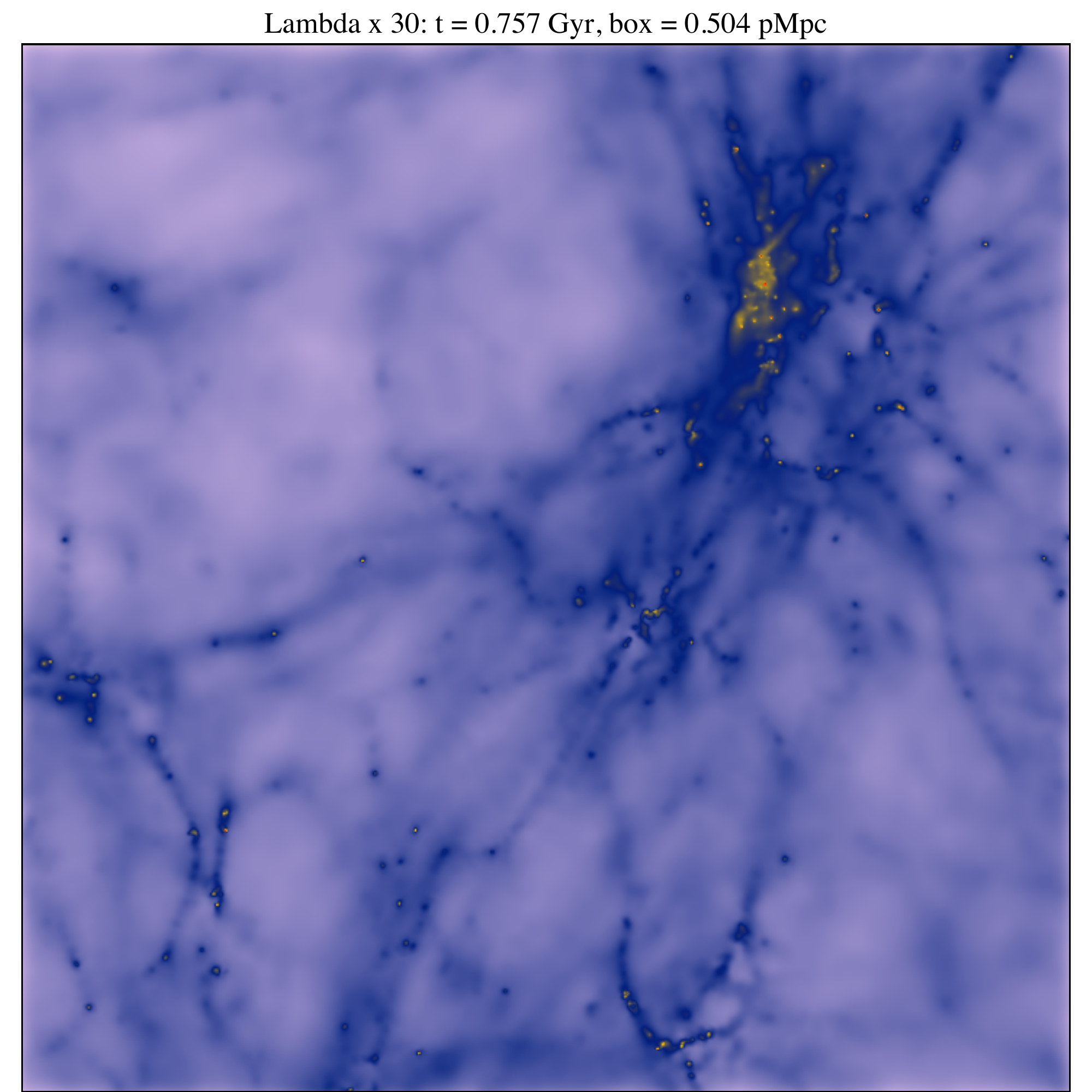}
	\end{minipage}
    \begin{minipage}{0.4\textwidth}
		\includegraphics[width=\textwidth]{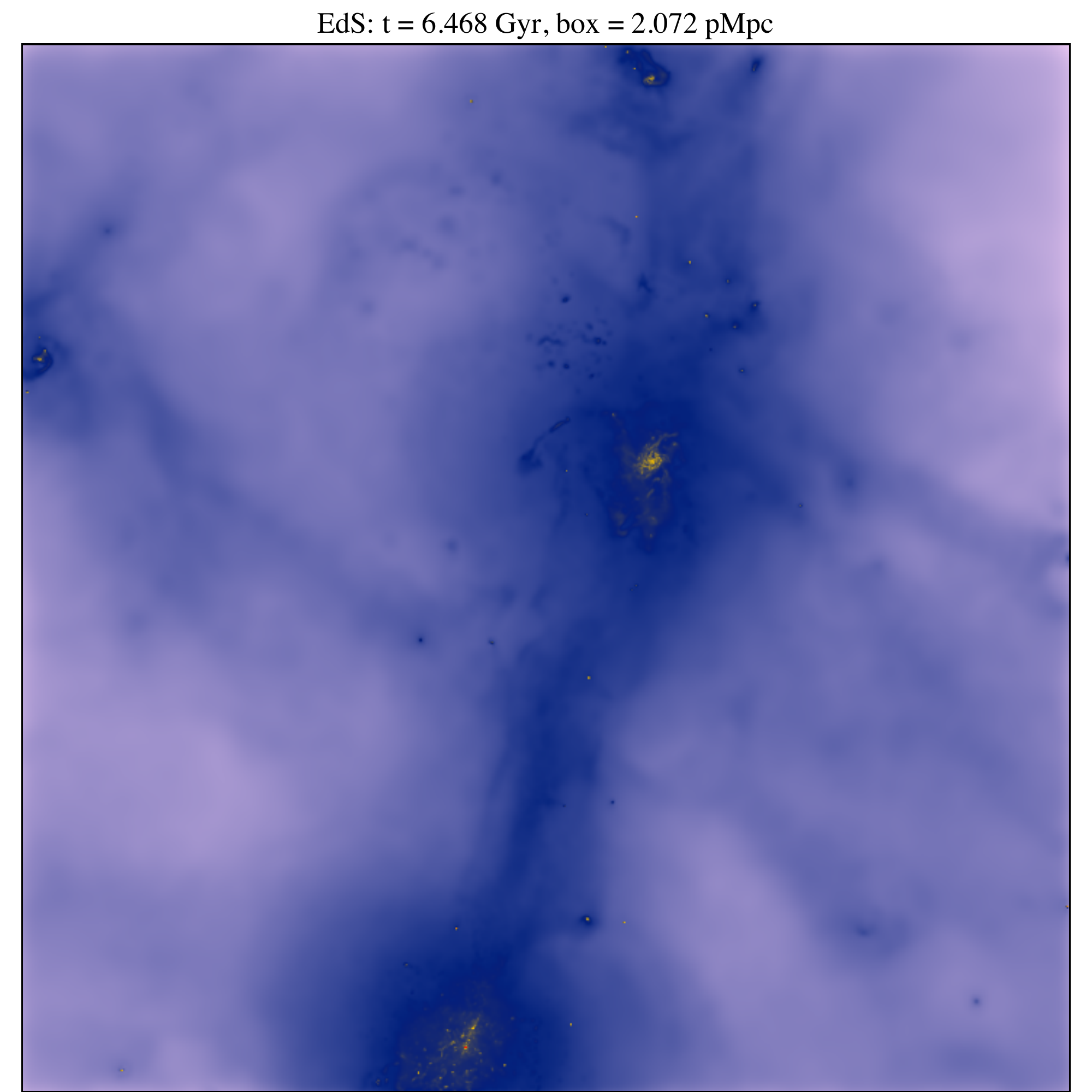}
	\end{minipage}
	\begin{minipage}{0.4\textwidth}
		\includegraphics[width=\textwidth]{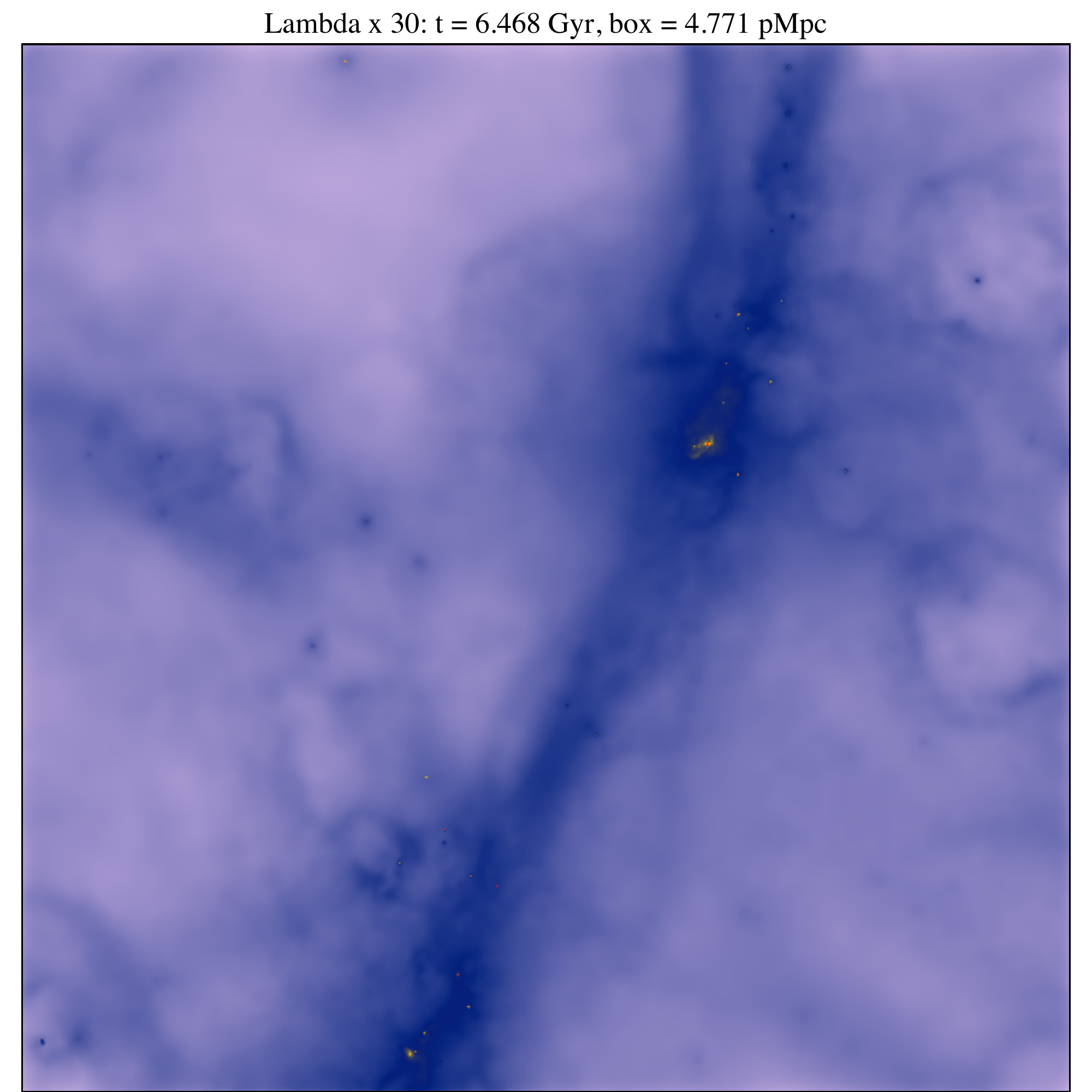}
	\end{minipage}
    \begin{minipage}{0.4\textwidth}
		\includegraphics[width=\textwidth]{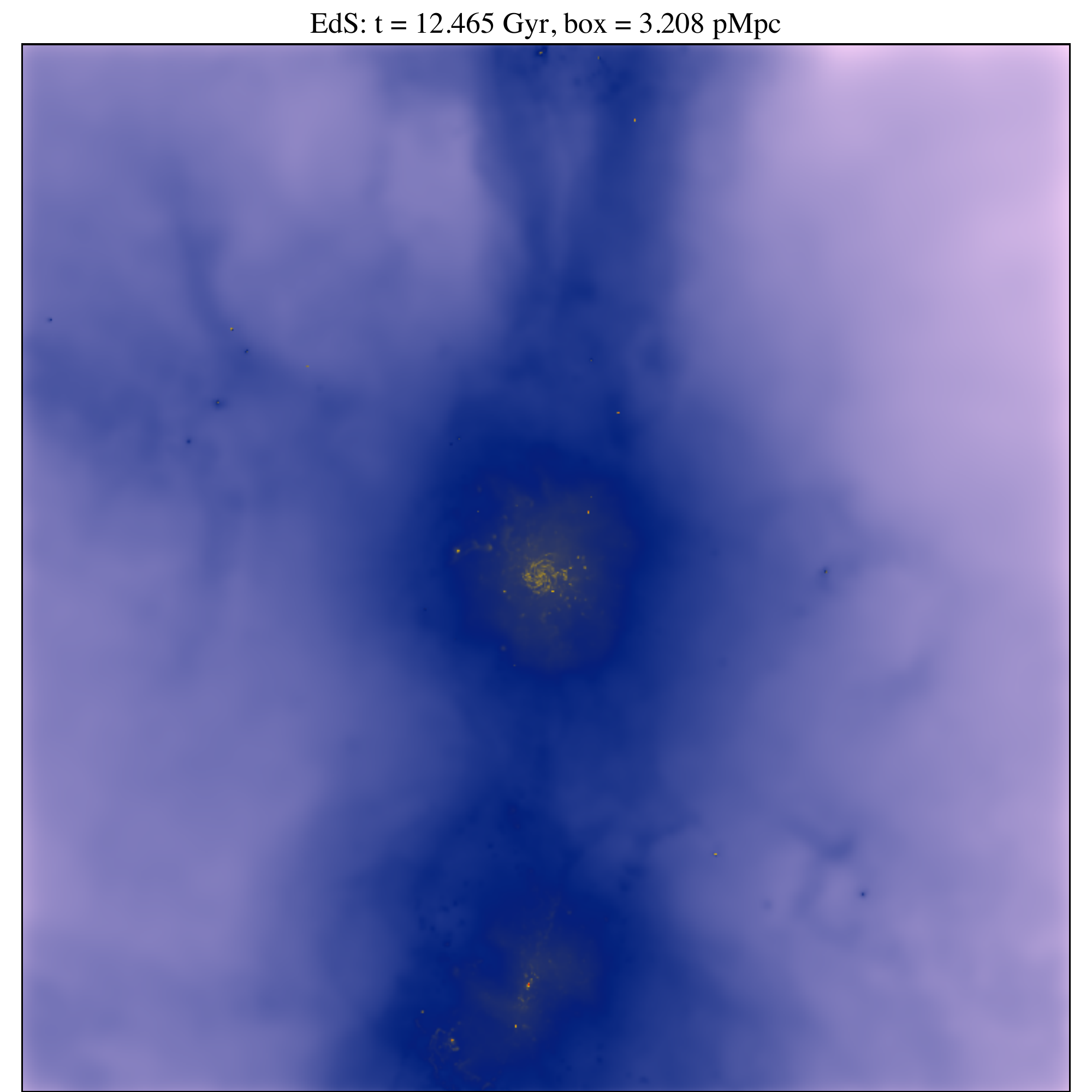}
	\end{minipage}
	\begin{minipage}{0.4\textwidth}
		\includegraphics[width=\textwidth]{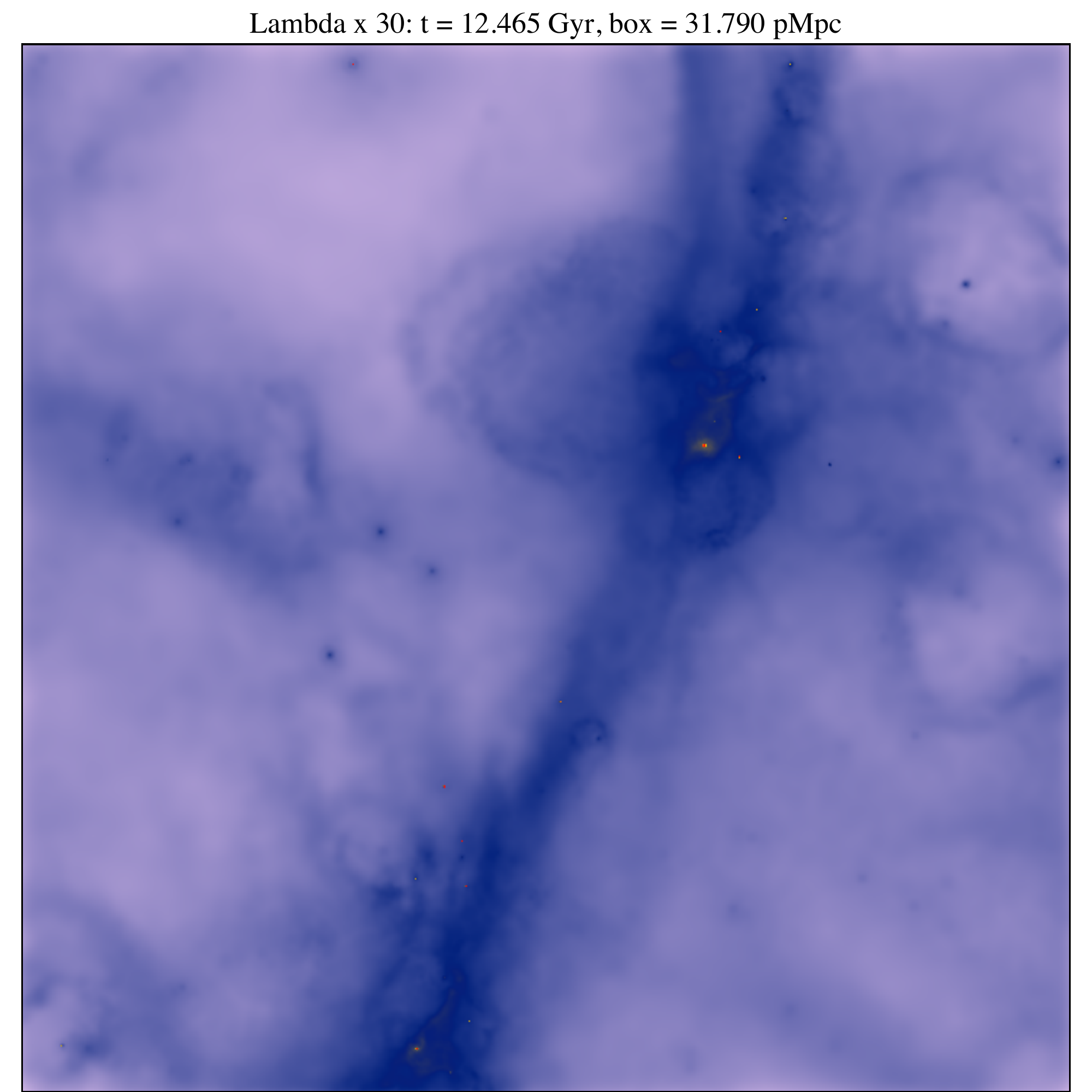}
	\end{minipage}
			\caption{The evolution of the projected gas density of a comoving region of space that, in our universe, evolves into a Milky Way-mass halo by the present day. The comoving size is 4 Mpc in our Universe. The proper time and proper size of the region are shown above each panel. \emph{Left:} an EdS universe ($\Lambda = 0$);  \emph{Right:} a $\Lambda_0 \times 30$ universe. The colour scaling on each row is held constant.} \label{fig:halo_com}
\end{figure*}

In this section, we will consider the evolution of a comoving region of space that, in our universe, evolves into a Milky Way-mass halo ($2 \ten{12} \Msol$) by the present day. Figure \ref{fig:halo_com} shows the projected gas density in a comoving region around the halo equal to 4 Mpc today in our universe; the cosmic time and proper size of the region are shown above each panel. The \emph{left} panels show an EdS universe ($\Lambda = 0$); the \emph{right} panels show a $\Lambda_0 \times 30$ universe.

The top two panels show this region of the universe at cosmic time $t = 0.757$ Gyr, while the $\Lambda_0 \times 30$ is still in its early decelerating phase. The proper sizes of the boxes are within 1\% of each other, and the distributions of matter are very similar. We see the usual picture of small haloes collapsing and hierarchically merging into larger haloes.

The middle two panels show this region of the universe at cosmic time $t = 6.5$ Gyr. The $\Lambda_0 \times 30$ is undergoing accelerating expansion, so the proper size of the region is 2.3 times larger than in the EdS universe, and the linear growth factor is 33\% smaller. The large central halo in the EdS simulation has drawn in a more matter from its surroundings, and is still being drawn towards a second halo at the bottom of the panel.

The bottom two panels show this region of the universe at cosmic time $t = 12.5$ Gyr. The accelerating expansion of the $\Lambda_0 \times 30$ means that the proper size of the comoving region is 10 times larger than in the EdS universe. The typical Newtonian force between two masses in the region is thus 100 times smaller, and the linear growth factor is 2.3 times smaller. The difference in the distribution of matter is quite dramatic: in the $\Lambda_0 \times 30$ universe, there has been little evolution of the structure of the universe since $t = 6.5$ Gyr. The matter in the vertical filaments has not fallen into the large halo, starving the galaxy of gas. In the EdS universe, the halo has been drawn closer to the second halo at the bottom of the panel; the filament of matter between them has largely fallen into one of the haloes.

\begin{figure*} \centering
    \begin{minipage}{0.4\textwidth}
		\includegraphics[width=\textwidth]{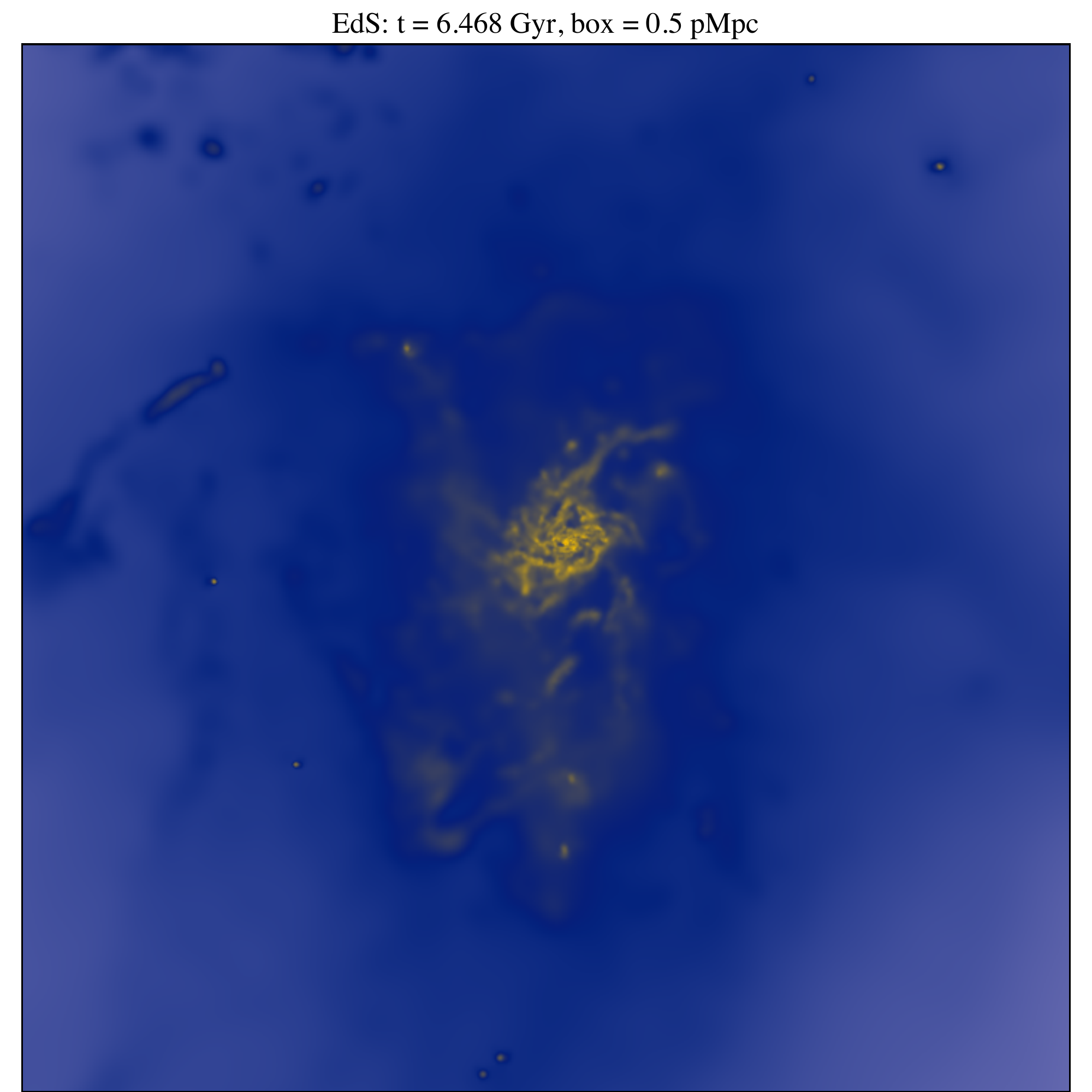}
	\end{minipage}
	\begin{minipage}{0.4\textwidth}
		\includegraphics[width=\textwidth]{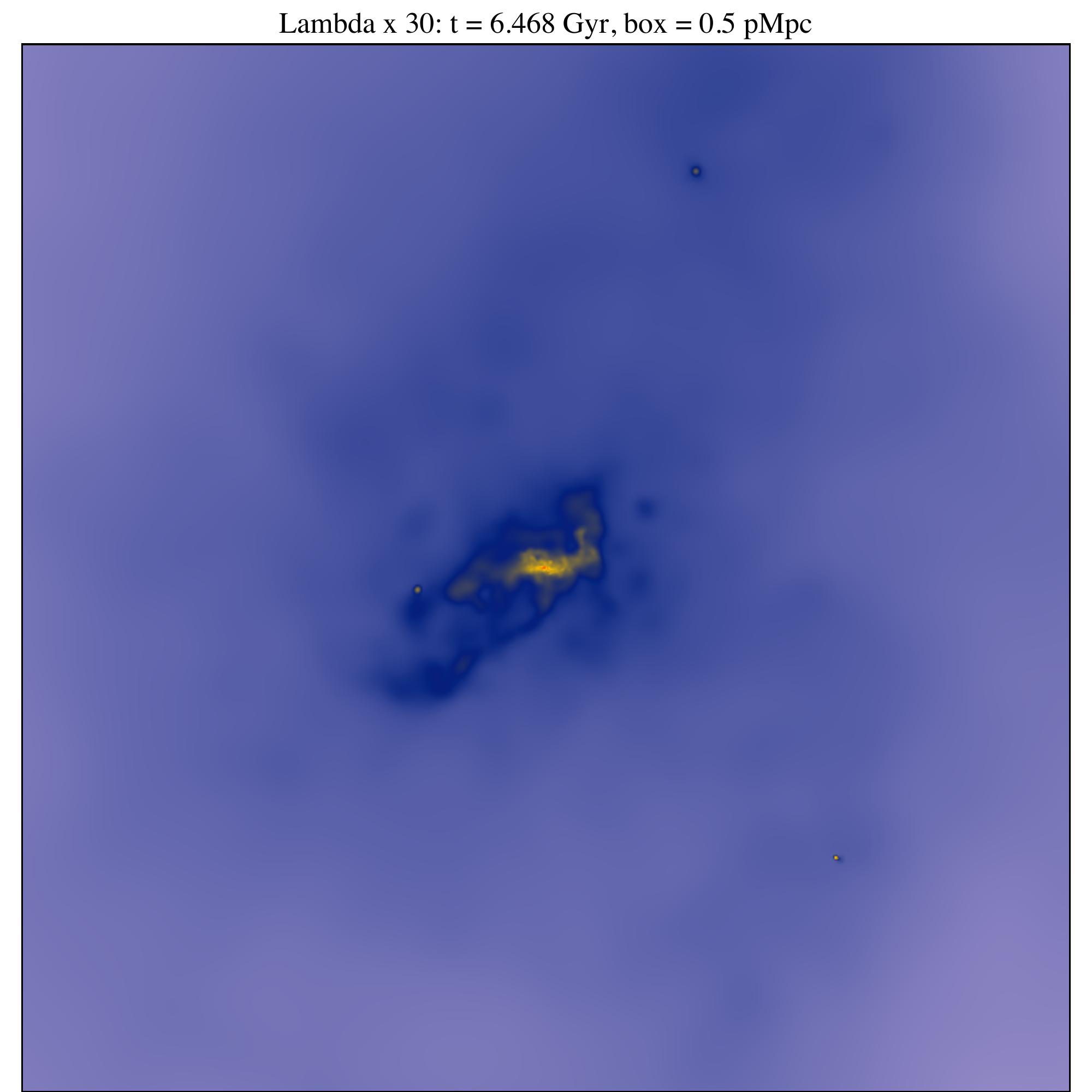}
	\end{minipage}
    \begin{minipage}{0.4\textwidth}
		\includegraphics[width=\textwidth]{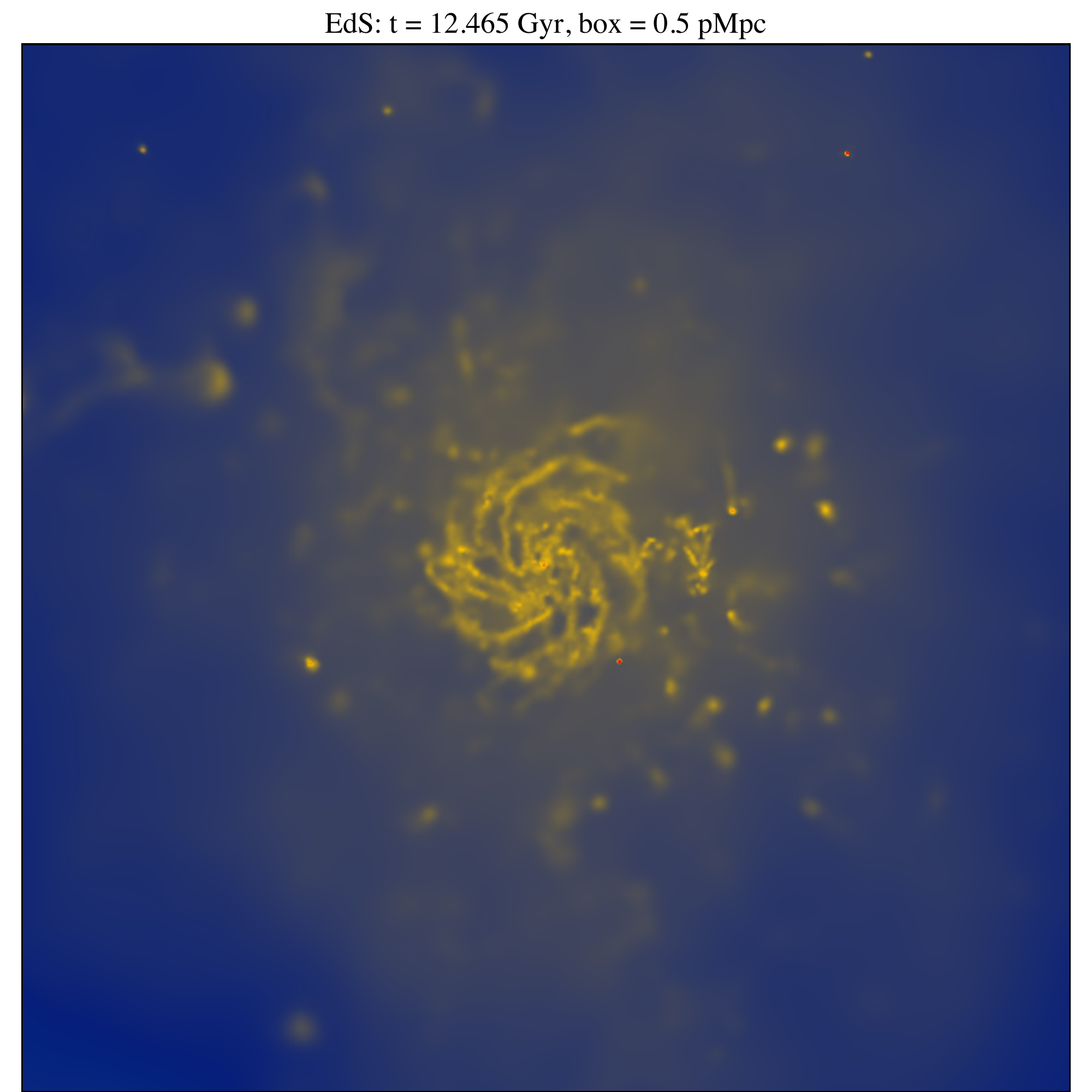}
	\end{minipage}
	\begin{minipage}{0.4\textwidth}
		\includegraphics[width=\textwidth]{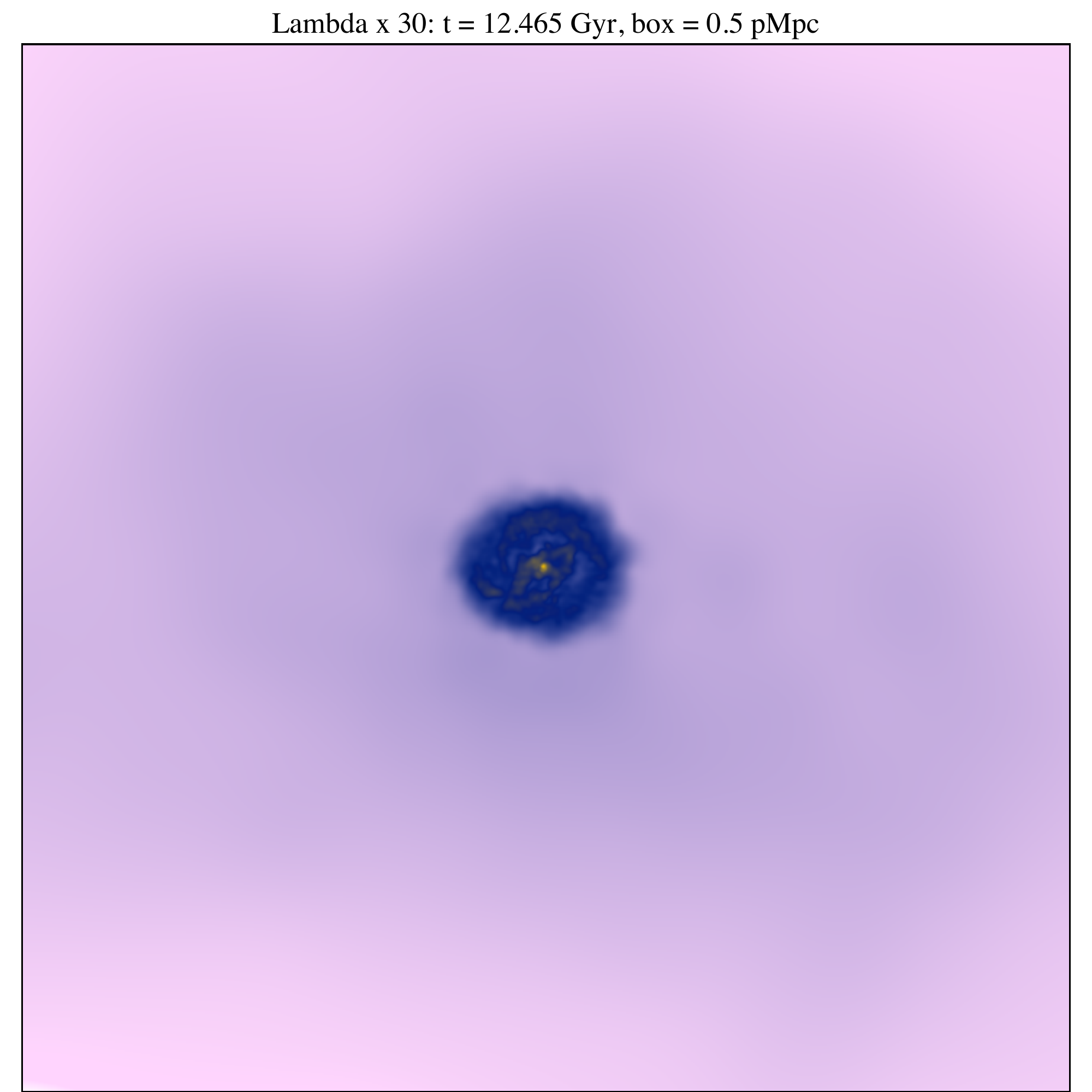}
	\end{minipage}
			\caption{The evolution of the projected proper gas density in a region of fixed proper size (0.5 Mpc) around the central galaxies in the panels in Figure \ref{fig:halo_prop}. The proper time is shown above each panel. \emph{Left:} an EdS universe ($\Lambda = 0$);  \emph{Right:} a $\Lambda_0 \times 30$ universe.} \label{fig:halo_prop}
\end{figure*}

To highlight the difference between the final states of the galaxies at the centre of the halo, Figure \ref{fig:halo_prop} shows a region of constant proper size (0.5 Mpc) around the central galaxy in the regions shown in Figure \ref{fig:halo_prop}. The colour scaling in all four panels is held constant.

The top two panels show this region of the universe at cosmic time $t = 6.5$ Gyr. Both show a galaxy in formation, being fed by streams of gas. But already we can see that the EdS galaxy (\emph{left}) is larger, and is surrounded by a much higher density circumgalactic medium. In the $\Lambda_0 \times 30$ universe (\emph{right}), the free fall time from the edge of the isolated region around the galaxy is a few Gyr, and so the halo accretes as much material as is available on this timescale. Accordingly, the total mass of the halo only grows by only $\sim$ 1\% in the 6 Gyr between the two snapshots shown in Figure \ref{fig:halo_prop}, to a final mass of $8 \ten{11} \Msol$. In this isolation, the gas collapses into one monolithic disk. In the same time in the EdS simulation, the halo has doubled in mass to $4 \ten{12} \Msol$ at $t = 12.5$ Gyr, and is still growing.

\section{Implications for multiverse models} \label{S:multiverse}

\subsection{The measure of the multiverse}
We can use our calculations to make predictions from multiverse models. Given a model that predicts an ensemble of universes with a distribution of values for the cosmological constant, we can ask what fraction of \emph{observers} will inhabit a universe with a particular value of $\Lambda$.

If the model in question predicts a \emph{finite} ensemble of universes, inhabited by a finite number of observers, then this calculation is straightforward. Scientific theories are tested by predicting observations, and so all observers are treated as of equal importance for the purposes of calculating the likelihood\footnote{We will ignore the complication of asking: what exactly counts as an observer? We cannot predict the occurrence of observers in sufficient detail to make any difference. That is, we might wonder whether any complex life form counts as an observer (an ant?), or whether we need to see evidence of communication (a dolphin?), or active observation of the universe at large (an astronomer?). Our model does not contain anything as detailed as ants, dolphins or astronomers, so we are unable to make such a fine distinction anyway. In any case, such a distinction is unlikely to bias our calculation toward any particular value of the cosmological constant.}. We thus use a counting metric to calculate the likelihood,
\begin{align}
p_\ro{obs}(\Lambda|MB) \df \Lambda = \frac{n_\ro{obs}([\Lambda,\Lambda + \dd \Lambda])}{n_\ro{obs}} ~,
\end{align}
where $M$ is the multiverse model, $B$ is any relevant background information (which should not give away any clues about the properties of the actual universe), $n_\ro{obs}([\Lambda,\Lambda + \dd \Lambda])$ is the number of observers (or observer-moments) that exist in a universe with cosmological constant in the range $[\Lambda,\Lambda + \dd \Lambda]$, and $n_\ro{obs}$ is the total number of observers in the multiverse.

To evaluate these quantities, we calculate (at least approximately) the rate at which observers are produced per unit time per unit comoving volume, for a given set of cosmic and fundamental parameters: $\df^2 n_\ro{obs} / \dd t \dd V$. So long as the universe has finite age, or if the rate at which observers are produced approaches zero quickly enough into the future, then the integral over cosmic time of this rate will be finite. Then, the likelihood of the cosmological constant is,
\begin{align} \label{eq:pobsfinite}
p_\ro{obs}(\Lambda|MB) \df \Lambda = 
\frac{\int_0^{t_\ro{max}} V(t;\Lambda) ~ p_V(\Lambda|t) \frac{\df^2 n_\ro{obs}} {\df t \df V} \df t \df \Lambda}
{\iint_0^{t_\ro{max}} V(t;\Lambda) ~ p_V(\Lambda|t) \frac{\df^2 n_\ro{obs}} {\df t \df V} \df t \df \Lambda} ~,
\end{align}
where $t_\ro{max}$ is the maximum age of the universe (possibly infinite), $V(t;\Lambda)$ is the total comoving volume of the universe, $p_V(\Lambda|t) \df \Lambda$ is the fraction of the universe by comoving volume at time $t$ in which the value of the cosmological constant is in the range $[\Lambda,\Lambda + \dd \Lambda]$. The comoving volume depends on the arbitrary normalization of $a(t)$, but this cancels in the equation above.

However, most proposed multiverses are not finite. In eternally-inflating universes, for example, it is argued that not only does the multiverse consist of an infinite number of universes, but each universe is infinitely large \citep{1997PhRvD..55..548V,2001PhRvD..64d3511G,2003physics...2071K,2006JHEP...03..039F,2007JPhA...40.6811G,2009GReGr..41.1475E}. Thus, the number of universes with a given value of $\Lambda$, times the average number of observers in those universes, divided by the total number of observers in the multiverse, is $\infty \times \infty / \infty$.

\begin{figure*} \centering
    \begin{minipage}{0.33\textwidth}
		\includegraphics[width=\textwidth]{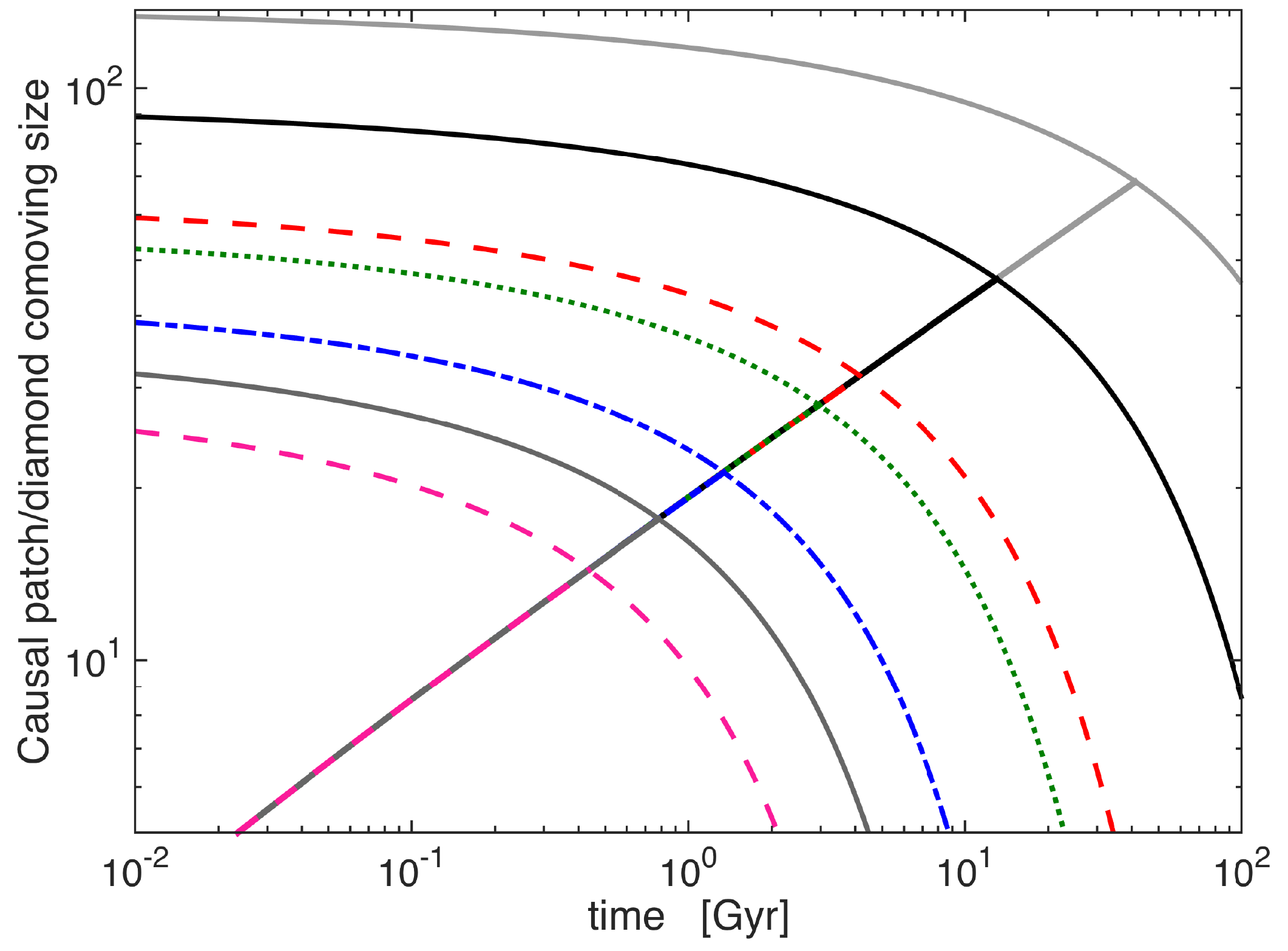}
	\end{minipage}
	\begin{minipage}{0.33\textwidth}
		\includegraphics[width=\textwidth]{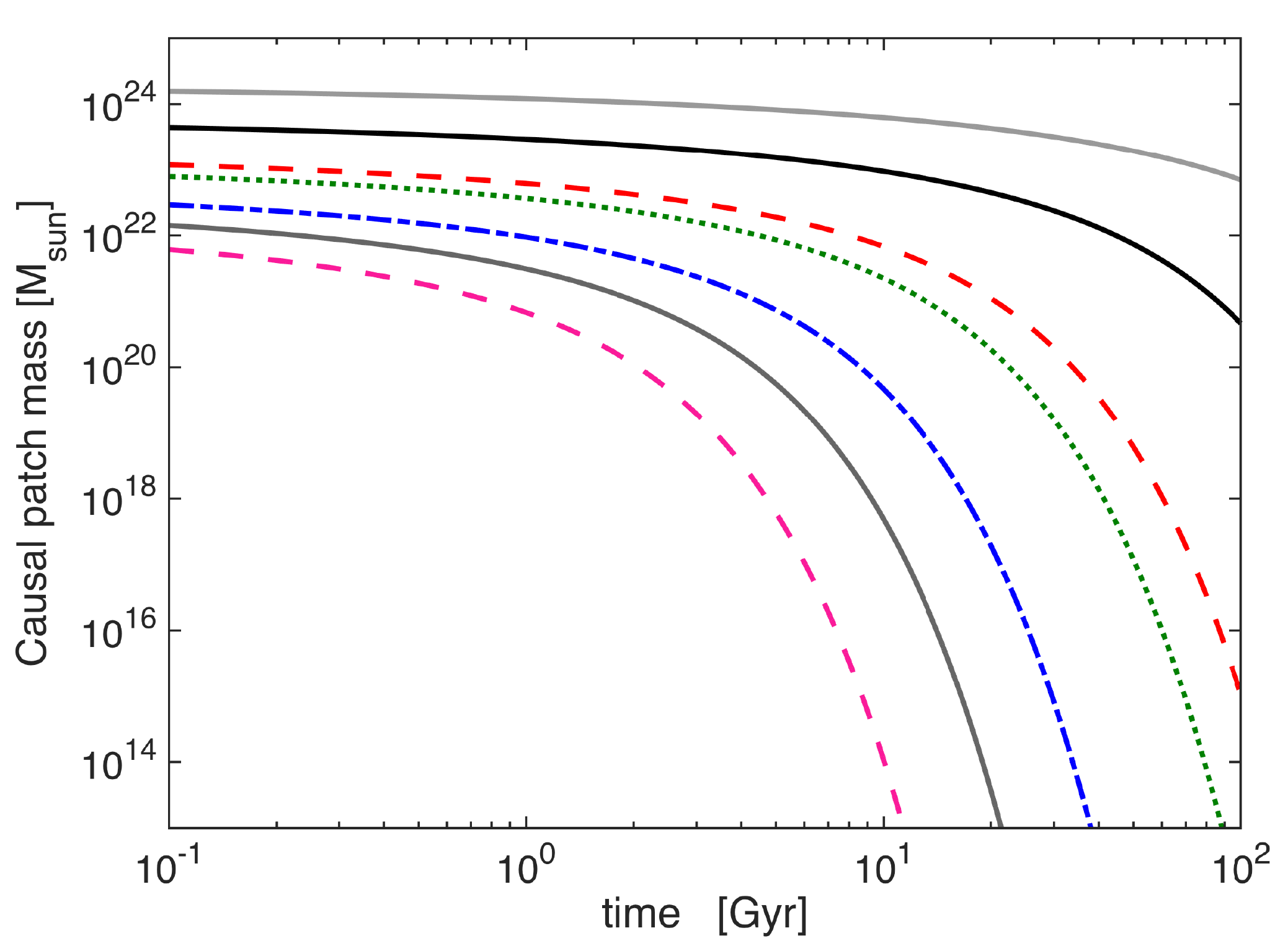}
	\end{minipage}
    \begin{minipage}{0.33\textwidth}
		\includegraphics[width=\textwidth]{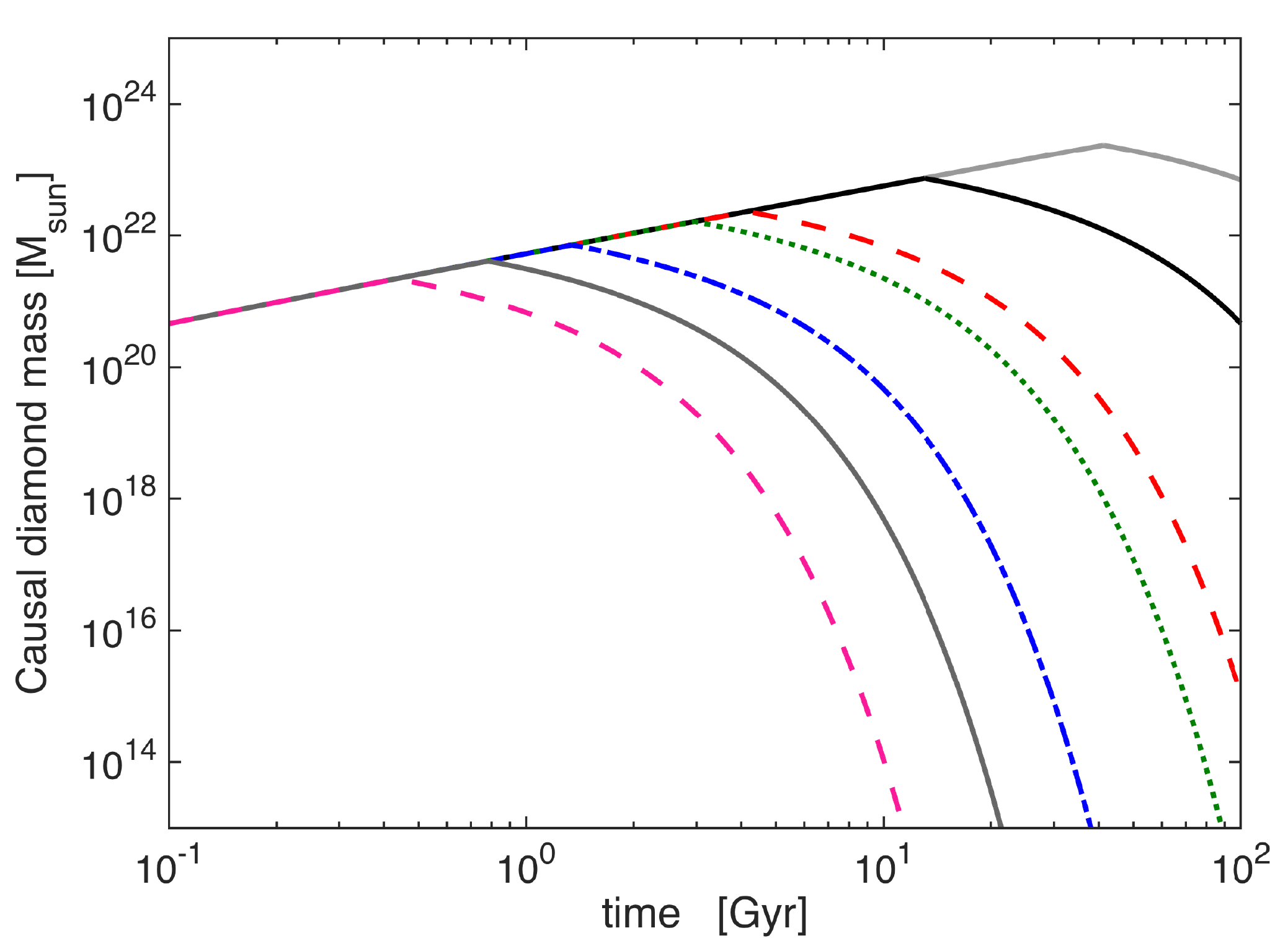}
	\end{minipage}
			\caption{\emph{Left:} Comoving causal patch and causal diamond vs proper time (Gyrs) for different values of the cosmological constant shown in the legend. The decreasing curves are the causal patch. Increasing (overlapping) curves are the quantity $\nu(t)$ from Equation \eqref{eq:diamond}; the causal diamond is the minimum of these two curves at a given time. The comoving distance is relative to the chosen normalisation of $a(t)$, as described in Section \ref{Ss:cosmparam}. Also shown are the physical mass inside the causal patch (\emph{middle}) and causal diamond (\emph{right}) as a function of cosmic time, which are independent of the normalisation.} \label{fig:chi_pd}
\end{figure*}

These infinities need to be managed with a \emph{measure}; see, among others, \citet{1995PhRvD..52.3365V,2006JCAP...01..017G,2007PhRvD..75l3501A,2007JHEP...01..092V,2007JPhA...40.6777V,2008PhRvD..77f3516G,2008JCAP...10..025P,2009PhRvD..79f3513B,deSimone2010,2011CQGra..28t4007F,2012PhRvD..85d5007B,2013JCAP...05..037G,2017arXiv170800449P}.
Simplistically, this measure can be thought of in two ways. Firstly, a multiverse model could motivate confining our attention to a finite region of the universe with volume $V(t;\Lambda)$ (as a function of time and $\Lambda$). Then, we can use the finite calculation for the likelihood (Equation \ref{eq:pobsfinite}). Secondly, the measure could specify the fraction of the volume of the universe in which cosmic parameters are in a given range, even though the total volume of the universe is infinite. This is used to weight the integral, effectively ``cancelling'' the infinite quantity $V(t;\Lambda)$ from the numerator and denominator of Equation \eqref{eq:pobsfinite}, which gives,
\begin{align} \label{eq:pobsmeasure}
p_\ro{obs}(\Lambda|MB) \df \Lambda = 
\frac{\int_0^{t_\ro{max}} p_V(\Lambda|t) \frac{\df^2 n_\ro{obs}} {\df t \df V} \df t \df \Lambda}
{\iint_0^{t_\ro{max}} p_V(\Lambda|t) \frac{\df^2 n_\ro{obs}} {\df t \df V} \df t \df \Lambda} ~.
\end{align}
Here, rather than focus on a specific multiverse model, we will consider three measures. Following \citet{1987PhRvL..59.2607W,1995MNRAS.274L..73E,2007MNRAS.379.1067P,2010PhRvD..81f3524B}, we note that nothing in fundamental physics picks out a value of the cosmological constant as privileged, including the value zero. This, in particular, rules out the use of a logarithmic prior. In the range of $\Lambda$ that we consider, which is very small compared to the Planck scale, we approximate the distribution as flat on a linear scale. The difference between the measures is the quantity with respect to which the distribution is flat.
\begin{itemize}
\item[1.] \emph{Mass-weighted:} there is a uniform probability that a given mass element in the universe will inhabit a region with a given value of the cosmological constant. Note that, for reasons discussed in Section \ref{S:global}, specifying that there is uniform probability with respect to \emph{comoving} volume is not sufficient, as there is no universal `today' relative to which we can define volume\footnote{Put another way, we are free to renormalise $a(t)$, but this normalisation could depend on $\Lambda$. This will not cancel out in equation \eqref{eq:pobsmeasure}, making the calculated probability arbitrary.}. We use the constraint of constant mass to define comoving volumes between universes.
\item[2.] \emph{Causal patch:} This measure was proposed to solve the quantum xeroxing paradox in black holes \citep{1993PhRvD..48.3743S,2006PhRvL..97s1302B,2006PhRvD..74j3516B}, treating the de-Sitter horizon in a universe with $\Lambda$ analogously to a black hole horizon. We ask: what is the volume of the region of the universe at time $t$ that can causally affect a given comoving world line in the future of $t$? The comoving extent of the region is,
\begin{equation} \label{eq:patch}
\chi_\ro{patch}(t) = \int_t^{t_\ro{max}} \frac{\df t}{a(t)} ~.
\end{equation}
Then, for the spatially flat universes that we consider here, the volume is $V = (4\pi/3) \chi^3(t)$, which goes in Equation \eqref{eq:pobsfinite}. Note that Equation \eqref{eq:patch} depends on the arbitrary normalisation of $a(t)$, but this is cancelled out when it is multiplied by the observer creation rate: $\df^2 n_\ro{obs} / \df t \df V$. The comoving size of the causal patch is shown in Figure \ref{fig:chi_pd} (\emph{left}, relative to the normalisation of $a(t)$ from Section \ref{Ss:cosmparam}). Also shown (\emph{middle}) is the physical mass contained within the causal patch (which is not relative to the normalisation of $a(t)$).
\item[3.] \emph{Causal diamond:} This measure is based on the principle that spacetime regions that are causally inaccessible should be disregarded \citep{2006PhRvL..97s1302B,2007PhRvD..76d3513B}. We consider a comoving world line in a universe, extending from the end of inflation (reheating at $t = t_\ro{rh}$) to the distant future. What is the volume (at time $t$) of the region of the universe that is enclosed by a photon that departs the world line at its beginning and returns at the end?  We can write,
\begin{align} \label{eq:diamond}
\chi_\ro{diamond}(t) &= \ro{min} \{\chi_\ro{patch}(t),\eta(t) \} ~, \\
\ro{where} \quad \eta(t) &= \int_{t_\ro{rh}}^{t} \frac{\df t}{a(t)}
\end{align}
As for the causal patch, the volume $V = (4\pi/3) \chi^3(t)$ is used in Equation \eqref{eq:pobsfinite}. The causal diamond is shown in Figure \ref{fig:chi_pd}. Also shown is the physical mass contained within the causal diamond (\emph{right}).
\end{itemize}

We stress, however, that the measure is not a ``degree of freedom'' in a multiverse model. It must not be inferred from or fit to observations, and the fact that a particular measure gives good agreement with observations is \emph{no} reason to prefer that measure. The reason is that any value of $\Lambda$ can be made practically certain with an appropriately jerry-rigged measure. If a model \emph{derives} its prediction from observations, then its predictions cannot then be tested by those same observations. A multiverse model is supposed to tell us about the global structure of the universe. There should not be any assumptions that need to be added ``on top'', because there are no physical facts left to specify, at least on relevant cosmological scales. The measure should follow naturally --- in some sense --- from the multiverse model\footnote{To put this another way, suppose a multiverse model specified the global structure of the universe in painstaking detail: the value of cosmic parameters and properties at every place and time. What would it mean to apply two different measures to this model, to derive two different predictions? How could all the physical facts be the same, and yet the predictions of the model be different in the two cases? What is the measure \emph{about}, if not the universe? Is it just our own subjective opinion? In that case, you can save yourself all the bother of calculating probabilities by having an opinion about your multiverse model directly.}.

\subsection{Models of observers}
\begin{figure} \centering
		\includegraphics[width=0.45\textwidth]{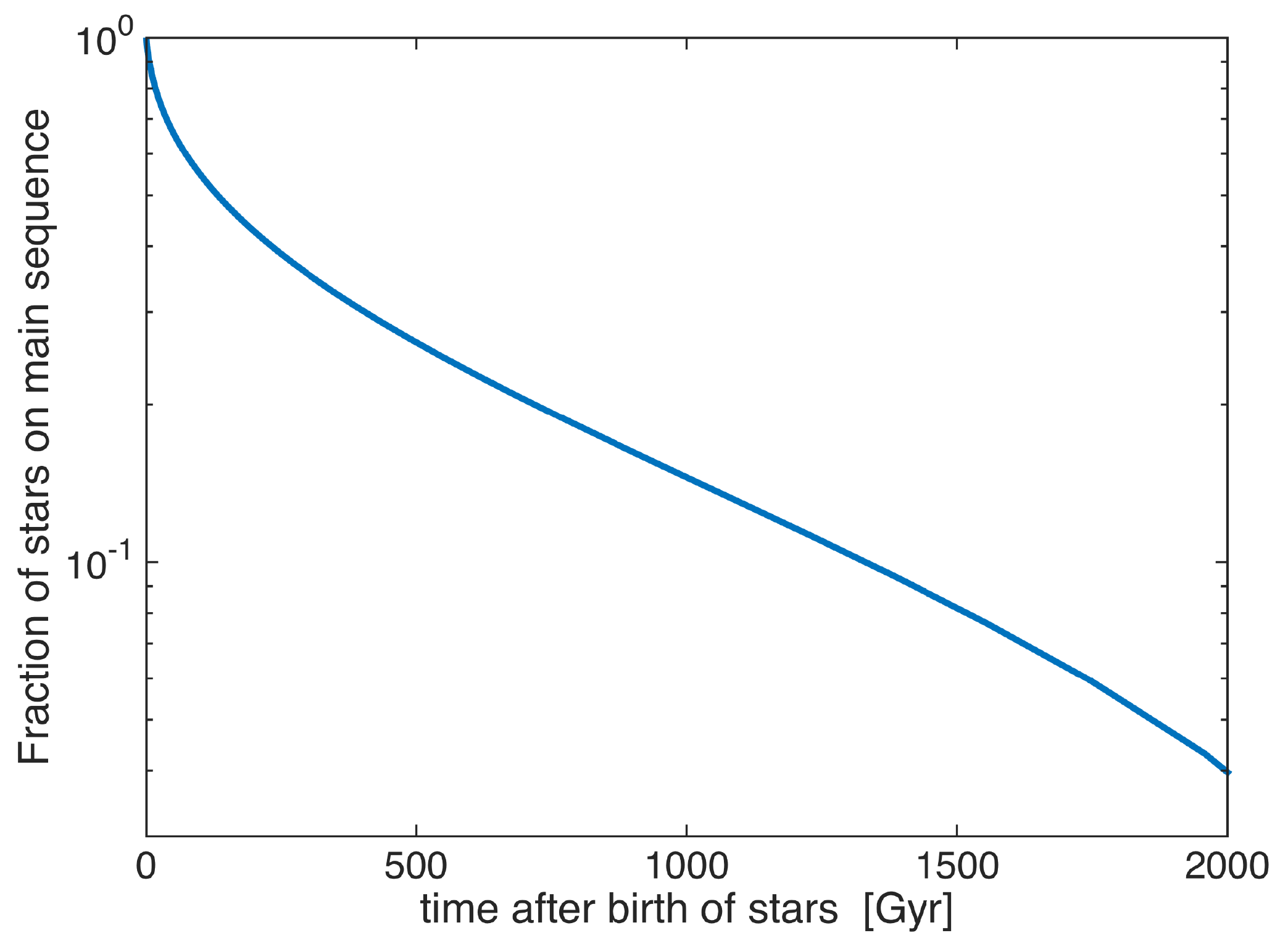}
		\caption{The fraction of stars by number that are still on the main sequence of their evolution after time $\Delta t$. We assume a \citet{Chabrier2003} IMF, and a relationship between mass and main-sequence lifetime from \citet{2008JCAP...08..010A}, normalised to $t_\ro{ms} = 10$ Gyr at Solar mass. To be consistent with the IMF used to  calibrate the \eagle simulations, we choose the minimum and maximum stellar masses to be: $M_\ro{min} = 0.1$ \Msol and $M_\ro{max} = 100$ \Msol.} \label{fig:frac_ms}
\end{figure}

We need to connect the presence of observers to local conditions in our simulations. This will, inevitably, be a combination of approximation and guesswork. Note that any constant factor in the observer creation rate will cancel in Equations \ref{eq:pobsfinite} and \ref{eq:pobsmeasure}, so an absolute rate is not required. We consider three models of observers, linked to the production of energy and chemical elements. 
\begin{itemize}
\item[1.] \emph{Star formation + fixed delay:} Following \citet{2010PhRvD..81f3524B}, we consider a model in which observers follow the formation of a star with a fixed time delay of 5 Gyr. We also considered a time delay of 10 Gyr, but it made minimal difference to our conclusions. This is inspired by the time taken for intelligent life to form on Earth after the birth of the Sun.
\item[2.] \emph{Star formation + main-sequence lifetime:} As first argued by \citet[][see also \citealt{1986acp..book.....B}]{Carter1983}, if the formation of life is extremely improbable --- that is, if the average timescale for its formation is much longer than the lifetimes of stars ---  then it will form at the last available moment, so to speak. Most stars will host lifeless planets, but where life forms it will do so at a time that is of order of the main-sequence lifetime of the star. As a first approximation, we assume that there is a constant probability per unit time of life forming around stars of all masses. The observer creation rate for each star population that forms is proportional to the fraction of stars (by number) that are still on the main sequence after time $\Delta t$,
\begin{equation}
f_\ro{ms}(\Delta t) = \frac{\int\theta(t_\ro{ms} (M) - \Delta t) ~ \xi(M) \df M}
{\int  \xi(M) \df M} ~,
\end{equation}
where $\xi(M)$ is the stellar initial mass function (IMF), $t_\ro{ms} (M)$ is the main-sequence lifetime of a star of mass $M$, the the limits of the integral are the minimum and maximum stellar masses, and $\theta(x)$ is the Heaviside step function, so that only those stars whose main-sequence lifetimes are longer than the time since the population was born contribute. We use the \citet{Chabrier2003} initial mass function, and a simple relationship between mass and main-sequence lifetime drawn from the analytic model of \citet{2008JCAP...08..010A} normalised to $t_\ro{ms} = 10$ Gyr at Solar mass; this broadly consistent with \citet{Portinari1998}. Of particular importance are the maximum and minimum stellar masses. To be consistent with the IMF used to  calibrate the \eagle simulations, we choose the minimum and maximum stellar masses to be: $M_\ro{min} = 0.1$ \Msol and $M_\ro{max} = 100$ \Msol. The resulting main-sequence fraction is shown in Figure \ref{fig:frac_ms}.

Folding in the star formation (birth) rate density ($\dot{\rho}_\ro{star}$), we calculate the global observer creation rate. A stellar population that formed at time $\Delta t$ before the present time $t$ provides a relative contribution of $f_\ro{ms}(\Delta t)$ to the observer creation rate,
\begin{equation}
\frac{\df^2 n_\ro{obs}} {\df t \df V}(t) \propto \int_0^t \dot{\rho}_\ro{star}(t') f_\ro{ms} (t - t') \df t' ~.
\end{equation}
Note that, since the time at which observers exist is irrelevant to the mass-weighted measure, the``Star formation + fixed delay'' and ``Star formation + main-sequence lifetime'' models give identical results. This is not the case for the causal patch and causal diamond measures --- a later observer at the same comoving position may be outside the patch/diamond, and so does not contribute to the integral in Equation \eqref{eq:pobsfinite}.
\item[3.] \emph{Star formation + metals:} The raw materials for life are the product of stellar-nucleosynthesis, and in particular metals that have been ejected from stars and returned to the interstellar medium. Planets, it is believed, form from the debris disks around newly-formed stars, and stars with higher metallicity are known to be more likely to have giant planets \citep{1997MNRAS.285..403G, 2005ApJ...622.1102F}. However, this result is less clear for smaller rocky planets \citep{2015ApJ...808..187B,2015AJ....149...14W}. There must, of course, be some metallicity dependence, since the probability of a rocky planet forming in a zero-metallicity debris disk is zero. We make the simple assumption that the probability of a rocky planet forming around a star is proportional to the metallicity of the star-forming gas, $Z_\ro{SF}$, so that the observer creation rate at time $t$ is proportional to the number of planets that exist around main-sequence stars,
\begin{equation}
\frac{\df^2 n_\ro{obs}} {\df t \df V}(t) \propto \int_0^t Z_\ro{SF}(t') ~ \dot{\rho}_\ro{star}(t') f_\ro{ms} (t - t') \df t' ~.
\end{equation}
where $Z_\ro{SF}(t')$ is the metallicity of star-forming gas at at time $t'$.
\end{itemize}

\subsection{Extrapolation} \label{Ss:extrap}

The integral in Equations \ref{eq:pobsfinite} and \ref{eq:pobsmeasure} is over all of cosmic time, but our simulations only extend to a finite time. They capture the initial burst of star formation in our universe, and so are converging thanks to the isolation of haloes by the acceleration of the expansion of space. There will, however, be a trickle of star formation into the future in our galaxies, which our simulations do not capture. Looking at the decline of star formation in the $\Lambda \ge \Lambda_0 \times 10$ simulations, we extrapolate our simulations using an exponential decrease in star formation (rate) efficiency (SFE) with time $[ \ro{SFE} = a ~ \ro{exp}(-b t) ]$, for constants $a$ and $b$ that are derived from the final few Gyr of the simulation. For our simulations, $Z_\ro{SF}(t)$ has converged; extrapolating by fitting an exponential makes only a negligible difference.

We also use the $\Lambda = 0$ simulation to calculate the relevant quantities for $0< \Lambda < \Lambda _0$. For $\Lambda < \Lambda_0 \times 0.1$, the time at which the universe begins to accelerate is greater than the limits of our simulation, at which time the star formation rate efficiency has peaked and is declining. By using the $\Lambda = 0$ simulation, there will be no difference in the observer model, but there will be a difference in the causal patch and causal diamond because these depend on $\Lambda$. Note that the both of these measures diverge for $\Lambda = 0$, so we only consider universes with $\Lambda > 0$.

As noted in Section \ref{Ss:sims}, in the far future of our simulations, a variety of very slow (on Gyr timescales) processes may become relevant but are not captured by our simulations. In particular, in the distant future of the $\Lambda =0$ cosmology, the growth function grows without limit and it is likely that close to 100\% of the mass in the universe is found in galaxies. However, in an old, extremely diffuse galactic disk, baryons may be more likely to accrete directly onto dead stellar remnants and black holes than collapse into a fresh star. What is the long-term fate of the interstellar medium in our isolated galaxies? Further modelling may be able to derive the expected fraction of baryons that will form stars into the distant future, and in particular, stars that are likely to host planets. Here, in the absence of such a model, we will simply extrapolate the simulations.

\subsection{Predicting the cosmological constant}

\begin{figure*} \centering
    \begin{minipage}{0.33\textwidth}
		\includegraphics[width=\textwidth]{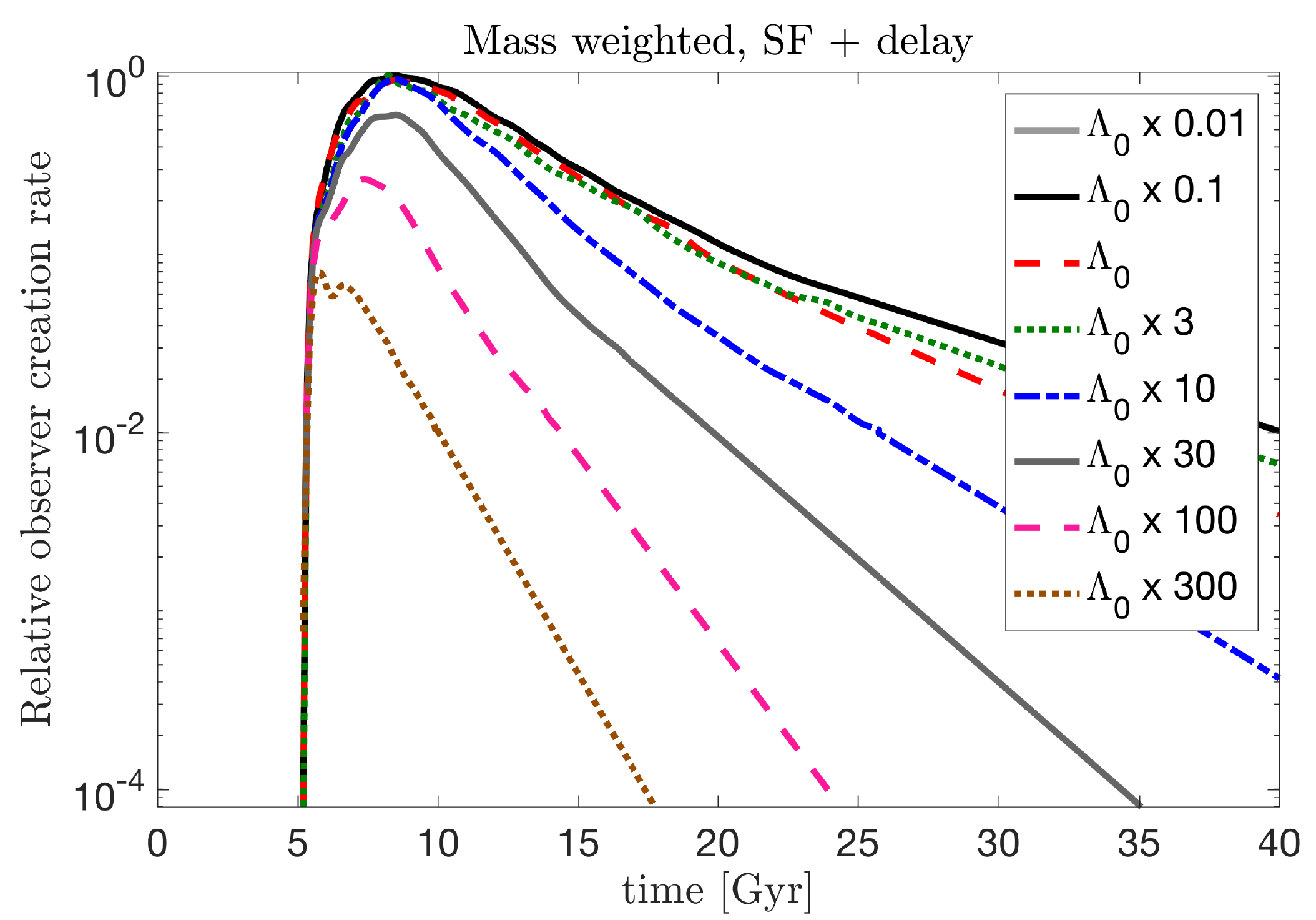}
	\end{minipage}
	\begin{minipage}{0.33\textwidth}
		\includegraphics[width=\textwidth]{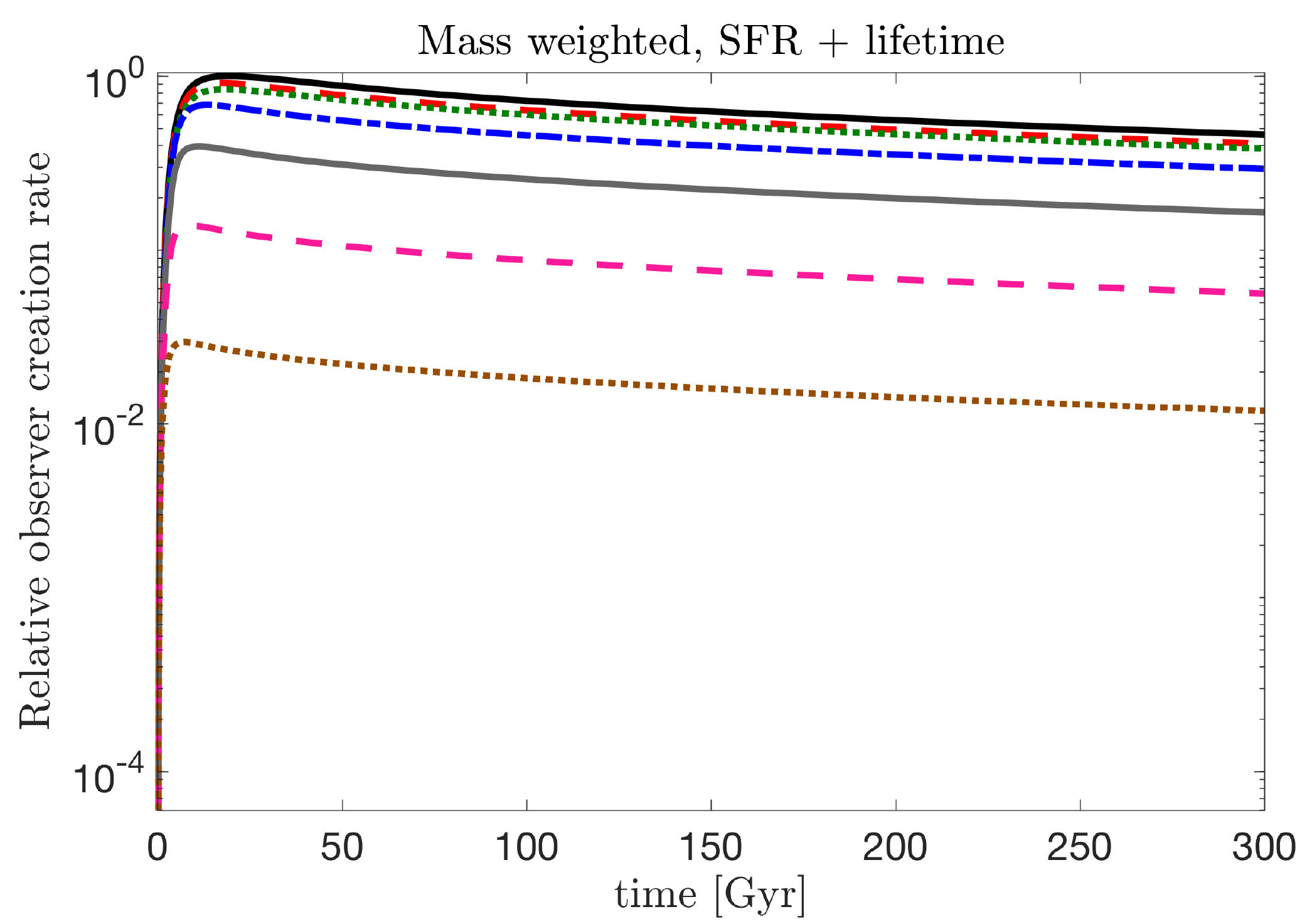}
	\end{minipage}
    \begin{minipage}{0.33\textwidth}
		\includegraphics[width=\textwidth]{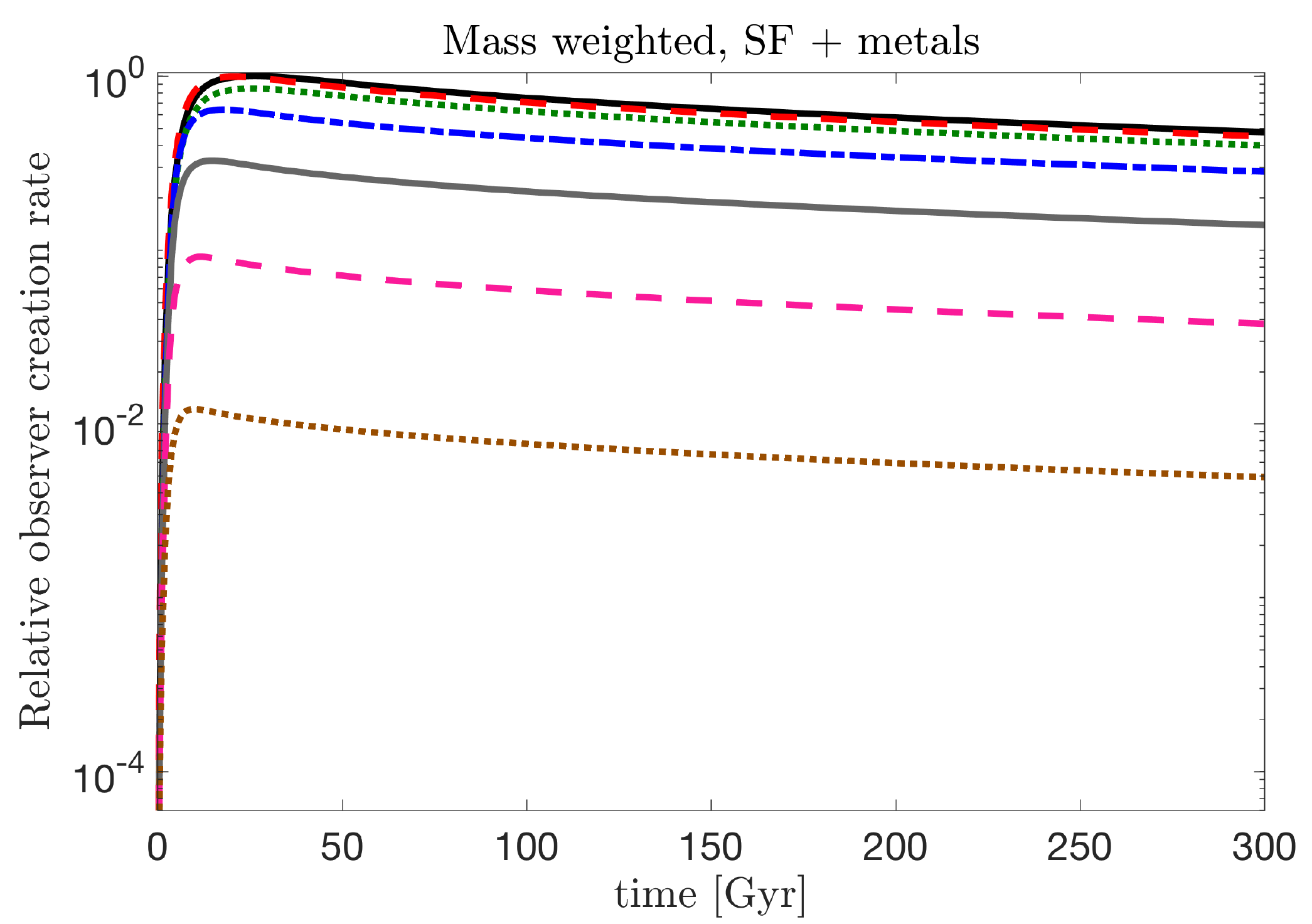}
	\end{minipage}
	    \begin{minipage}{0.33\textwidth}
		\includegraphics[width=\textwidth]{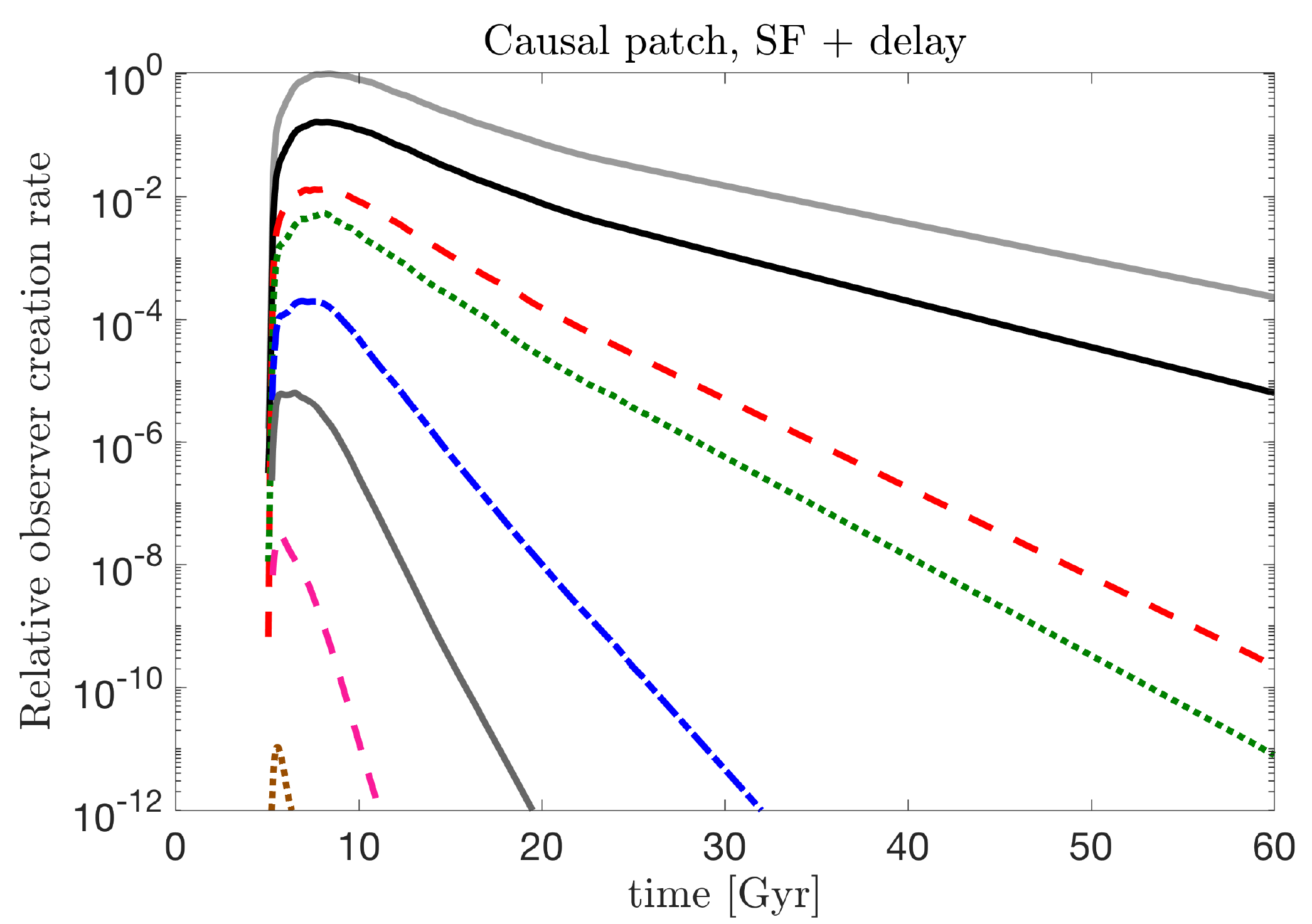}
	\end{minipage}
	\begin{minipage}{0.33\textwidth}
		\includegraphics[width=\textwidth]{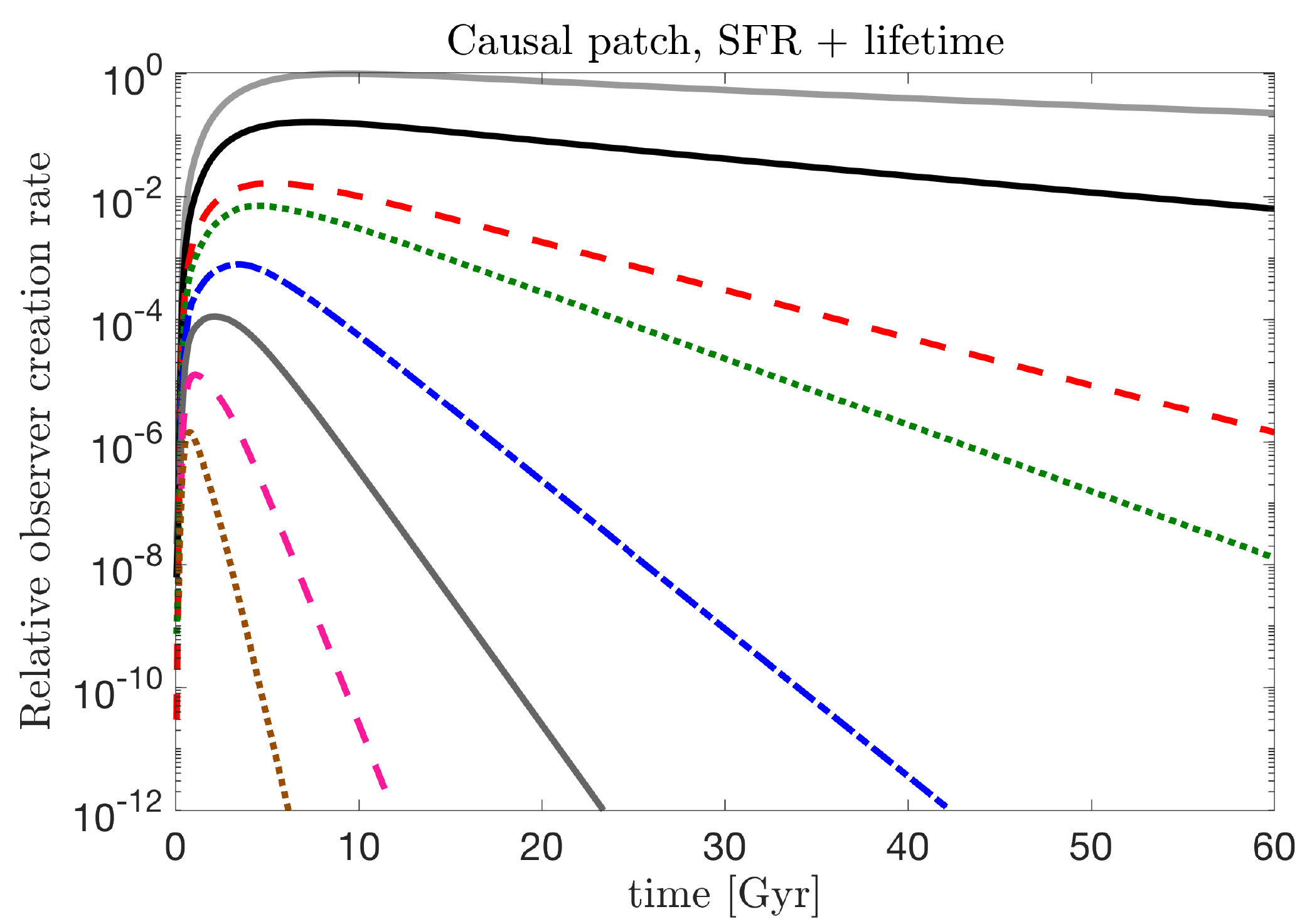}
	\end{minipage}
    \begin{minipage}{0.33\textwidth}
		\includegraphics[width=\textwidth]{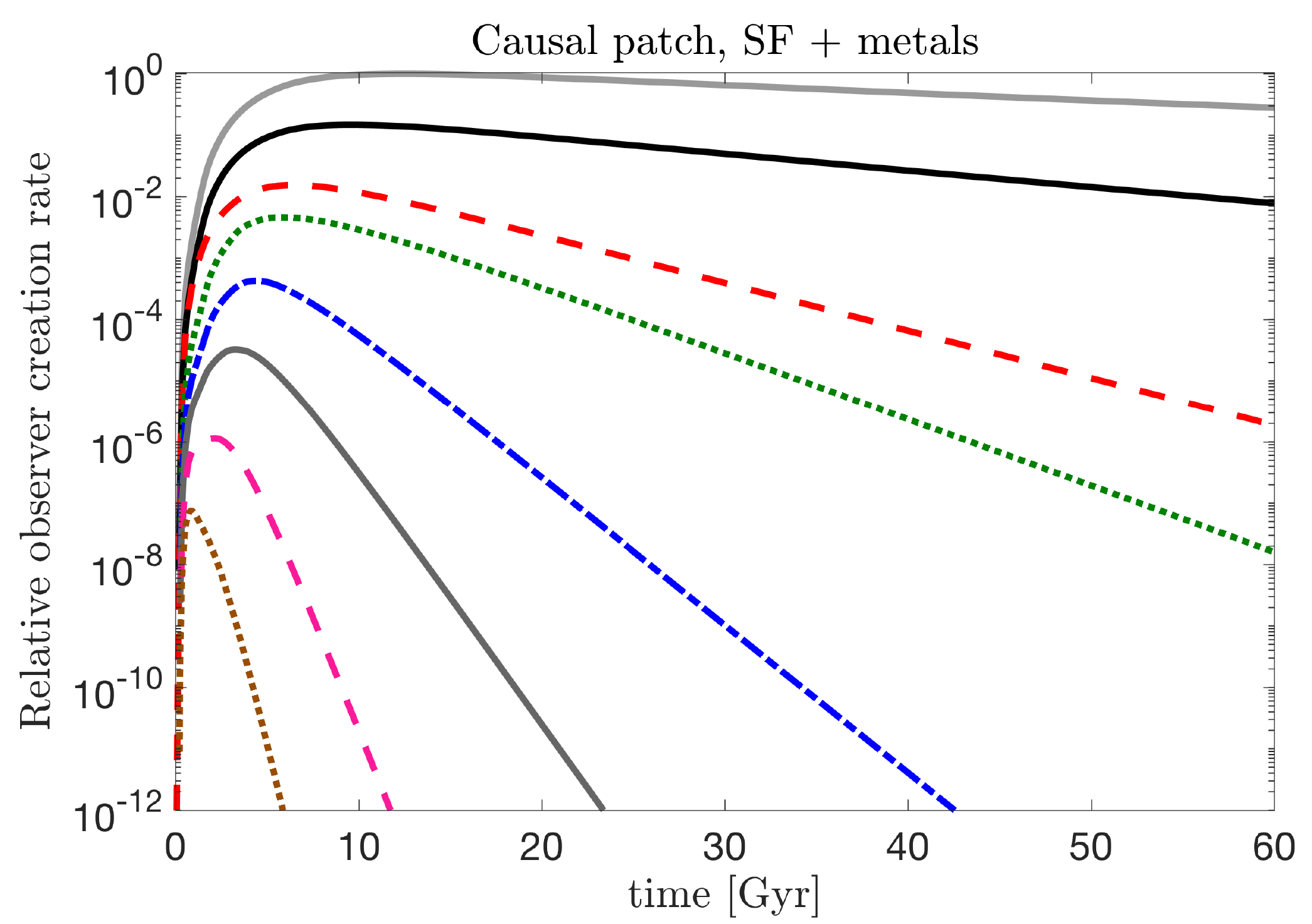}
	\end{minipage}
	    \begin{minipage}{0.33\textwidth}
		\includegraphics[width=\textwidth]{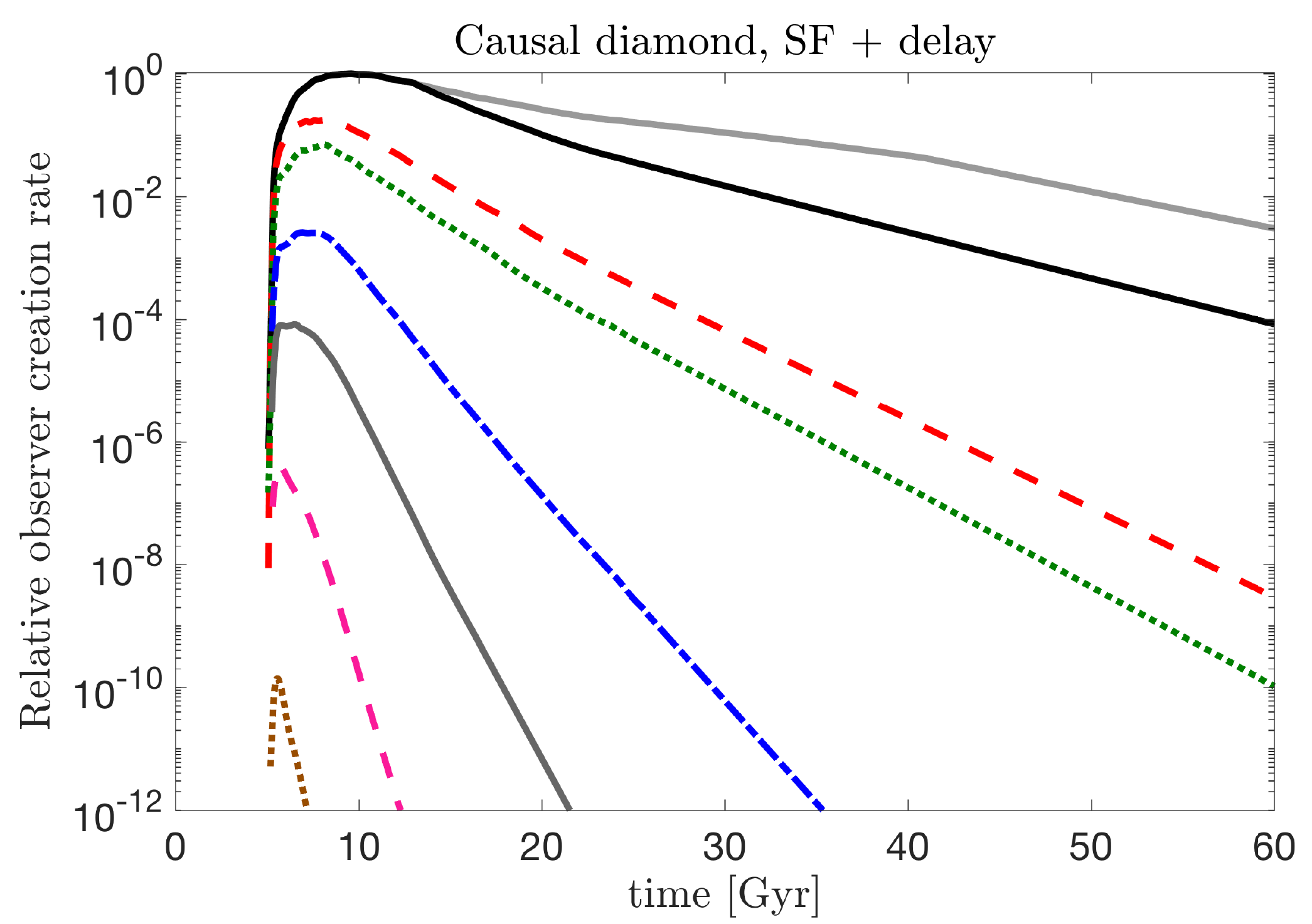}
	\end{minipage}
	\begin{minipage}{0.33\textwidth}
		\includegraphics[width=\textwidth]{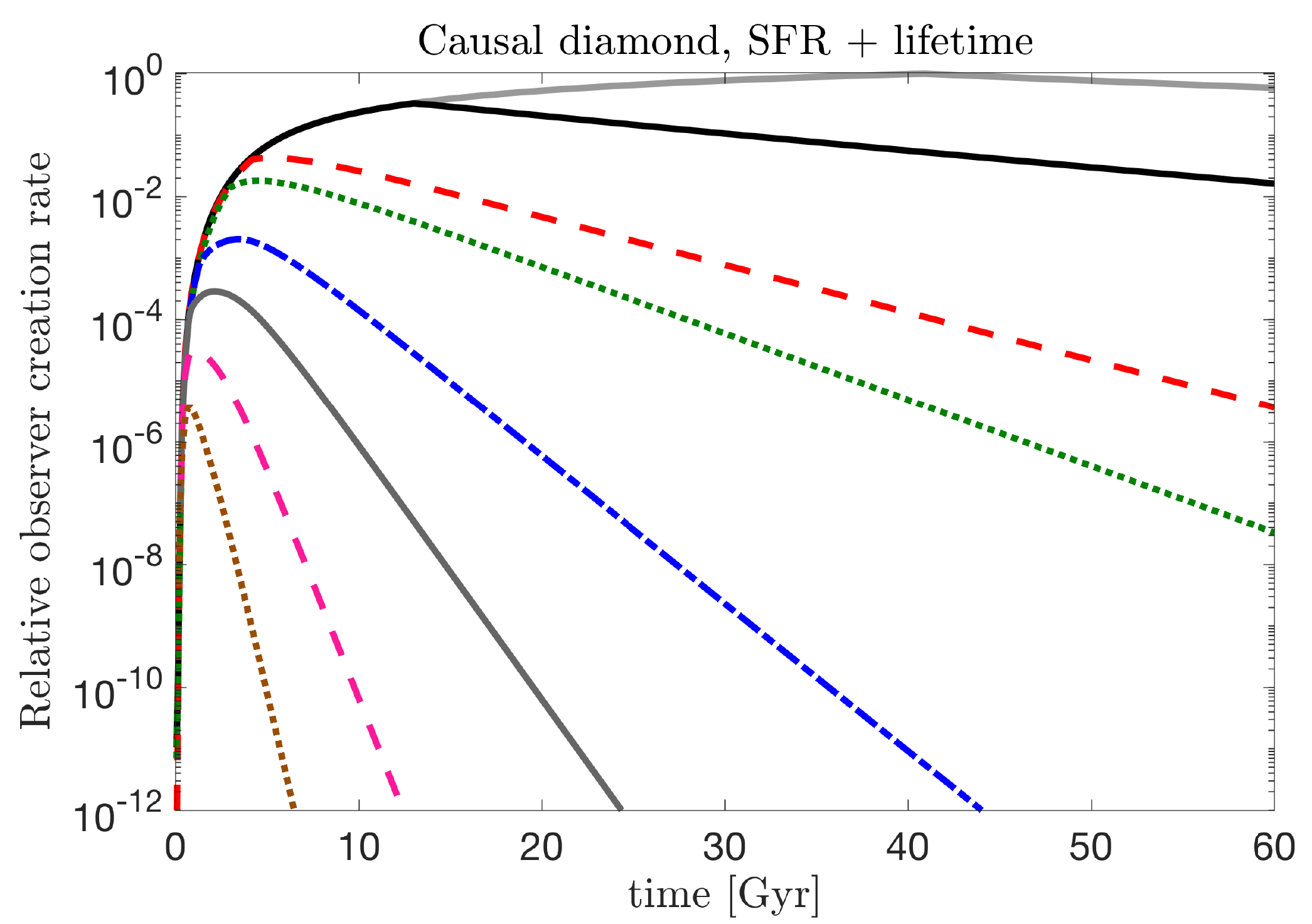}
	\end{minipage}
    \begin{minipage}{0.33\textwidth}
		\includegraphics[width=\textwidth]{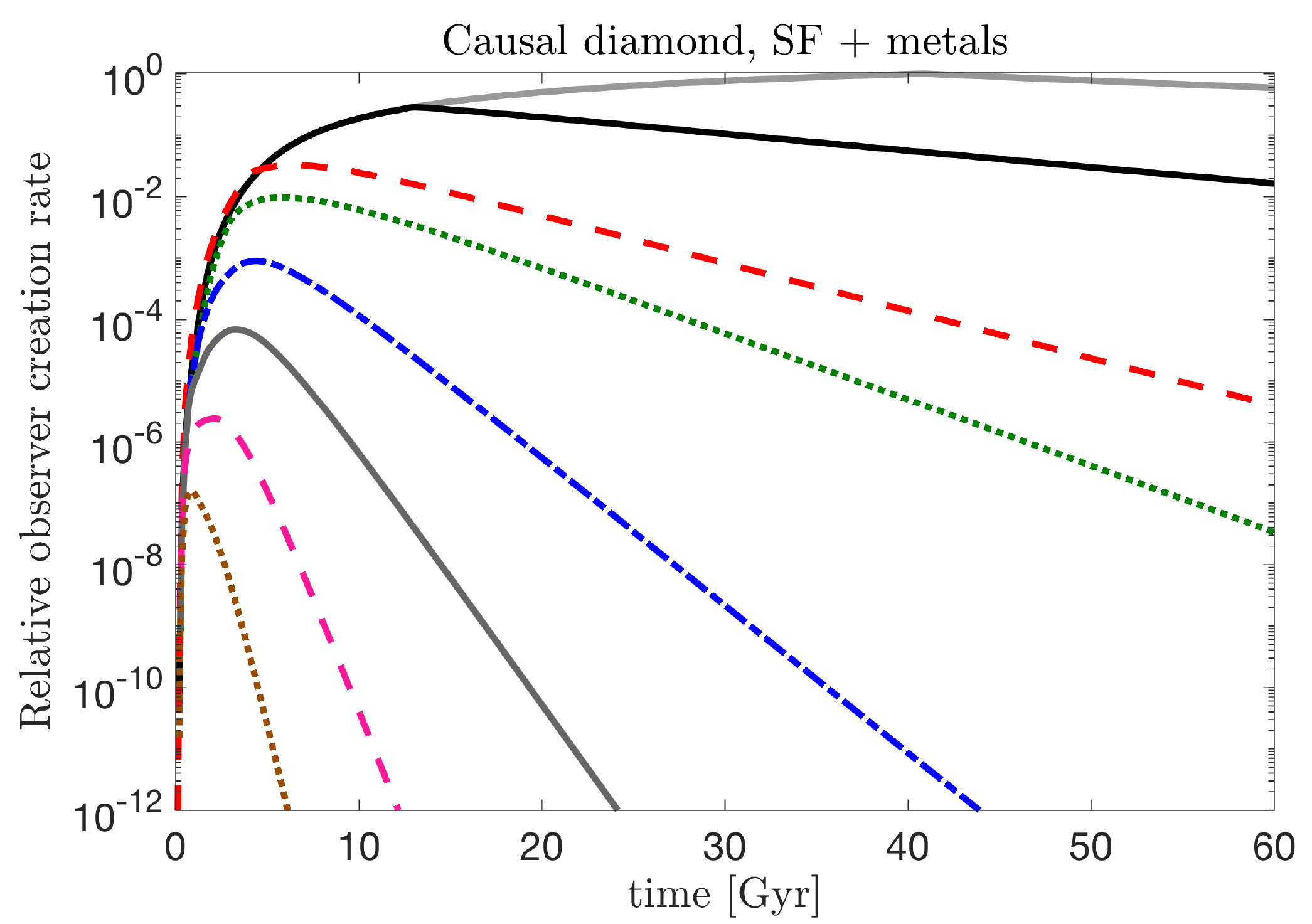}
	\end{minipage}
			\caption{The `relative observer creation rate', which is the integrand of Equation \ref{eq:pobsfinite} or \ref{eq:pobsmeasure}. The value of $\Lambda$ is shown in the legend in the top left panel. The first row shows the mass-weighted measure, derived directly from the star formation rate efficiencies in Figure \ref{fig:accr_star_cuml} and metal fraction in Figure \ref{fig:metgas_cuml}. The second row shows the causal patch measure, and the third row shows the causal diamond measure. The columns show the different observer models. The first column shows the star formation + fixed delay model, the second column shows the star formation + main-sequence lifetime model, and the third column shows the star formation + metals model.} \label{fig:dndt}
\end{figure*}

Figure \ref{fig:dndt} shows the integrand in the numerator of Equation \ref{eq:pobsfinite} or \ref{eq:pobsmeasure}, which combines the observer creation rate, the measure, and the chosen comoving volume (if needed). We will call this the `relative observer creation rate'. The value of $\Lambda$ is shown in the legend in the top left panel. The first row shows the mass-weighted measure, derived directly from the star formation rate efficiencies in Figure \ref{fig:accr_star_cuml} and metal fraction in Figure \ref{fig:metgas_cuml}. The second row shows the causal patch measure, and the third row shows the causal diamond measure.

The columns show the different observer models. The first column shows the star formation + fixed delay model, the second column shows the star formation + main-sequence lifetime model, and the third column shows the star formation + metals model.

The first row (mass weighted) shows most directly the effects of the different observer models. The grey $\Lambda_0 \times 0.01$ model is indistinguishable from the $\Lambda_0 \times 0.1$ model because they both use the results of the $\Lambda = 0$ simulation, as described above. In the second column, we can see the effect of folding in the main-sequence stellar lifetime. The decline in the observer creation rate follows the decline in the main-sequence fraction, as the initial burst of stars formed in the first 10 Gyr after the big bang grow old. Because of the abundance of small, long-lived stars, observers are created even at very late times in the universe \citep{2016JCAP...08..040L}. The addition of metal-weighting further diminishes observer creation in large $\Lambda$ universes, as they have fewer stars and fewer bound metals to make planets around their stars.

The second and third rows show the causal patch and causal diamond measures. Note that the $\Lambda_0 \times 0.01$ and $\Lambda_0 \times 0.1$ curves are distinguishable because of the difference in the comoving volume $V(t;\Lambda)$. For each observer model, these measures show similar trends. The smaller comoving volume at earlier times in the causal diamond moves the peak to slightly later times, but otherwise the two measures are very similar. The main effect of these measures is to decrease the relative observer creation rate exponentially once the expansion of the universe begins to accelerate. This somewhat cancels out the effect of the longer main-sequence lifetimes in the second and third observer models. 

\begin{table*}
\begin{center}
\begin{tabular}{|l|c|c|c|}
\hline
median $\Lambda / \Lambda_0$ $\pm$ 68\%  & Mass weighted & Causal patch & Causal diamond \\
\hline   
\rule{0pt}{3.5ex}  SF + delay   		&  59$^{+135}_{-49}$ & 0.34$^{+0.62}_{-0.3}$ ~~(0.37)		 & 0.65$^{+2.5}_{-0.52}$  ~~(0.68) \\
\rule{0pt}{3.5ex}  SF + lifetime   	&  59$^{+135}_{-49}$ & 0.089$^{+0.76}_{-0.08}$ ~~(0.095) & 0.25$^{+0.71}_{-0.24}$ ~~(0.28) \\
\rule{0pt}{3.5ex}  SF + metals   	&  45$^{+118}_{-37}$ & 0.07$^{+0.71}_{-0.066}$ ~~(0.072) & 0.17$^{+0.7}_{-0.16}$ ~~(0.18) \\
\hline
\end{tabular}
\caption{Median and ``one-sigma'' (68\%) probability limits of the cosmological constant for the three multiverse measures and three observer models. For the causal patch and causal diamond mesures, the value in brackets shows the median value of the cosmological constant using the observer creation rate (per unit mass) from the $\Lambda = 0$ simulation. This illustrates the effect of these measures.} \label{tab:expLam}
\end{center}
\end{table*}

\begin{figure*} \centering
    \begin{minipage}{0.33\textwidth}
		\includegraphics[width=\textwidth]{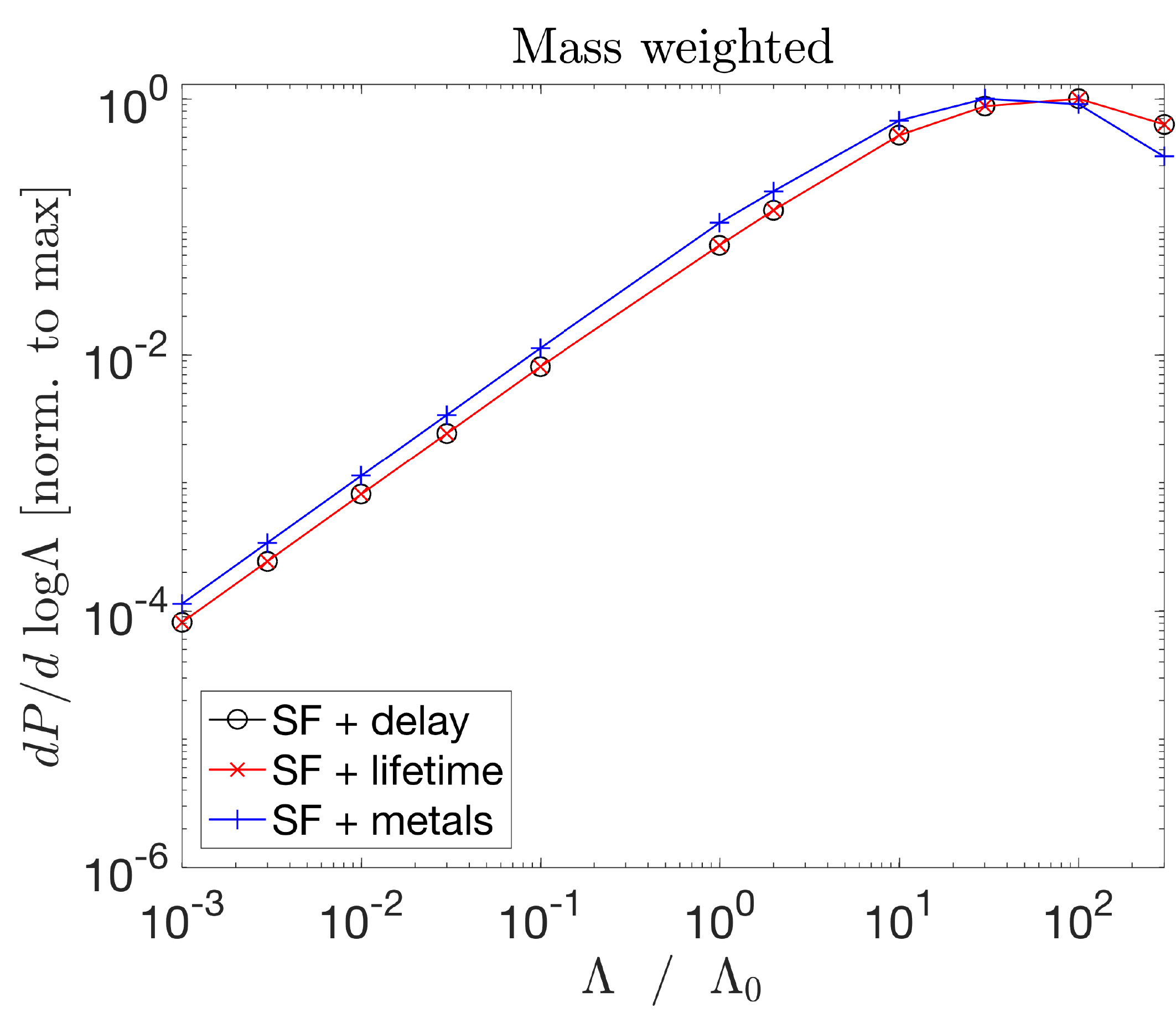}
	\end{minipage}
	\begin{minipage}{0.33\textwidth}
		\includegraphics[width=\textwidth]{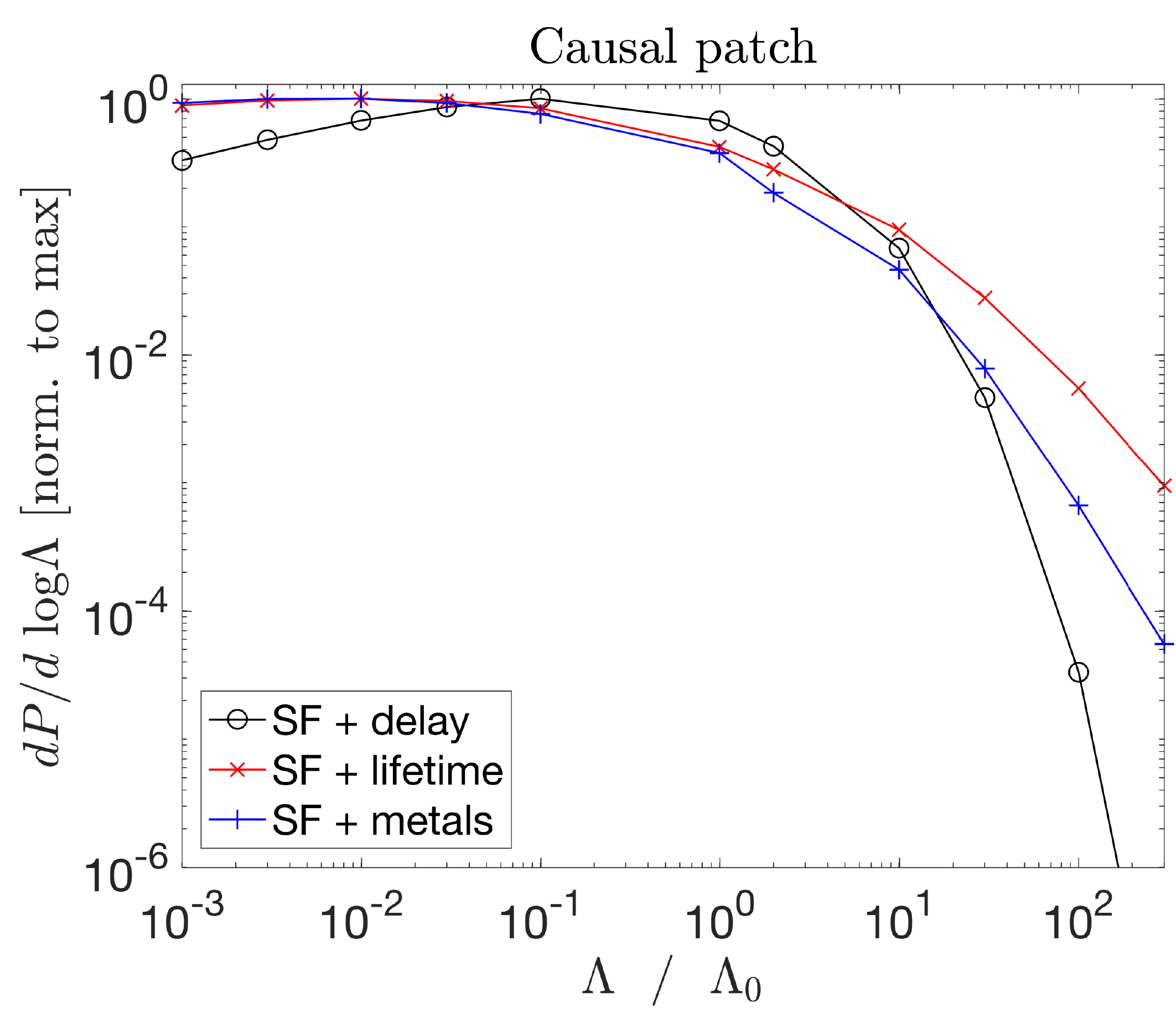}
	\end{minipage}
    \begin{minipage}{0.33\textwidth}
		\includegraphics[width=\textwidth]{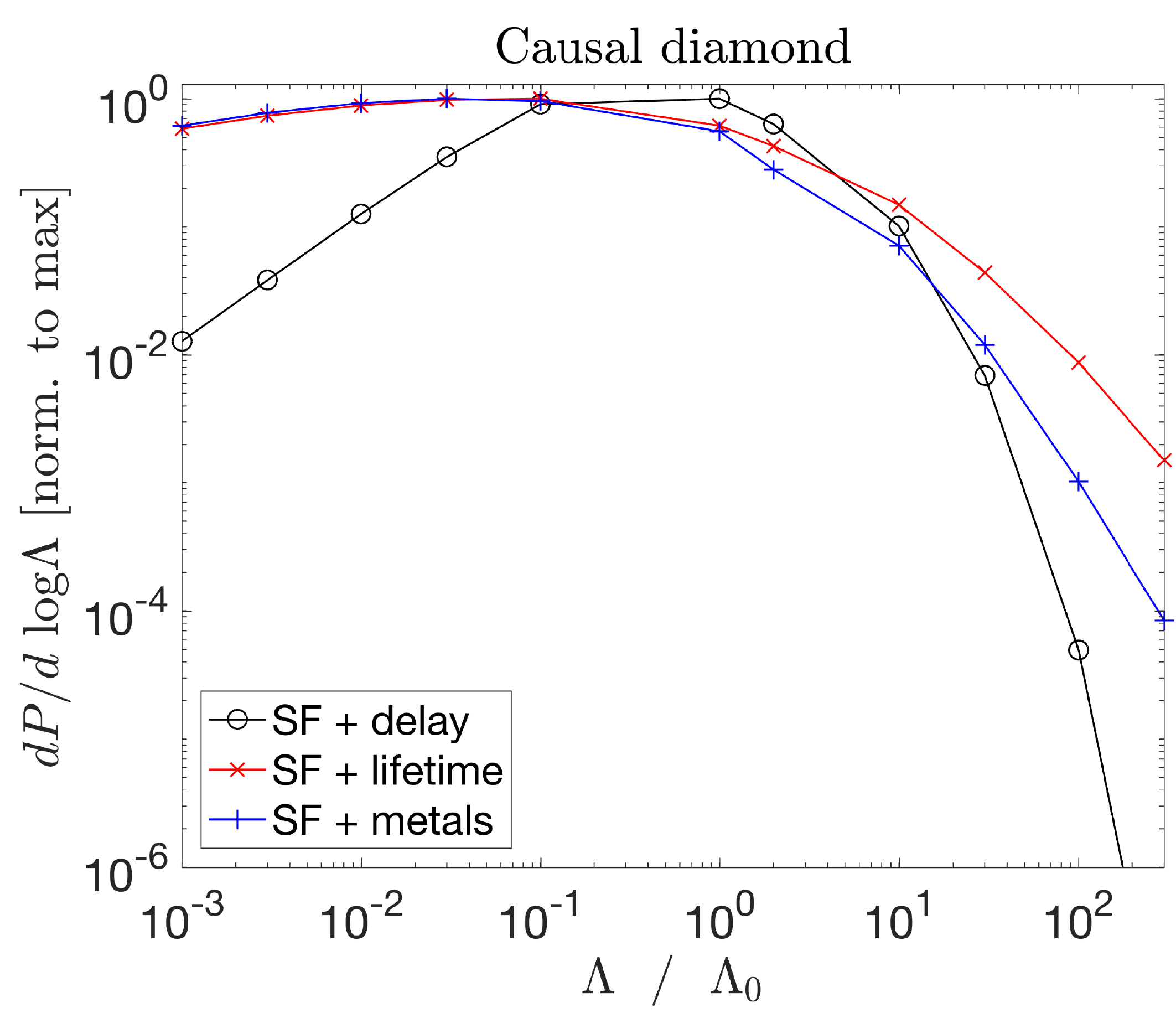}
	\end{minipage}
			\caption{The relative probabilities per unit log $\Lambda$ from Equation \ref{eq:pobsfinite} or \ref{eq:pobsmeasure}; each line integrates a panel of Figure \ref{fig:dndt} over cosmic time. The \emph{left} panel shows the mass weighted measure the \emph{middle} panel shows the causal patch measure, and the \emph{right} panel shows the causal diamond measure.} \label{fig:prob_lam}
\end{figure*}

Figure \ref{fig:prob_lam} shows the relative probabilities $p(\Lambda|MB)$ from Equation \ref{eq:pobsfinite} or \ref{eq:pobsmeasure}; each line integrates a panel of Figure \ref{fig:dndt} over cosmic time. We plot the probability per unit log $\Lambda$, and normalise by setting the maximum value to one, rather than integrating over the limited range of $\Lambda$. Note that in the \emph{left} plot (mass weighted measure), the `SF + delay' and `SF + lifetime' curves are indistinguishable --- the integral over cosmic time is not affected by the 5 Gyr delay, and cancels out the effect of $f_\ro{ms}$. For the range of $\Lambda$ we consider, the median and ``one-sigma'' (68\%) values are shown in Table \ref{tab:expLam}.

As we have noted previously, the decline in star formation in our universe after $t = 3.5$ Gyr is not due to the effect of the cosmological constant \citep{Salcido2017}. Universes without a cosmological constant show a similar decline. The initial burst of star-formation in the universe, then, is not dramatically affected by moderate increases in $\Lambda$. Only for $\Lambda \gtrsim \Lambda_0 \times 30$ do we see a significant effect on the total number of stars in the universe. Thus, in the mass-weighted measure, the probability distribution for $\Lambda$ is reasonably flat to large values ($\sim \Lambda_0 \times 30$). The median value in this case is 60 times larger than the observed value. Adding metal-weighting to the observer model increases the suppressing effects of $\Lambda$, but the median value is still $\sim$45 times larger than the observed value. While these distributions are broad, most of the probability is at large values of $\Lambda$. Table \ref{tab:probLam} shows the the probability that the cosmological constant observed by a typical observer is less than or equal to the value in our universe ($\Lambda_0$) for the three multiverse measures and three observer models. For the mass weighted measure, this probability is small (2\%).

Note that the results above are for the parameter range $\Lambda_0 \times 0.01 < \Lambda < \Lambda_0 \times 300$. The results for small values of $\Lambda$ have converged, but increasing the upper limit increases the median value. If we extrapolate the probability distribution to larger values of $\Lambda$, we find that the median value for $\Lambda$ is $\sim$200 $\Lambda_0$ for the mass-weighted measure.

For the causal patch and causal diamond measures, the fact that the comoving volume in the measure decreases with time suppresses large values of the cosmological constant, independently of the effect on the observer creation rate $\df^2 n_\ro{obs} / \df t \df V$. This leads to the $\gtrsim 50\%$ probabilities for small values of the cosmological constant (Table \ref{tab:probLam}). To illustrate the effect of the measure, we calculate the median value of the cosmological constant  using the observer creation rate (per unit mass) from the $\Lambda = 0$ simulation. The results are shown in brackets in Table \ref{tab:expLam}. These values are consistently larger than the actual median value, but only by a small factor. Thus, the causal patch and causal diamond measures are playing the dominant role in setting the expected value of the cosmological constant. The predicted value of $\Lambda$ is set by the time at which the star formation efficiency peaks in universes with small values of $\Lambda$, which is set by other cosmological and physical parameters. The decline of star formation efficiency with $\Lambda$ plays a secondary role.

Our results are broadly consistent with the analytic model of \citet{2010PhRvD..81f3524B}, who find that for fixed values of the primordial inhomogeneity $Q$ and spatial curvature, and for $\Lambda > 0$, the causal patch and causal diamond measures predict a value of $0.1 \lesssim \Lambda / \Lambda_0 \lesssim 10$, depending on the model for observers. As noted there, the suppression of structure formation by accelerating expansion is only important for a cosmological constant of order $\Lambda_0 \times 100$. Thus, the agreement between our calculations is due to the ``geometric'' effects of the causal patch and causal diamond measures; the astrophysics of galaxy formation does not prefer values of the cosmological constant less than $\Lambda_0 \times 100$.

\begin{table*}
\begin{center}
\begin{tabular}{|l|c|c|c|}
\hline
Prob $\Lambda \leq \Lambda_0$  & Mass weighted & Causal patch & Causal diamond \\
\hline   
SF + delay   &1.9\% & 86\% & 73\% \\
SF + lifetime   &1.9\% & 90\% & 86\% \\
SF + metals   &2.5\% & 93\% & 90\% \\
\hline
\end{tabular}
\caption{The probability that the cosmological constant observed by a typical observer is less than or equal to the value in our universe ($\Lambda_0$) for the three multiverse measures and three observer models. For the causal patch and causal diamond measures, these probabilities are greater than 50\%, but the value for the mass-weighted measure is small.} \label{tab:probLam}
\end{center}
\end{table*}

\section{Conclusions}

Models of the very early universe, including inflationary models, are argued to produce varying universe domains with different values of fundamental constants and cosmic parameters. In such models, predicting observations \emph{necessarily} involves understanding where observers are created in the multiverse. In particular, this anthropic approach has been used to predict the value of the cosmological constant.

Using the cosmological hydrodynamical simulation code from the \eagle collaboration, we have investigated the effect of the cosmological constant on the formation of galaxies and stars. This SPH code follows the gravitational collapse of matter in an expanding universe, incorporating sub-grid recipes for radiative cooling for 11 elements, star formation, stellar mass loss, energy feedback from star formation, gas accretion onto and mergers of supermassive black holes, and AGN feedback. We simulate universes with values of the cosmological constant ranging from $\Lambda = 0$ to $\Lambda_0 \times 300$, where $\Lambda_0$ is the values of the cosmological constant in our Universe. For larger values of the cosmological constant, the time at which the expansion of the universe begins to accelerate declines as $t_\Lambda \propto (\Lambda / \Lambda_0)^{-1/2}$.

Our Universe shows a peak in the global star formation rate at $t = 3.5$ Gyr, coming after the peak in the halo matter accretion rate at $t = 1$ Gyr. By the time the expansion of our Universe begins to accelerate (at $t = 7.6$ Gyr), the global halo mass accretion rate has dropped to about 10\% of its earlier maximum, and most of the mass that will ever accrete into haloes has already accreted. As a result, increases in $\Lambda$ of even an order of magnitude have a small effect on the star formation efficiency of the universe.

One interesting effect that affects the raw materials of life is stellar and AGN feedback. In our Universe, these processes slow star formation by sending baryons back into the outer parts of the halo and the local intergalactic medium. This material is largely recycled into the galaxy after $\sim$ 1 Gyr, and forms a later generation of stars. But in universes with $\Lambda \gtrsim \Lambda_0 \times 10$, much of this material is  lost to the intergalactic medium, carried away by the accelerating expansion of the universe rather than reaccreting. The net baryon accretion rate becomes negative as more material is lost to galactic winds than is accreted/reaccreted.

In universes with larger values of $\Lambda$, galaxies quickly become isolated from their cosmic surroundings. The familiar ecosystem of galaxies in our universe, which balance accretion, major and minor merging, galactic cannibalism, star formation, galactic winds, and reaccretion, is reduced to a closed box, as galaxies become island universes, surrounded by vacuum and isolated from the rest of the matter in the universes. They burn through their finite matter supply, forming stars at a decreasing rate.

We use our simulations to predict the observed value of the cosmological constant, given a measure of the mulitiverse. We considered three simple but plausible models for where we would expect observers to be created in our simulations, and three measures of the multiverse.

In the mass-weighted measure, with a uniform probability that a given mass element in the universe will inhabit a region with a given value of the cosmological constant, the predicted size of $\Lambda$ is determined by the decline in the star formation efficiency of the universe. For the reasons described above, this is relatively flat as a function of $\Lambda$, and so the predicted (median) value is $50-60$ times larger than the observed value. The probability of observing a value as small as our cosmological constant $\Lambda_0$ is $\sim 2$\%. In this case, an anthropic argument for value of $\Lambda$, while doing much better than the famous 120 orders-of-magnitude discrepancy from quantum field theory, is not a particularly successful prediction.

For the causal patch and causal diamond measures, which consider a subset of the universe that depends on $\Lambda$, the predicted value is within a factor of a few of the observed value. But, this has very little to do with the decline in the star formation efficiency (and so, presumably, observer creation rate) with $\Lambda$. It is a result of the rapid decrease in the size of the causal patch/diamond with increasing cosmological constant. 

\emph{We stress again: this is no reason to prefer the causal patch and causal diamond measures.} This is not an observational test of these measures. A specific multiverse model must justify its measure on its own terms, since the freedom to choose a measure is simply the freedom to choose predictions ad hoc.

We conclude that the impact of the cosmological constant on the formation of structure in the universe does not straightforwardly explain the small observed value of $\Lambda$. The prediction depends crucially on the measure. If the observer creation rate had been sufficiently sharply peaked at values near $\Lambda_0$, the measure would not much matter. But in fact, in the absence of a multiverse model that can convincingly justify a measure, it is not clear whether the anthropic prediction $\Lambda$ is successful. Future work will consider varying more cosmological and fundamental parameters, to shed more light on which kind of universe is to be expected from a multiverse.


\section*{Acknowledgements} 
We are grateful to all members of the Virgo Consortium and the \eagle collaboration who have contributed to the development of the codes and simulations used here, as well as to the people who helped with the analysis. This work was supported by the Science and Technology Facilities Council (grant number ST/P000541/1); European Research Council (grant number GA 267291 “Cosmiway”) and by the Netherlands Organisation for Scientific Research (NWO, VICS grant 639.043.409).

This work used the DiRAC Data Centric system at Durham University, operated by the Institute for Computational Cosmology on behalf of the STFC DiRAC HPC Facility (www.dirac.ac.uk). This equipment was funded by BIS National E-infrastructure capital grant ST/K00042X/1, STFC capital grants ST/H008519/1 and ST/K00087X/1, STFC DiRAC Operations grant  ST/K003267/1 and Durham University. DiRAC is part of the National E-Infrastructure. We acknowledge PRACE for awarding us access to the Curie machine based in France at TGCC, CEA, Bruyeres-le-Chatel.

This work used the NCI National Facility at the Australian National University and SwinStar on Green II at Swinburne University, thanks to a generous allocation from the Australian Supercomputer Time Allocation Committee. 

LAB is supported by a grant from the John Templeton Foundation. This publication was made possible through the support of a grant from the John Templeton Foundation. The opinions expressed in this publication are those of the author and do not necessarily reflect the views of the John Templeton Foundation. PJE is supported by the Australian Research Council Centre of Excellence for All Sky Astrophysics in 3 Dimensions (ASTRO 3D), through project number CE170100013. Jaime Salcido gratefully acknowledges the financial support from the Mexican Council for Science and Technology (CONACyT), fellow no. 218259. RAC is a Royal Society University Research Fellow.

We thank the anonymous referee for their helpful comments.

\bsp 
\label{lastpage}
\end{document}